\documentclass[a4paper,fleqn,usenatbib]{mnras}
\usepackage{newtxtext,newtxmath}

\usepackage[T1]{fontenc}
\usepackage{ae,aecompl}

\usepackage{graphicx}	
\usepackage{amsmath}	
\usepackage{amssymb}	
\usepackage{breqn}
\usepackage{subcaption}
\usepackage{threeparttable}
\usepackage{ltablex}
\usepackage{hyperref}
\usepackage{pdflscape}
\usepackage{soul}
\title[Cold Molecular Outflows in the Local Universe]{Cold Molecular Outflows in the Local Universe and Their Feedback Effect on Galaxies}

\author[A Fluetsch et al.]{A. Fluetsch$^{1,2}$, R. Maiolino$^{1,2}$, S. Carniani$^{1,2}$, A. Marconi$^{3,4}$, C.
Cicone$^{5}$, M. A. Bourne$^{2, 6}$, \newauthor
T. Costa$^{7}$,  A. C. Fabian$^{6}$,  W. Ishibashi$^{6}$,
G. Venturi$^{1,2,3,4}$ \\
 \\
$^{1}$University of Cambridge, Cavendish Laboratory, Cambridge CB3 0HE, UK\\
$^{2}$University of Cambridge, Kavli Institute for Cosmology, Cambridge CB3 0HE, UK\\
$^{3}$ Dipartimento di Fisica e Astronomia, Università degli Studi di Firenze, Firenze, Italy \\
$^{4}$ INAF Osservatorio Astrofisico di Arcetri, Firenze, Italy\\
$^{5}$ INAF Osservatorio Astronomico di Brera, Via Brera 28, I-20121, Milano, Italy\\
$^{6}$ Institute of Astronomy, Madingley Road, Cambridge CB3 0HA, UK\\
$^{7}$ Leiden Observatory, Leiden University, PO Box 9513, NL-2300 RA Leiden, the Netherlands\\
}

\date{Accepted XXX. Received YYY; in original form ZZZ}

\pubyear{2018}

\begin{document}
\label{firstpage}
\pagerange{\pageref{firstpage}--\pageref{lastpage}}
\maketitle

\begin{abstract}
We study molecular outflows in a sample of 45 local galaxies, both star forming and AGN,
primarily by using CO data from the ALMA archive and from the literature. For a subsample we also compare the 
molecular outflow with the ionized and neutral atomic phases.
We infer an empirical analytical function relating the outflow rate simultaneously to the SFR,
$L_{\rm AGN}$, and galaxy stellar mass; this relation is much tighter than the relations with the
individual quantities.
The outflow kinetic power shows a larger scatter than in previous, more biased studies,
spanning from 0.1 to 5~per cent of  $L_{\rm AGN}$, while the momentum rate ranges from 1 to 30 times $L_{\rm AGN}/c$, indicating that
these outflows can be both energy-driven, but with a broad range of coupling efficiencies with the ISM, and
radiation pressure-driven.
For about 10~per cent of the objects the outflow energetics significantly exceed the maximum theoretical values;
we interpret these as ``fossil outflows'' resulting from activity of a past strong AGN, which has now faded.
We estimate that, in the stellar mass range probed here ($>$ 10$^{10}~\rm M_{\odot}$),
less than 5~per cent of the outflowing gas escapes the galaxy. The molecular gas depletion time associated
with the outflow can be as short as a few million years in powerful AGN, however, the total gas (H$_2$+HI)
depletion times are much longer. Altogether, our findings
suggest that even AGN-driven outflows might be relatively ineffective in clearing galaxies of their entire gas content,
although they are likely capable of clearing and quenching the central region.

\end{abstract}

\begin{keywords}
galaxies: active --- galaxies: evolution --- quasars: general --- galaxies: ISM --- galaxies: star formation
\end{keywords}



\section{Introduction}

Galactic outflows driven either by active galactic nuclei (AGN) or starbursts may be capable of expelling ionized, 
atomic neutral and molecular gas from galaxies and thereby regulate or even shut down star formation. As a consequence, outflows may provide
the (negative) feedback effect that is invoked to explain several key observable properties of galaxies. For instance,
star formation suppression from AGN-driven outflows is thought to play a key role in accounting for
the the local population of massive passive galaxies and the lack of over-massive galaxies \citep[e.g.][]{Scannapieco2004,
doi:10.1111/j.1365-2966.2005.09238.x, doi:10.1111/j.1365-2966.2007.12153.x, doi:10.1093/mnras/sts243, doi:10.1093/mnras/stu1536, Beckmann2017}.
Furthermore, this process may offer an explanation for the tight correlations
between the masses of the central supermassive black holes (SMBHs) and the stellar masses or velocity dispersions of their host galaxy bulges 
\citep[e.g][]{Fabian2012,doi:10.1146/annurev-astro-082214-122316}. On the other hand, starburst-driven outflows are thought
to play a key role in self-regulating star formation in low-mass galaxies and also to be responsible for the
chemical enrichment of the circumgalactic medium \citep[e.g.][]{Erb2015,Chisholm2017}.

In the last few years, extensive observing programmes have been dedicated to the detection and characterisation
of galactic outflows, especially powerful outflows that are driven by AGN.
Several studies have investigated the warm ionized phase of outflows, finding velocities up to several 1000~km s$^{-1}$ and radii up to several kpc \citep[e.g.][]{Westmoquette2012, Harrison2014, Arribas2014, Rupke2017}. In high-$z$ 
quasars (QSOs) such ionized outflows are seen to spatially anti-correlate with star formation in the host galaxy, which has
been regarded as direct evidence for quasar-driven outflows quenching star formation in galaxies \citep{Cano-Diaz2011,
Carniani2016, Carniani2017}. Numerous studies have confirmed the
presence of prominent neutral atomic outflows in local galaxies \citep{Morganti2005, Rupke2005, Cazzoli2016, Morganti2016, Rupke2017}.
However, among all the gas phases involved in galactic outflows, the molecular phase is of particular interest, because it is
found to dominate the outflow mass
\citep{Feruglio2010, Rupke2013, Cicone2014, Garcia-Burillo2015, Carniani2015, Fiore2017}. Furthermore, molecular gas is the phase out of which stars form, hence molecular outflows directly affect star formation.

Molecular outflows have been detected through P-Cygni profiles of FIR OH transitions {\citep{Fischer2010, Sturm2011, Veilleux2013, Spoon2013, Stone2016, 
Gonzalez-Alfonso2016} and through broad wings seen in interferometric observations of molecular transitions such as low-\textit{J} CO lines \citep{Feruglio2010,
Cicone2012, Combes2013, Sakamoto2014, Garcia-Burillo2015} as well as higher density tracers such as HCN \citep{Aalto2012,Aalto2015,Walter2017}.
Using CO line mapping, \cite{Cicone2014} found that starburst galaxies have outflow
mass-loading factors ($\eta$ = $\dot{M}_{\rm outf}(\rm H_2)$/SFR) of 1-4, but the presence of an
AGN dramatically increases $\eta$. Depletion time-scales due to the outflow, i.e. $ \tau_{\rm dep,outf}(\rm H_2) \equiv$ $M(\rm H_2)$/$\dot{M}_{\rm outf}(\rm H_2)$,
were found to anti-correlate with $L_{\rm AGN}$, which further indicates that AGN boost galactic outflows.
In a study of AGN wind scaling relations including molecular and ionized winds, \cite{Fiore2017} observed that molecular
outflow mass rates correlate with AGN luminosity  as $\dot{M}_{\rm outf}(H_2)$ $\propto$ $L_{\rm AGN}^{0.76}$,
while the ionized outflow mass
rates has a steeper dependence of the form $\dot{M}_{\rm outf}(\rm ion)$ $\propto$ $L_{\rm AGN}^{1.29}$,
suggesting that at high
luminosities the ionized phase may contribute significantly to the mass-loss rate. However, it should be noted that these results were achieved by
comparing outflow phases observed in different samples of galaxies, hence these results are potentially subject to (differential)
selection effects among samples selected to investigate different phases.

The purpose of this work is not to provide a census of molecular outflows in galaxies (which would require high sensitivity
millimetre data for a large, volume--limited or mass--limited sample of galaxies), but it is to explore the
scaling relations between molecular outflows and galaxy properties. This will shed light on the driving
mechanisms of outflows and their effect on the host galaxies. We improve relative to previous studies
by significantly increasing the statistics with a sample size of nearly 50 galaxies (
which is more than twice that of previous molecular outflow studies using CO data) and by
tackling some of the biases and selection
effects.
We use interferometric CO measurements that allow us to determine the velocity and spatial extent of the outflows. 
We specifically investigate the relations between outflow and galaxy properties such as star formation rate, stellar
mass and AGN luminosity. Furthermore, we include data from the ionized and
atomic phase of the outflow for those galaxies in our sample that have this information available, and we investigate their
relationship with the molecular phase. This is crucial since galactic outflows are multiphase and by focussing only on one phase 
the total impact of galactic winds on the ISM might be underestimated \citep[e.g.][]{2018NatAs...2..176C}.


Throughout this work, a \textit{H$_{\rm 0}$} = 70~km s$^{-1}$ Mpc$^{-1}$, $\Omega_{\rm M}$ = 0.27 and $\Omega_{\Lambda}$ = 0.73 cosmology is adopted.

\section{Sample and Data Analysis}

\subsection{Sample Selection}

We have characterised molecular outflows both by
collecting data from the literature and from an extensive analysis of ALMA archival data. We set an upper limit of $z$$<$0.2 on the 
redshifts of the targets, since beyond this redshift the angular resolution of most ALMA archival observations ($>$0.3$''$) probes
scales too coarse ($>$1~kpc) to enable a proper characterisation of outflows. We search the ALMA archive for low-\textit{J}
transitions (i.e. CO(1-0), CO(2-1) and CO(3-2)) of all local galaxies observed in these
transitions and with publicly available data in the archive as of April, 1st, 2018. As a result we have
analysed about 100 galaxies from the ALMA archive. However, most of these data have turned out to have sensitivities too low to enable the
detection of putative faint broad CO transitions associated with outflows. However, we have detected
outflow signatures in seven of these galaxies, according to the procedure described in Sect. \ref{sec:identification_outflows}.

We generally do not use the ALMA observations for which there
is no outflow detection to set upper limits on the outflow properties (e.g. outflow rate, kinetic power, momentum rate)
since these would need knowledge of both outflow size and velocity, which is not known a priori. Yet, we can infer tentative
upper limits in three cases for which the outflow is detected in other phases (in particular the ionized phase) by assuming that
the (undetected) molecular outflow has the same size and velocity as those observed in the detected outflow phases. 
As a consequence, from the ALMA archive we have obtained molecular outflow information for a total of 10 galaxies (7 detections
and 3 upper limits).

For what concerns the literature sample,
we have searched for published molecular outflows at \textit{z} < 0.2 obtained through the analysis of
the CO(1-0) and CO(2-1) emission lines. We have compiled a total of 31 galaxies with published 
molecular outflows (five of which are upper limits).

We also include four ULIRGs from \cite{Gonzalez-Alfonso2016} in our sample. In these cases the molecular outflow properties
have been determined based on the far-infrared transitions of OH observed through the Herschel/PACS spectrometer. Their outflow mass
rates are calculated assuming a single expulsion of gas, which is analogous to what we assume in this paper (as it will be described in Sect. \ref{sec:calc_outf_prop}).
For an additional four galaxies of the \cite{Gonzalez-Alfonso2016} sample the molecular
outflow rates inferred from OH have been measured also through CO observation and in these cases they are in reasonable agreement
(typically within a factor of two).

The total sample used in this work consists of 45 galaxies whose properties, such as redshift, luminosity distance, optical
classification, star formation rate, AGN luminosity, AGN contribution to the bolometric luminosity ($\alpha_{\rm bol}$ =
$L_{\rm AGN}$/$L_{\rm bol}$), molecular and atomic gas content and radio parameter $q_{\rm IR}$
are listed in Table \ref{tab:sample_basic_properties}. This sample is homogenised as described in Sections \ref{sec:calc_outf_prop} and \ref{sec:anc_info}, 
i.e. the properties of the host galaxy and the outflow are calculated in a consistent way across the entire sample.
We stress that even though we have not used any other selection criteria, our sample is still heavily
biased, as most of the ALMA observing programmes (as well as results from the literature) have primarily
targeted samples with enhanced star formation (ULIRGs or, more generally, starbursts) or with a known AGN.
Nevertheless, we 
have significantly enlarged the sample relative to previous CO outflow studies by more than 
doubling its size and by including galaxies that are more representative of the massive star forming galaxy population, 
as they feature also lower velocity outflows and much less extreme objects than in previous studies.
In particular, we have included targets
from the ALMA archive culled from observing programmes that were not aimed at extreme classes of galaxies
(starbursts or AGN), and this has resulted in a less biased sample than in previous studies.

However, we emphasize, once more, that the goal of this paper is not to provide an unbiased census of the
occurrence of molecular outflows in galaxies. 
The primary goal of this paper is to explore the relations
between molecular outflows and galactic properties by sampling the broadest possible range of galactic properties,
such as star formation rate, mass, activity type, AGN luminosity.

To illustrate the range of SFRs and galaxy stellar masses
Fig. \ref{fig:sdss_contours} shows the galaxies in our sample in the stellar mass - star formation rate plane.
We also over-plot the contours of the distribution of galaxies from the SDSS DR7 release, which shows the
obvious biases affecting our sample.
The grey points indicate galaxies from a previous study on molecular outflows by \cite{Cicone2014},
the blue points show the new,
additional galaxies added in this work. Different symbols indicate different optical spectral classification, as discussed
more in detail later on. The sample spans about two orders of magnitude in stellar mass
and nearly four orders of magnitude in SFR. Clearly the galaxies in our sample are not distributed uniformly over these ranges and do not even follow
the distribution traced by SDSS galaxies. As a consequence of the selection biases our sample
is skewed towards massive galaxies and mainly sampling galaxies above the main sequence, hence (probably merger-driven) `starbursts'.
However, our sample also probes the main sequence and a few galaxies located in the green valley.
Unfortunately, quiescent galaxies are not included by our sample.

In addition, the fraction of AGN in our sample ($\approx$ 50~per cent) is higher than in other local complete surveys, where about
10--20~per cent are unambiguously AGN (although the actual number might be anything up to 40~per cent, depending
on the AGN luminosity threshold and the selection band) \citep{Maiolino95,Miller2003}. However, this enables us to properly probe different level
and types of AGN activity. The AGN in our sample probe a wide range of bolometric AGN luminosities, from very weak AGN ($~10^{40}$~erg~s$^{-1}$),
to powerful AGN in the quasar regime ($~$10$^{46}$~erg~s$^{-1}$).



\begin{figure}
\centering 
\includegraphics[width=\columnwidth]{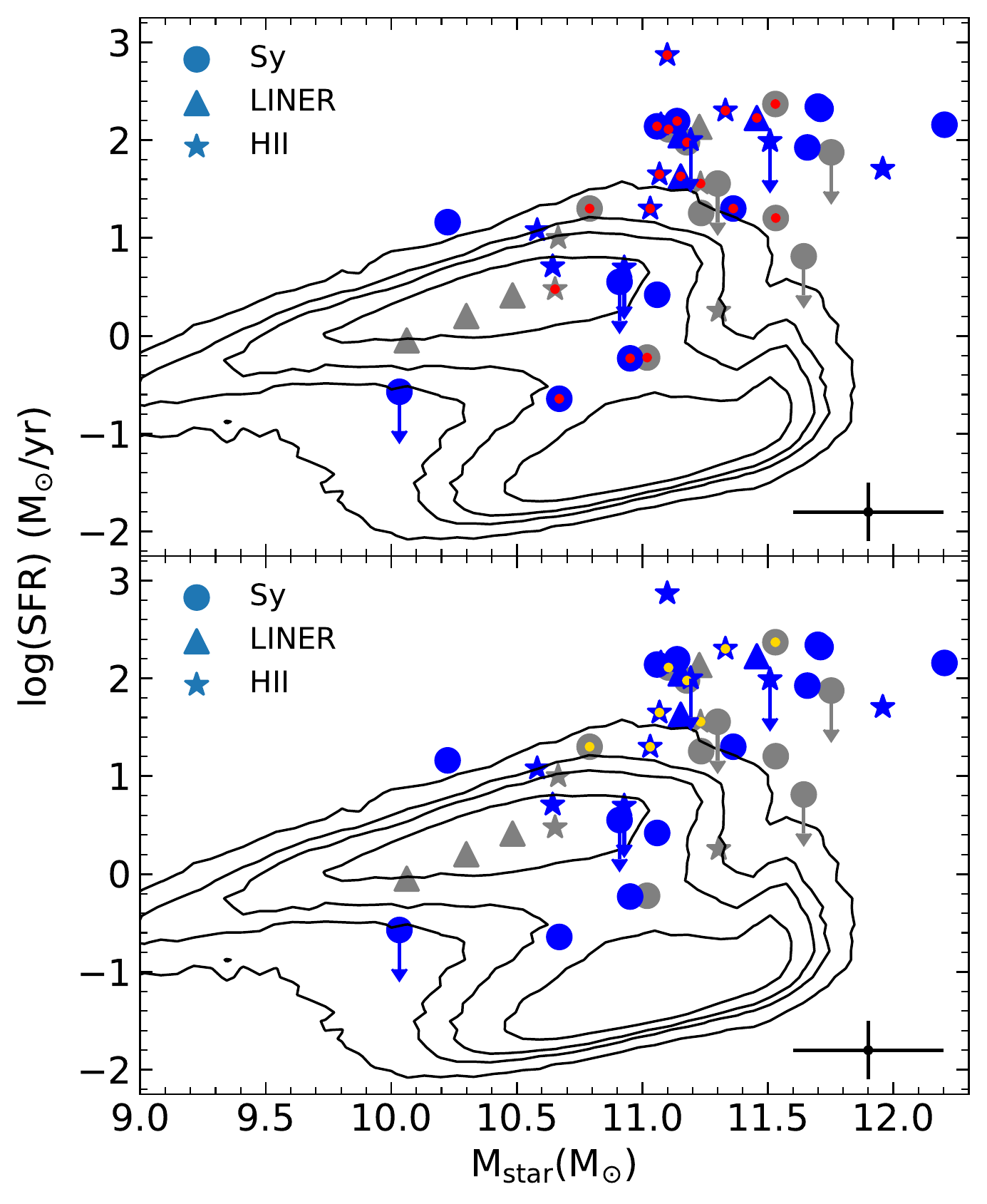}
\caption{Distribution of our sample in the stellar mass-SFR diagram, compared to the SDSS galaxies (contours). The grey points show the galaxies in the sample of \protect\cite{Cicone2014}, the blue galaxies represent the additional galaxies added to that sample. The SDSS contours show the levels 100, 300, 500, 1000 galaxies. The red dot in the upper panel marks galaxies with measurements of ionized outflows, the yellow dot in the lower panel galaxies with observed neutral outflows.}
\label{fig:sdss_contours}
\end{figure}

\subsection{Identification of Outflows}
\label{sec:identification_outflows}

The ALMA archival data have been calibrated and imaged using the CASA software version 4.7 \citep{McMullin2007}. We have ensured that the data cubes have a
spectral window broad enough to find possible wings (covering at least 1500km s$^{-1}$).
We analyse the ALMA CO data initially by searching for outflow signatures by fitting a single or a double Gaussian profile to the CO
emission integrated over the whole galaxy. Whether only one or two Gaussians are
required, is determined by comparing the reduced
chi-square ($\chi_{\rm red}^{2}$) value of their respective fits. If two Gaussians lead to a decrease in $\chi_{\rm red}^{2}$ of 10~per cent or more, 
then we consider this as an initial clue for the possible presence of an outflow.
In these cases we also visually verify whether
we can clearly distinguish a narrow ($\sigma_{\rm narrow}$ $\lesssim$ 100~km s$^{-1}$) and a broad component ($\sigma_{\rm broad}$ ranging from
$\sim$100~km s$^{-1}$ to several 100~km s$^{-1}$, depending on the galaxy). In these candidate cases we tentatively identify the broad component
as emission from an outflow as a first clue.

We then verify the presence of outflows by inspecting the position-velocity (pv) diagram and producing a map of the line wings.
Position-velocity diagrams are generated by extracting a 2D spectrum along a pseudo-slit
(with a typical width of about 0.6~arcsec) placed along the major and minor axes of the galaxy
and plotting the velocity as a function of the position along the pseudo-slit. Rotation-dominated galaxies show a characteristic S-shape in the 
pv diagram (along the major axis), whereas outflows are identified by an excess of high-velocity gas on top of rotation. Line wings maps are also produced 
by integrating over the spectral range where the broad component (i.e. outflow component) is dominant. We determine the root mean square (RMS) 
of the line maps and identify the wings as significant when they are detected at a significance level of $> 5\sigma$. The line wings are identified as due to 
outflows if they have velocities in excess of two times the width of the narrow component and are not located in the direction of rotation. In the Appendix \ref{sec:figures_ALMA}, 
we show for each galaxy the spectrum integrated over the whole galaxy including the narrow and the broad component, the pv diagrams along the major and minor axes 
and the line maps of the wings.
%

\subsection{Outflow Properties}
\label{sec:calc_outf_prop}
We calculate the outflow mass based on the flux of the broad line component, which can be converted into $L'_{\rm CO}$, defined as
 \citep{Solomon2005}:
\begin{eqnarray}
L'_{\mathrm{CO}} = 3.25 \times 10^{7} S_{\mathrm{CO}} \Delta v \nu_{\mathrm{obs}}^{-2} D_{\mathrm{L}}^{2} (1+z)^{-3}
\label{eq:CO_line_lumino}
\end{eqnarray}
where $S_{\rm CO}\Delta v$ is the integrated flux in Jy km s$^{-1}$, $\nu_{\rm obs}$ is the observed frequency of the CO transition (in GHz), $D_{L}$ the luminosity distance (in Mpc) and \textit{z} the redshift.
$L'_{\rm CO(1-0)}$ can in turn be converted into molecular mass of the outflow ($M_{\rm outf}(\rm H_2)$) via $M_{\rm outf}(\rm H_2)$ = $\alpha_{\rm CO}$$L'_{\rm CO}$, where $\alpha_{\rm CO}$ is
the CO-to-H$_{2}$ conversion factor. For outflows we conservatively assume a CO-to-H$_{2}$ conversion factor of 0.8~M$_{\odot}$/(K km s$^{-1}$ pc$^{2}$) to
for consistency with previous work. This is the value typically adopted for the molecular ISM of ULIRGs \citep{Bolatto2013}.
The excitation in the wings and the core in Mrk 231, a ULIRG hosting the closest QSO and a well studied outflow, were found to be very similar and hence the
conversion factor in the non-outflowing and outflowing components are likely to be similar \citep{Cicone2012}. In some outflows
the conversion factor has been studied in detail \citep[see e.g.][]{Weiss2005, Cicone2018} yielding values closer to $\alpha_{\rm CO}$ $\approx$ 2~M$_{\odot}$/(K km s$^{-1}$ pc$^{2}$).

For outflows observed in higher-$J$ transitions
we assume that the CO emission is thermalised and optically thick, hence $L'_{\rm CO(3-2)}$ = $L'_{\rm CO(2-1)}$ = $L'_{\rm CO(1-0)}$. 
This is consistent, within the errors, with what was found in Mrk 231 \citep{Feruglio2015}.  The double component fitting allows us to directly 
estimate the outflow velocity ($v_{\rm outf}$) using the prescription of \cite{Rupke2005}: $v_{\rm outf}$ = FWHM$_{\rm broad}$/2 + |$v_{\rm broad}$
- $v_{\rm narrow}$|, where FWHM$_{\rm broad}$ is the full width at half maximum of the broad component and $v_{\rm broad}$ and $v_{\rm narrow}$ 
are the velocity centroids of the broad and narrow components, respectively. The spatial extent of the outflow is calculated based 
on the line maps of the broad wings. We fit a 2D-Gaussian profile to the wing map and use the beam-deconvolved major axis (FWHM) 
divided by two as the radius of the outflow.

The mass outflow rate, $\dot{M}_{\rm outf}(\rm H_2)$, is calculated assuming time-averaged thin expelled shells or clumps \citep{Rupke2005b}:

\begin{eqnarray}
\dot{M}_{\rm outf}(\rm H_2) = \frac{\it v_{\rm outf}(\rm H_2)\it M_{\rm outf}(\rm H_2)}{\it r_{\rm outf}(\rm H_2)}.
\end{eqnarray}
where $v_{\rm outf}(\rm H_2)$, $r_{\rm outf}(\rm H_2)$ and  $M_{\rm outf}(\rm H_2)$ are the velocity, radius and molecular gas mass of the outflow, respectively.
This description allows us a better comparison with models and is more realistic than the assumption of spherical (or multi-conical) volume with uniform filling
factor \citep{Cicone2015,Pereira-Santaella2016,Veilleux2017}. It can be shown that between the two scenarios there is a difference of a factor of three 
in the estimates of the outflow rate (and derived quantities such as kinetic power and momentum rate), which does not alter our conclusions significantly. 
Projection effects certainly plague the estimation of the outflow radius and velocity. However, as discussed in \cite{Cicone2015}, since the 
orientations of outflows are distributed randomly, it can be shown that the resulting average correction factor is unity, hence statistically the (unknown)
projection correction factors cancel out on average, though they certainly introduce scatter. By combining all sources of uncertainty, we infer that the average 
uncertainty on the mass outflow rate is about 0.3~dex. The errors on the associated outflow properties (kinetic power, momentum rate) is estimated to be as
large as 0.5 dex.

\subsection{Ancillary information} 
\label{sec:anc_info}

In this section we provide ancillary information on the host galaxy, which are summarized
in Table \ref{tab:sample_basic_properties}.

\subsubsection{Optical classification}

In terms of activity classifications, we refer to galaxies as `star forming', `Seyfert' and `LINER' 
based on their optical spectroscopic classification, and in particular through the BPT-[SII] diagram \citep{Kewley2006}. The nature
of galaxies classified as `LINER' is not always clear, and this classification appears to include a mixed population. It has been
shown that the `LINER' emission can extend on kpc-scales across a large fraction of passive
and green valley galaxies
(hence renaming this 
class as `LIER', i.e. dropping the `N' which stands for `Nuclear' in the original acronym)
and correlates with the old stellar population, and this can be explained
in terms of excitation by the hard radiation field produced by evolved post-AGB stars \citep[e.g.][]{doi:10.1111/j.1365-2966.2009.16039.x, doi:10.1093/mnras/stw1234}.
However, in the nuclear regions, LI(N)ER-like emission can also be associated with excitation by weak, radiatively inefficient AGN \citep[e.g.][]{1993ApJ...417...63H}.
Yet, in LIRGs, ULIRGs, and other galaxies characterised by prominent outflows, which are most of the LINER-like galaxies in our sample, 
LI(N)ER-like diagnostics are likely associated which shock excitation \citep[e.g.][]{MonrealIbero2006}. 
Many authors broadly group Seyfert and LI(N)ER-like diagnostics into a generic `AGN' category. As discussed above, this rough classification can
be misleading as in many galaxies the LINER classification is not associated with an AGN at all; however, in the case of our sample it is true
that many LINER-like galaxies do host an AGN based on the X-ray or mid-IR properties; therefore in a few instances in the paper (e.g. Sect. \ref{sec:scaling_relations}) 
we will adopt this classification as well. Regardless of the optical classification, the role of the AGN, if present, will be clarified by the AGN fractional contribution to the bolometric
luminosity, as discussed in the following.

\subsubsection{AGN luminosity}

AGN bolometric luminosities were derived from the hard X-ray flux (2-10~keV) by using the relation given in \cite{Marconi2004}: log[$L_{\rm AGN}$/$L$ (2-10~keV)] = 1.54 +
0.25$\mathcal{L}$ + 0.012$\mathcal{L}^{2}$ - 0.0015$\mathcal{L}^{3}$, where $\mathcal{L}$ = (log $L_{\rm AGN}$  -12) and $L_{\rm AGN}$ is the AGN bolometric luminosity in units of
\textit{L}$_{\odot}$. Typically, X-ray-based AGN luminosities have a scatter of $\sim$ 0.1~dex \citep{Marconi2004}. In a few cases where no X-ray
data are available, or the source is Compton-thick, we used the
[OIII]$\lambda$5007 luminosity. In this case the AGN luminosity is inferred from the relation $L_{\rm AGN}$ $\sim$ 3500~\textit{L}$_{\rm [OIII]}$
\citep{Heckman2004}. In some cases for which [OIII]$\lambda$5007 is not available, or which are heavily obscured in the optical, we estimated the AGN luminosity by using
the AGN contribution to the bolometric luminosity $\alpha_{\rm bol}$ as inferred from various mid-IR diagnostics in the literature
\citep{Veilleux2009,Nardini2009,Nardini2010}. \cite{Nardini2009} and \cite{Nardini2010} use spectral features in the wavelength range 5-8 $\mu$m that allow
them to disentangle AGN and starburst contribution. \cite{Veilleux2009} use six different IR-based methods, as for instance the equivalent width of the PAH feature at
7.7~$\mu$m and the continuum ratio of $f_{\rm 30}$/$f_{\rm 15}$, and average them to calculate the AGN contribution. Using the AGN fraction, we can then calculate the
AGN luminosity via $L_{\rm AGN}$ = $\alpha_{\rm bol}$$L_{\rm bol}$, where in most cases $L_{\rm bol}$ $\approx$ $L_{\rm IR}$ (although
for ULIRGs $L_{\rm bol}$ $\sim$ 1.15 $L_{\rm IR}$ \citep{Veilleux2009}). In the rest of the paper $\alpha_{\rm bol}$ = $L_{\rm AGN}$/$L_{\rm bol}$ refers to the AGN contribution 
to the total IR luminosity, which generally dominates in most of our galaxies, although in a few more quiescent galaxies the stellar optical/NIR light may contribute significantly. 
The uncertainty of the IR luminosity consists of the contribution from uncertainties on the IR fluxes at 12, 25, 60 and 100~$\mu$m, which are generally below 10~per cent and the scatter in the calculation of the total IR luminosity, $L_{\rm IR}$, based on these IRAS fluxes, which is about 10-20~per cent \citep{Takeuchi2005}. In total, we therefore conservatively assume 30~per cent uncertainty on $L_{\rm IR}$.

\subsubsection{Star Formation Rate}

To compute the total star formation, we use the $L_{\rm IR}$-SFR relation given in \cite{Kennicutt2012}, which assumes a Chabrier IMF and 
the total infrared luminosity from 8 to 1000~$\mu$m, corrected for the AGN contribution through the $\alpha_{\rm bol}$ factor.
The uncertainty in star formation rates stems from uncertainties on $L_{\rm IR}$ (which is discussed above and amounts to $\sim$ 30~per cent) and on $\alpha_{\rm bol}$. $\alpha_{\rm bol}$ has similarly values using various techniques and they usually agree within 10-15~per cent \citep{Veilleux2009}. We assume this as the typical error. The conversion of infrared luminosity to SFR comes with a 30~per cent calibration uncertainty \citep{Kennicutt1998}.
For star formation rate estimates, we therefore infer conservatively a typical uncertainty of 0.3~dex.

\subsubsection{Gas content}

The molecular gas mass in the host galaxy is inferred from the CO(1-0) (narrow) line luminosity $L'_{\rm CO}$, as discussed above.
The CO-to-H$_{2}$ conversion factor is one of the major uncertainties in the calculation of the molecular gas mass and depends heavily on the
metallicity and physical state of the molecular ISM \citep{Bolatto2013}.
We adopt three different CO-to-H$_{2}$ conversion factors depending on the type of galaxy. For ULIRGs, we adopt $\alpha_{\rm CO}$ = 0.8~M$_{\odot}$/(K km s$^{-1}$ pc$^{2}$), for LIRGs we use $\alpha_{\rm CO}$ = 1.2~M$_{\odot}$/(K km s$^{-1}$ pc$^{2}$) and for all other galaxies we use a Milky Way-type conversion factor of 4.4~M$_{\odot}$/(K km s$^{-1}$ pc$^{2}$) \citep{Bolatto2013}.

For about half of the galaxies 21cm HI, single dish, observations are also available which provide the atomic gas mass 
in the host galaxy.

\subsubsection{Stellar mass}

Stellar masses are calculated for all galaxies in this sample by using the \textit{K}-band magnitude and a colour correction (e.g. \textit{B}-\textit{V}) \citep{0067-0049-149-2-289}. $K$-band magnitudes are taken from the extended source catalogue of 2MASS \citep{2006AJ....131.1163S}.

However, the presence of an AGN can potentially contaminate the observed fluxes.
For Seyfert 2 galaxies, the direct continuum radiation from the accretion disc is obscured along our line of sight, but the hot dust heated by the AGN can still contribute significantly
to the light observed in the \textit{K}-band. Therefore, in the case of Seyfert 2 galaxies, in order
to avoid the latter issue,
we use \textit{J}-band magnitudes, that are not affected by AGN-heated circumnuclear dust emission, and estimate the
$K$-band magnitude by assuming $J$--$K$=0.75, which is the average colour (with little scatter)
found by \cite{Mannucci2002}. 

For Seyfert 1 galaxies in our sample a contamination by the AGN might be very high also in the \textit{J}-band and optical bands (because the radiation from the accretion
disc is directly observable) and, therefore, we need to use a different approach. For Mrk 231 and IRAS F11119+3257, the contribution of the AGN to the total magnitudes has been estimated in
\cite{Veilleux2002} and we simply subtract this nuclear contribution to estimate the stellar masses in these
two galaxies. For the other three Seyfert 1 galaxies, we compute the stellar mass by using the \textit{H}-band magnitude of the host (which 
does not include nuclear contribution by the AGN) inferred by \cite{Zhang2016} and the mass-to-light correction given in their paper.

For non-type 1 AGN,
the colours for the mass-to-light ratio correction are obtained from the literature. We combine different colours, \textit{u}-\textit{g} for galaxies with SDSS photometry,
\textit{B}-\textit{V} from VERONCAT, the Veron Catalogue of Quasars and AGN \citep{2010A&A...518A..10V}, or from the GALEX survey \citep{dePaz2007} and \textit{B}-\textit{R} from the
APM catalogue\footnote{http://www.ast.cam.ac.uk/$\sim$mike/apmcat/}. In a few cases where no information about colours is available, we assume an average
logarithmic mass-to-light correction of -0.08 \citep{0067-0049-149-2-289,Zhang2016}. 
Our final errors on the stellar mass comprises errors on the photometry of the host galaxy ($H$, $J$ or $K$-band) and the uncertainty of the $M_{\star}$/$L$ ratio.
The $K$-band magnitude are estimated to have an 0.1 mag uncertainty based on comparison between different samples \citep{0067-0049-149-2-289}. In the $J$ and $H$-band, the typical uncertainty is 0.2 mag \citep{Zhang2016}. Our estimates of $M_{\star}$/$L$ ratios have a typical systematic error of
about 25~per cent which stems from uncertainties in galaxy age, dust extinction and the impact of SF bursts on the
star formation history \citep{0067-0049-149-2-289}. 
Furthermore, for AGN host galaxies, additional uncertainties might be introduced by the corrections applied here. Therefore, we
conservatively obtain an average error of $\pm$0.2 dex on the stellar masses. Although the use of different
colours for some of the galaxies may potentially be a matter of concern, \cite{Bell2001} and \cite{Taylor2011} have shown that 
there is no systematic uncertainties on the inferred stellar masses when different colours and different (infrared/red)
bands are used.

\subsubsection{Radio emission}
\label{sec:radio_data}

In order to investigate the potential link between outflows and radio jets, we have also collected data about the radio power
in galaxies at 1.4~GHz, mostly by using the database provided by NED. Since in normal star forming galaxies the radio luminosity
simply scales with the SFR as traced by the infrared luminosity \citep[e.g.][]{Yun2001, Ivison2010}, the contribution from a radio jet can be inferred
in terms of excess relative to the radio-to-infrared ratio observed in normal star forming galaxies. Therefore, in Table
\ref{tab:sample_basic_properties} we provide the quantity $q_{\rm IR}$ which is the ratio between the rest frame 8-to-1000~$\mu$m flux and the 1.4~GHz monochromatic radio flux \citep{Ivison2010}.




\subsection{Ionized Outflows}
\label{sec:ion_outflow}
We complement our results on molecular outflows with data on the ionized outflow phase. For each galaxy, we search whether a reliable estimate of the ionized outflow mass is provided
in the literature. 16 of our sources, i.e. about 1/3 of the sample
have measurements of the ionized outflow mass, velocity and radius.  \cite{Rupke2013} provide ionized gas masses for four galaxies (IRAS F08572+3915, IRAS F10565+2448, Mrk 273 and 
Mrk 231) based on the H~$\alpha$ emission. \cite{Greene2012} estimate the ionized outflow mass for SDSS J1356+1026 using H~$\beta$. The other sources with ionized outflow rates are 
taken from \cite{Arribas2014} and are based on integral field spectroscopy (IFS) of H~$\alpha$.

We carefully homogenise the calculations of the ionized outflow properties. Outflow velocities are calculated in the same way as for molecular
outflow, i.e. 
$v_{\rm outf}(\rm ion)$ = FWHM$_{\rm broad}(\rm ion)$/2 + |$v_{\rm broad}(\rm ion) $-$v_{\rm narrow}(\rm ion)$| \citep{Rupke2005}. For the calculation of the outflow mass, $M_{\rm outf}(\rm ion)$, we assume an electron density of $n_{\rm e}$
= 315~cm$^{-3}$ as found in \cite{Arribas2014}. This value is also close to the electron density values found in other works \citep[e.g.][]{Perna2015, Bischetti2016}.
We calculate the ionized outflow mass rate as follows:

\begin{eqnarray}
\dot{M}_{\rm outf}(\rm ion) = \frac{\textit{v}_{\rm outf}(\rm ion)\textit{M}_{\rm outf}(\rm ion)}{\textit{R}_{\rm outf}(\rm ion)},
\end{eqnarray}
where $R_{\rm outf}(\rm ion)$ is the radius of the outflow.
\newline For the outflow extent, we generally assume the same value given in the corresponding paper. However, in \cite{Arribas2014}, an average fixed radius of 700~pc is assumed. 
Instead of using this fixed radius, we use the value inferred from \cite{Bellocchi2013} using the broad H~$\alpha$ maps obtained with VIMOS at the Very Large Telescope (VLT) and 
assuming a spherical geometry.

We note that the spatial extent of ionized outflow can be severely affected by beam smearing effects, because 
the observed spatial distribution is luminosity weighted, hence the central, compact regions dominate the outflow size measurement, even if the outflow
is much more extended. This results in an overestimation of the outflow rate. Indeed, the uncertainty on the outflow extent
(limited by the typical seeing of about 1$''$), often resulting into an error of 50\% on the outflow radius,
dominates the uncertainty on the outflow rate shown in the various figures.
\newline 
Measurements of the ionized outflow are available for only 1/3 of the sample.
Galaxies in our sample with ionized outflow measurement are indicated with a red dot
in the top panel of Fig. \ref{fig:sdss_contours}. Unfortunately, the part of the sample with available ionized
outflow information is restricted to the mass range $10.4<\log{(M_*/M_{\odot})}<11.6$, hence limiting our capability
to properly explore the full mass range probed by the molecular outflow sample. However, within this mass range, the
galaxies with ionized outflows sample the full SFR range.

\subsection{Neutral Outflows from Na~I~D}
\label{sec:neutral_of}
We also include data on neutral atomic outflows in the same way as for the ionized outflows. We crossmatch our sample with \cite{Rupke2013} and \cite{Cazzoli2016},
where the properties of the neutral outflow are inferred from the blue-shifted neutral sodium absorption doublet lines (Na~I~D) at 5890~\AA\  and 5896~\AA. The Na~I~D 
absorption method can only trace outflows towards those lines of sights that have enough stellar continuum light in the background; 
this often limits the use of this diagnostic to the central regions of galaxy disks. Additional issues affecting this diagnostics is that it has to be disentangled from the Na~I~D stellar
absorption, from possible Na~I~D emission and from the nearby He~I nebular emission.

With these caveats in mind we have taken the outflow rate inferred in the original papers. We have also attempted to re-determine the outflow
rate by taking the outflow mass given in the original papers and re-calculating the outflow by using our approach, based on size and outflow velocity,
although in this case the method is not fully applicable as the covering factor is another parameter that should be taken into account. 
In most cases we obtain values close to those reported in the original papers, however, for a couple of galaxies the difference is as high as a factor of
three. We will use both the atomic outflow rate reported in the original
papers and those recalculated by us.

\subsection{Neutral outflows from [CII]}

The fine-structure transition of C$^{+}$, [CII], is another tracer of cold neutral gas, which has been increasingly used to search outflows,
especially at high redshift \citep{Janssen2016,Maiolino2012,Cicone2015,doi:10.1093/mnras/stx2458}.  The majority
of [CII] emission is believed to stem from photon dominated regions (PDRs), where the bulk of the gas is in the neutral atomic phase. However about 20~per cent is generally
coming from CO-dark molecular gas and about 30~per cent can come from the partly ionized phase \citep{Pineda2014}. We have collected data on the [CII]-outflow
for our sample from \cite{Janssen2016}, who provide the atomic mass in the outflow based on the [CII] broad/narrow components decomposition,
assuming a temperature of 100~K and density of $10^5~\rm cm^{-3}$, which should be typical of the ULIRGs in their sample.
The outflow rate is then calculated by taking the radius estimated from the CO observations and consistently with the method used for molecular outflows, as
described above.
\newline 
As we will see in the next section a few targets have measurements of atomic neutral outflows both from Na~I~D and [CII].
In some of these cases the agreement between the measured mass-loss rates is within a factor of two, which is remarkable given all assumption, potential
systematics and issues discussed above. However, for one ULIRG hosting a powerful AGN (Mrk 273)
the difference is as large as an order of magnitude, hinting at the fact that for some extreme targets the uncertainties
in the estimation of the atomic neutral outflow can be very large.

Combining the measurements of atomic neutral outflows from Na~I~D and [CII] measurements (also taking into account
galaxies which have both measurements) only 12 galaxies have this information (about one fourth of the full sample).
These are indicated with a yellow dot
in the bottom panel of Fig. \ref{fig:sdss_contours}. They span the limited
mass range $10.6<\log{(M_*/M_{\odot})}<11.6$, and only probe galaxies with high SFR (generally higher than the main
sequence), implying that the atomic neutral outflows in our sample are hardly representative of the broader population
of galaxies (in terms of mass and SFR), probed by the full sample with molecular outflow information. The extrapolation of
the atomic neutral information to the full sample should be considered with great care, and possibly expanded in the future.

\section{Results}

In this section we report the main results obtained through our sample of molecular outflows, in combination with the ancillary data.
A more extensive analysis of the results and of their interpretation is given in Section \ref{sec:discussion}.

\begin{figure}
\centering 
\includegraphics[width=\columnwidth]{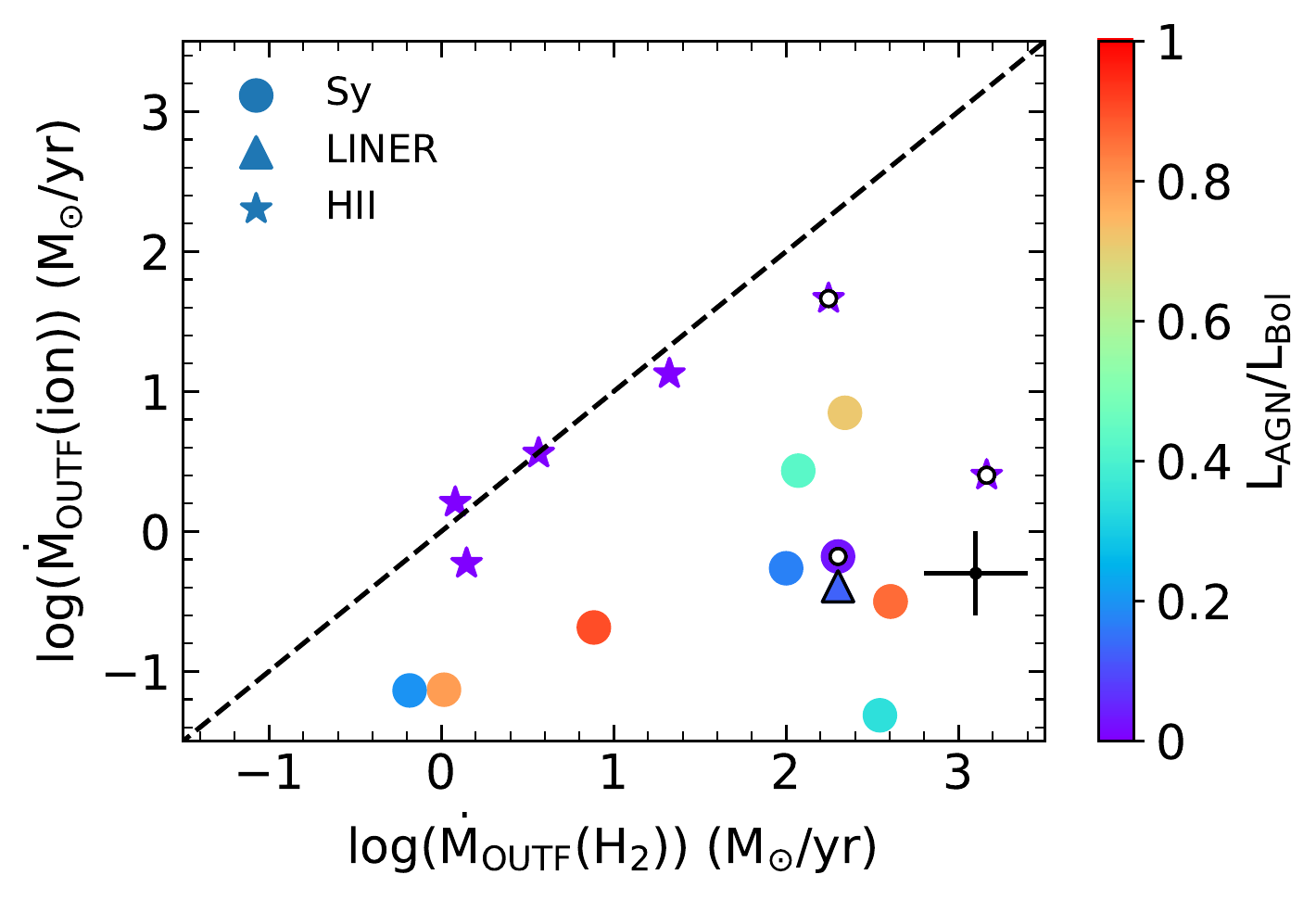}
\caption{Molecular outflow mass rate ($\dot{M}_{\rm outf}(\rm H_2)$) compared to the ionized outflow mass rate ($\dot{M}_{\rm outf}(\rm ion)$).
The dashed line shows the 1:1 relation. Circles indicate Seyfert host galaxies, LINERs are plotted as triangles and
purely star forming galaxies as stars. The data points are colour-coded according to their AGN contribution ($L_{\rm AGN}$/$L_{\rm bol}$), as given in the colour bar on the right. The data points with
black edges are molecular outflows inferred from OH measurements by \protect\cite{Gonzalez-Alfonso2016}. The symbols with a central white dot
are the candidate `fossil' outflows (see sect.\ref{sec:fossil}).
}
\label{fig:molec_ion}
\end{figure}

\begin{figure}
\centering 
\includegraphics[width=\columnwidth]{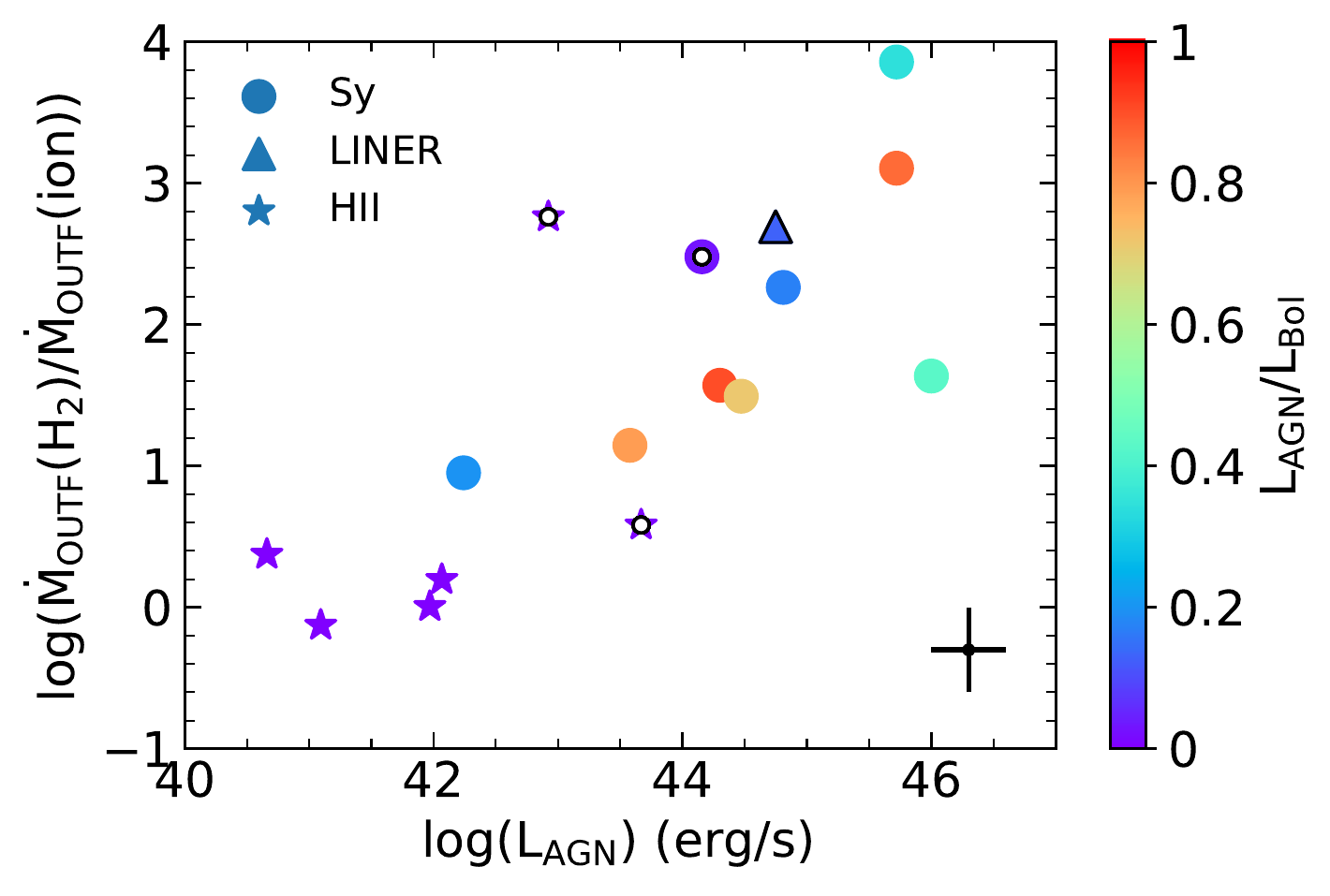}
\caption{Molecular to ionized mass outflow rate as a function of AGN luminosity. Colour-coding and symbols are as in Fig. \ref{fig:OFrate_SFR}.}
\label{fig:ratio_mol_ion_LAGN}
\end{figure}


%
\begin{figure}
\centering 
\includegraphics[width=\columnwidth]{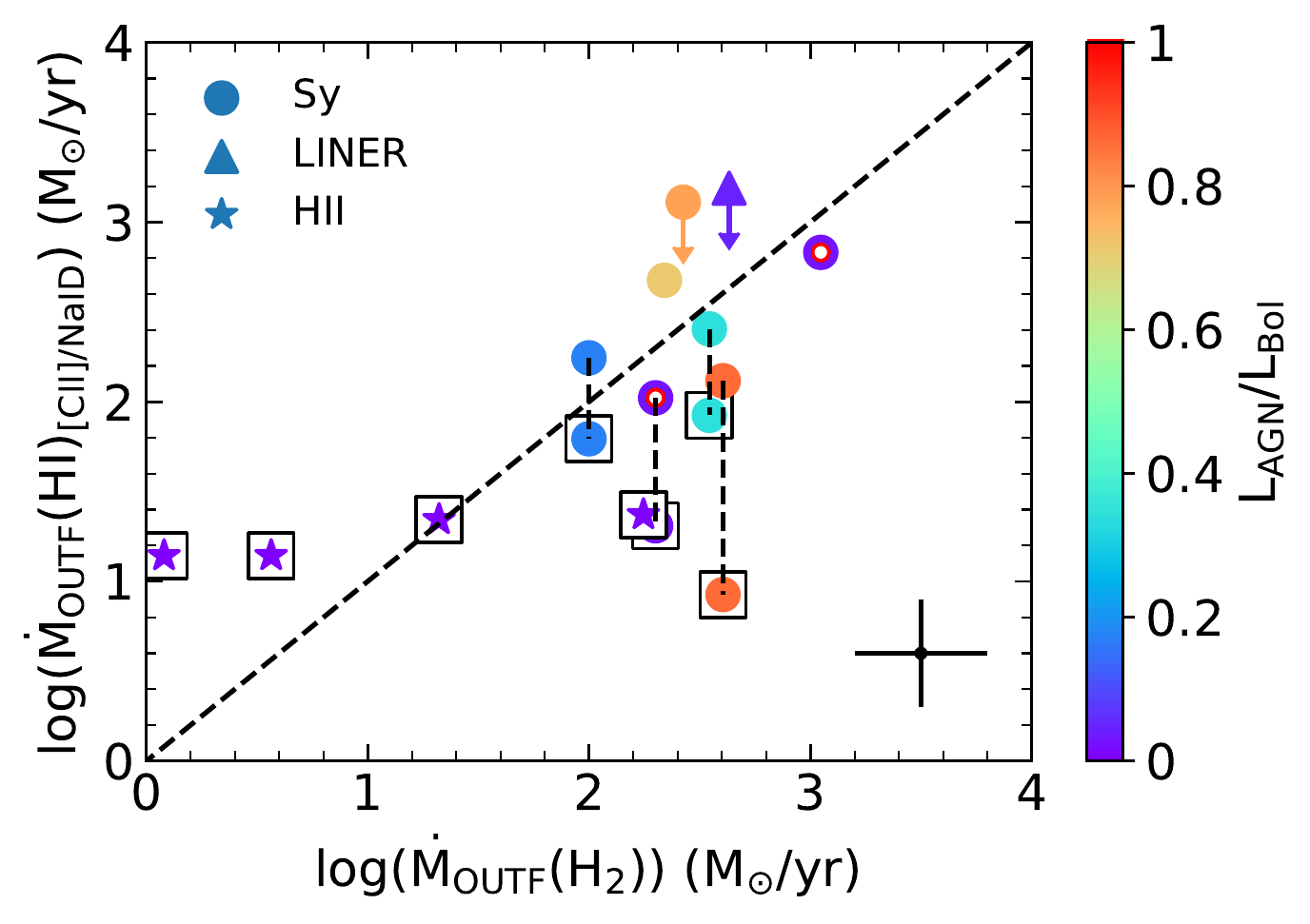}
\caption{Neutral atomic outflow rates inferred from the [CII] line or the Na~I~D absorption (outlined using black squares) as a function of the molecular outflow rates. Measurements of the same galaxy obtained with two different methods ([CII] or Na~I~D) are connected with a dashed black line. The diagonal dashed line gives the 1:1 relation. Colour-coding and symbols are as in Fig. \ref{fig:molec_ion}.}
\label{fig:molec_CIIoutflowrates}
\end{figure}

\subsection{Atomic neutral and ionized outflows in comparison with molecular outflows}
\label{sec:multiphase}

We start by investigating the
relation between molecular outflow rate and atomic neutral and ionized outflow rates for
those galaxies in our sample that have additional multi-wavelength data suited for such a study.

In Figure \ref{fig:molec_ion}, 
we plot the molecular outflow rate as a function of ionized outflow mass rate
of the same object, for those galaxies that have information on {\it both} outflow phases available. Star forming galaxies have
comparable ionized and molecular outflow rates. As we will see, the molecular 
outflow loading factor for star forming galaxies is
close to one, implying that also the ionized outflow loading factor is close to unity,
in agreement with independent studies focussed specifically on ionized outflows \citep{Heckman2015}.

In contrast, AGN host galaxies have much higher molecular outflow rates than ionized ones. This is in agreement with
previous studies \citep{Rupke2013, Garcia-Burillo2015, Carniani2015, Fiore2017}, which observed that molecular outflow rates are 2-3 magnitudes
higher than ionized outflow rates. \cite{Rupke2017} also investigate the multi-phase outflow in a few quasars; only two objects in their sample
have measurements in the molecular phase, and in these two cases the molecular phase dominate the outflow mass relative
to the atomic phase.

At higher AGN luminosities (above 10$^{46}$~erg s$^{-1}$), it has been suggested that the ionized winds have similar
mass outflow rates to molecular winds \citep{Fiore2017}; however, these previous studies were mostly based on the comparison
of galaxy samples which were observed in different gas phases. It is difficult to directly investigate the relation
between molecular and ionized gas in
luminous, distant quasars, as it generally very challenging to detect their molecular outflows through
their weak CO wings, which are in most cases still
below the detection limits, even for ALMA observations, for most high-$z$ QSOs. The few deep ALMA/NOEMA observations and studies reported so far
on some individual quasars are not conclusive yet. \cite{Brusa2017} have reported the detection of a molecular (CO) outflow
rate comparable with the ionized outflow rate in a quasar at $z$$\sim$1.5, while \cite{Toba2017} have reported the lack of molecular (CO) outflow in an AGN
at $z$$\sim$0.5. However, in both cases the sensitivity of the millimetre observation is still far from what would be required to match the optical/near-IR
observations, hence a significant amount of outflowing molecular gas may still be missed in these observations.
\cite{Carniani2017} have reported the detection of a molecular outflow in a 
quasar at $z$$\sim$2.3, having an outflow rate much larger than the ionized outflow rate.
\cite{Feruglio2017} have reported the detection of a fast and massive molecular outflow ($\dot{M}_{\rm outf}(\rm H_2)$ = 3-7$\times 10^3 ~\rm M_{\odot}  yr^{-1}$)
in a lensed quasar at $z$$\sim$4, but unfortunately in this case the ionized
outflow rate is not available for comparison. Although, it is not yet possible to make a direct, statistically sound
comparison of the ionized and molecular outflow rates at very high luminosities, we can at least investigate the relationship and trend within the
luminosity range probed by our sample.
Fig. \ref{fig:ratio_mol_ion_LAGN} shows the dependence of the ratio between molecular
and ionized outflow rate as a function of the bolometric luminosity of the AGN. The ratio $\dot{M}_{\rm outf}(\rm H_2)$/$\dot{M}_{\rm outf}(\rm ion)$ clearly
increases with AGN luminosity, which is the opposite trend of that obtained in the past based on disjoint samples of ionized and molecular outflows
\citep{Fiore2017}.
However, our sample is still small and
does not reach up to very high AGN luminosities (> 10$^{46}$~erg s$^{-1}$). Yet, overall our data confirm
the results of \cite{Fiore2017} and \cite{Carniani2015}
that, in our luminosity range ($L_{\rm AGN} <\: $10$^{46}$~erg s$^{-1}$), molecular outflows have outflow rates about two order of magnitude
larger than ionized outflow rates.

Neutral atomic gas can be measured through the Na~I~D absorption, but it is subject to
significant uncertainties due to the fact that it can be probed only where there is enough background stellar light. Moreover, disentangling the Na~I~D
absorption outflow feature from the Na~I~D stellar absorption and from the ISM absorption in the host galaxies, as well as from the He~I an Na~I~D nebular
emission, further increases the uncertainties. An alternative way to probe the atomic neutral outflow is to exploit the 
fine-structure line of C$^{+}$, [CII]$\lambda$157.74~$\mu$m, which, as discussed in Sect. 2.7, traces primarily atomic gas. \cite{Janssen2016} measured the 
outflowing mass of atomic gas in a sample of ULIRGs/LIRGs some of which are in our sample, and for which we have inferred the atomic outflow rate as discussed in Sect. 2.7.
In Fig. \ref{fig:molec_CIIoutflowrates} we compare the atomic outflow mass rate inferred from both [CII] and Na~I~D (surrounded by a black square) with the molecular outflow rate
from this work for galaxies with measurements of both tracers. The sample size is small so far. Despite this, the inferred atomic outflow rates of AGN host galaxies seem to be very similar to the molecular outflow rates
suggesting that outflows have similar contribution of atomic neutral and molecular gas. For star-forming galaxies the outflow masses are comparable in the atomic and molecular phase, but any possible trend is likely washed out by the above mentioned uncertainties. The discrepancy between the Na~I~D and the carbon measurements can be
explained in the uncertainties associated with Na~I~D discussed above, though more observations are needed to investigate these issues.

With the limited statistics available for the sub-samples with multiple outflow phases, it is not possible to provide accurate relationships among
the various phases, also because there is significant dispersion. However, it is useful to provide some indication on the rough relationship
between the molecular outflow
and other two phases, which can provide some guidance on how to obtain the {\it total} outflow rate (and other derived quantities) to
correct the molecular outflow rate by roughly accounting for the additional phases. Based on the results above we can roughly state that in starburst-driven
outflows the ionized and neutral atomic phases contribute, each of them, to the outflow rate at the same level as the molecular outflow rate.
In AGN-driven outflows the atomic neutral outflow rate is similar to the molecular outflow rate, while the ionized outflow rate is negligible. We will
adopt these simple recipes, when attempting to infer the global properties of outflows in the following sections.

\begin{figure}
\includegraphics[width=\columnwidth]{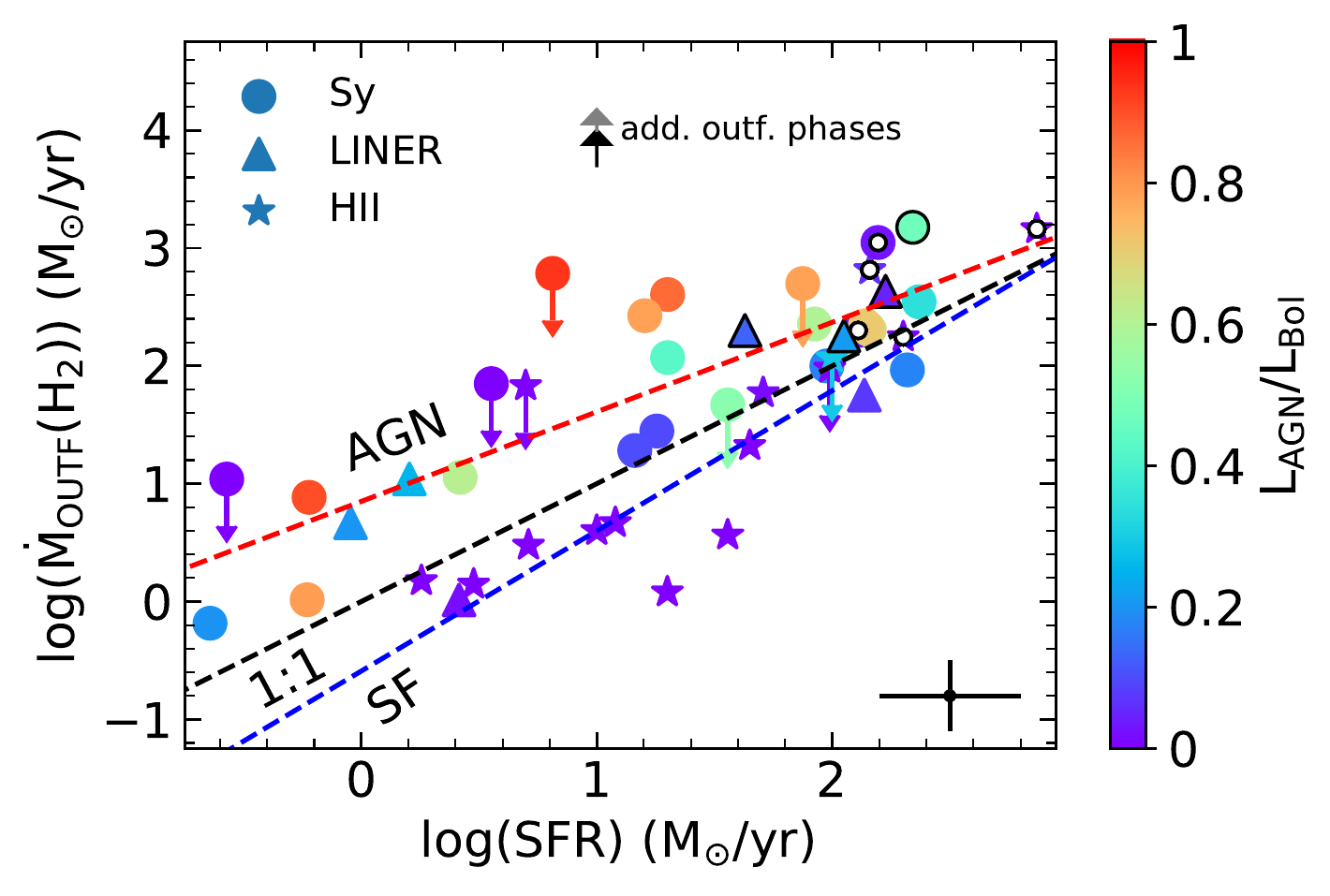}

\caption{Mass outflow rate as a function of star formation rate.  The black dashed line shows the relation for a outflow mass-loading factor
$\eta$ = $\dot{M}_{\rm outf}(\rm H_2)$/SFR = 1.  The black dashed line is the 1:1 relation between outflow rate and SFR, i.e. $\eta =1$.
The red and blue dashed lines represent the best fits to AGN hosts and star forming/starburst galaxies, respectively. The vertical black and
grey arrows indicate the average correction of the outflow rate, for AGN and star forming galaxies, respectively, once 
the atomic (ionized and neutral) phases average contributions to the of the outflow rate (as inferred in Sect. \ref{sec:multiphase}) are
included. Colour-coding and symbols are as in Fig. \ref{fig:molec_ion}.
}
\label{fig:OFrate_SFR}
\end{figure}

\begin{figure}
\centering
\includegraphics[width=\columnwidth]{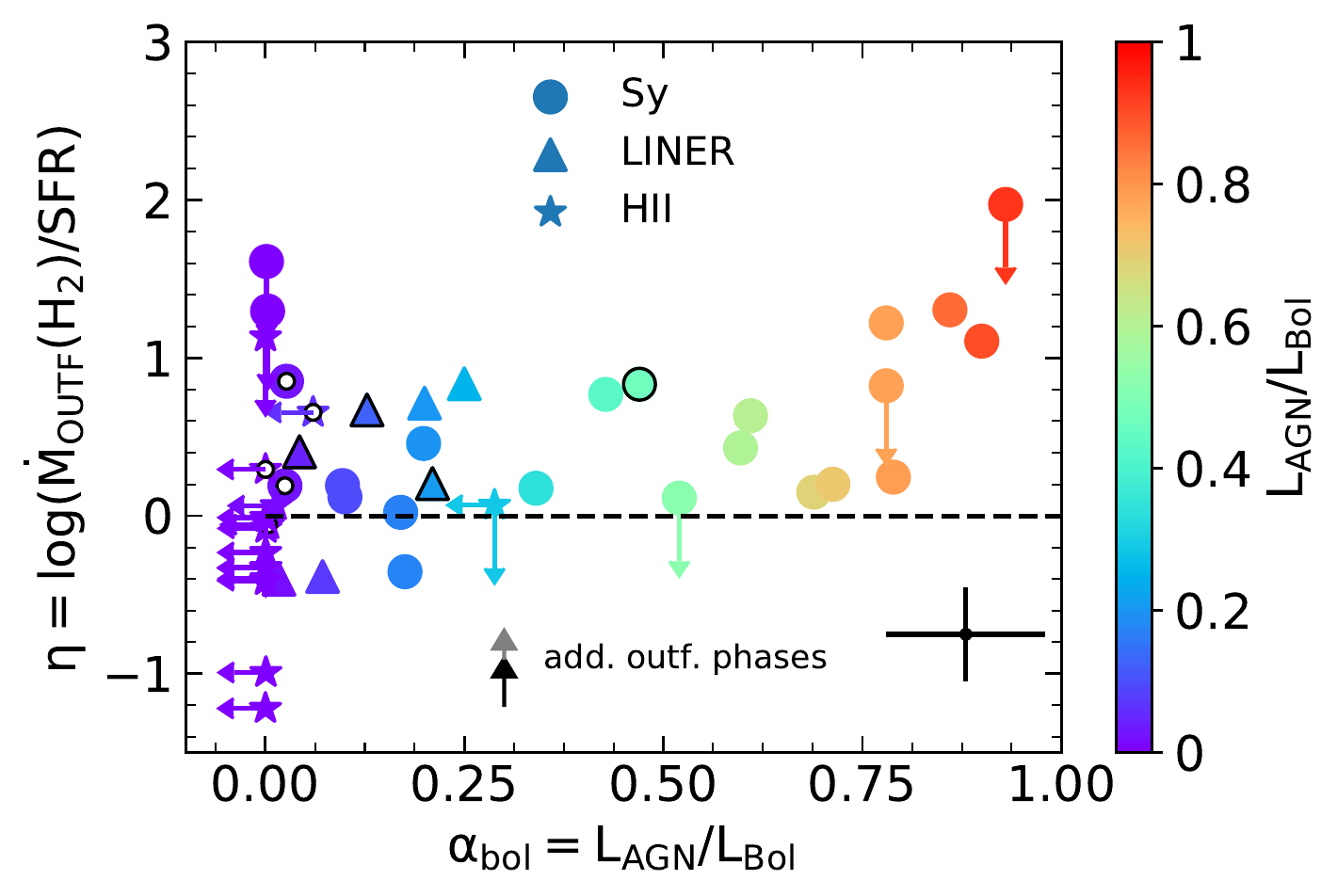}
\caption{Mass loading factor $\eta$ = $\dot{M}_{\rm outf}(\rm H_2)$/SFR
as a function of AGN fractional contribution to the bolometric luminosity, $L_{\rm AGN}$/$L_{\rm bol}$.
The black dashed line shows the relation for an outflow mass-loading factor $\eta$ = 1.
Colour-coding and symbols are as in Fig. \ref{fig:molec_ion}.}
\label{fig:loading_LAGN}
\end{figure}

\subsection{Mass outflow rate scaling relations}
\label{sec:scaling_relations}

In this section we start investigating the scaling relations between the molecular outflow rate and galaxy properties, with the goal of obtaining a first
indication of the driving mechanism in different regimes.

\subsubsection{Dependence on SFR and L$_{\rm AGN}$}

Figure \ref{fig:OFrate_SFR} shows the molecular mass outflow rate $\dot{M}_{\rm outf}(\rm H_2)$ as a function of the SFR, colour-coded by AGN contribution to the bolometric luminosity.
Similar to what was found in smaller samples in previous works \citep{Cicone2014,
Garcia-Burillo2015}, the star forming/starburst galaxies have a mass-loading factor $\eta$ = $\dot{M}_{\rm outf}(\rm H_2)$/SFR consistent with unity or slightly lower.
As we have discussed in the previous section, in star forming galaxies the contribution to the total mass-loss rate is similar for different gas phases
(ionized/neutral atomic and molecular). By including
all the gas phases, the total mass-loss rate increases roughly by 0.5 dex, which is indicated by the grey arrow, and which brings the total
loading factor closer to (or exceeding) unity for star forming
galaxies. However, for the moment we focus on the molecular outflow rate. The best-fit of the relation between molecular outflow rate and
SFR for SF galaxies (shown as a dashed blue line in Fig. \ref{fig:OFrate_SFR}) is log($\dot{M}_{\rm outf}(\rm H_2)$/(M$_{\odot}$ yr$^{-1}$)) =  
1.19$^{+0.16}_{-0.16}$log(SFR/(M$_{\odot}$ yr$^{-1}$)) -- 0.59$^{+0.28}_{-0.28}$. This and the following fits are performed by using linmix \citep{Kelly2007}, considering the error bars both in x and y
and including upper limits.

The AGN host galaxies have a mass-loading factor larger than unity, especially those 
that are AGN-dominated, and $\eta$ ranges from a factor of a few up to a hundred. The best-fit relation for AGN host galaxies is
log($\dot{M}_{\rm outf}(\rm H_2)$/(M$_{\odot}\: \rm yr^{-1}$)) = 0.76$^{+0.11}_{-0.11}$log{(SFR/(M$_{\odot}$ yr$^{-1}$))} +
0.85$^{+0.18}_{-0.18}$ and is shown with a red
dashed line in Fig. \ref{fig:OFrate_SFR}. As we have discussed in the previous subsection, in AGN host galaxies the atomic phase makes,
on average, a comparable
contribution to the outflow rate as the molecular phase, while the ionized phase is generally negligible, at least in the luminosity range probed
by us. The effect
of including the atomic component of the outflow for AGN is shown with a black arrow.

The outflow properties inferred in star forming galaxies are in good agreement with models predicting a mass-loading factor $\eta$ close to 1
\citep[e.g.][]{Finlator2008, Dave2011, Heckman2015}, where feedback from supernovae is the main outflow driver and required
to properly regulate star formation in galaxies.

Galaxies containing an AGN have loading factors larger than 1 indicating that gas is removed at a faster rate than stars are formed. In particular,
the presence of a strong AGN in the galaxy increases the (molecular) outflow mass loading-factor substantially.
In particular, the higher the AGN contribution ($\alpha_{\rm bol}$, see colour-coding in Fig. \ref{fig:OFrate_SFR}), the higher their mass-loading factor $\eta$.
This is illustrated even more clearly in Fig. \ref{fig:loading_LAGN}, where the relation between the outflow loading factor, i.e. $\eta = \dot{M}_{\rm
outf}(\rm H_2)$/SFR and $\alpha _{\rm bol}=L_{\rm AGN}$/$L_{\rm bol}$ is shown. However, a
correlation is only seen at $L_{\rm AGN}$/$L_{\rm bol}$ > 0.7, while at 0.1 < $L_{\rm AGN}$/$L_{\rm bol}$ < 0.7, the loading factor $\eta$ simply
scatters between 1 and 10 for AGN. As we will discuss further later
on, this is probably due to two effects: 1) additional contribution from star formation to the outflow rate (which, however, is expected to contribute only
with $\eta$ $\sim$ 1); 2) the fact that the outflow has much longer time-scale (> 10$^{6}$~yr) than the AGN accretion variability ($\sim$ 10
-10$^{5}$~yr) \citep{2000A&A...355..485G, doi:10.1093/mnras/stv1136}, hence the outflow is expected to generally outlast an AGN which has
recently switched off, or decreased in luminosity.

Figure \ref{fig:OFrate_LAGN} shows the correlation between the mass outflow rate and $L_{\rm AGN}$. 
The dashed line shows the best-fit to the AGN host galaxies (LINERs, Seyfert 1 and Seyfert 2), excluding purely star forming
galaxies (optically classified as star forming/starburst), which gives the following relation: log($\dot{M}_{\rm outf}$/(M$_{\odot}$ yr$^{-1}$)) = 0.68$^{+0.10}_{-0.10}$$\mathrm{log(}L_{\rm AGN}$/(erg s$^{-1}$)) - 28.5$^{+4.6}_{4.6}$. Comparing with the predictions
from chemo-hydrodynamic simulations \citep{Richings2018}, the observed values are about 1~dex higher at $L_{\rm AGN}$ = 10$^{44}$~erg s$^{-1}$, but are consistent with simulations
at $L_{\rm AGN}$ $\approx$ 10$^{46}$~erg s$^{-1}$ within the errors.
Although Seyfert galaxies show a correlation between AGN luminosity and molecular outflow mass rate, suggesting that these outflows are AGN-driven, this correlation is looser and with
larger scatter than previously
found in the literature \citep[e.g.][]{Cicone2014,Fiore2017}, probably as a consequence of our sample being less biased.
Nevertheless, this is still supportive of the scenario in which luminous AGN boost the outflow rate by a large factor and nearly proportionally to the AGN radiative power. 
It is interesting to note that Fig. \ref{fig:OFrate_LAGN} clearly shows the presence of a significant fraction of galaxies (indicated by symbols with a central white dot) 
with high outflow rates but little AGN contribution and, for those classified as AGN, clearly not following the correlation observed
for the bulk of luminous AGN host galaxies. As discussed above, 
 this is partly due to contribution by star formation, but the bulk of the effect may be due to `fossil' AGN-driven outflows as it will
 be clarified in
 the Section \ref{sec:discussion}.

\begin{figure}
\centering
\includegraphics[width=\columnwidth]{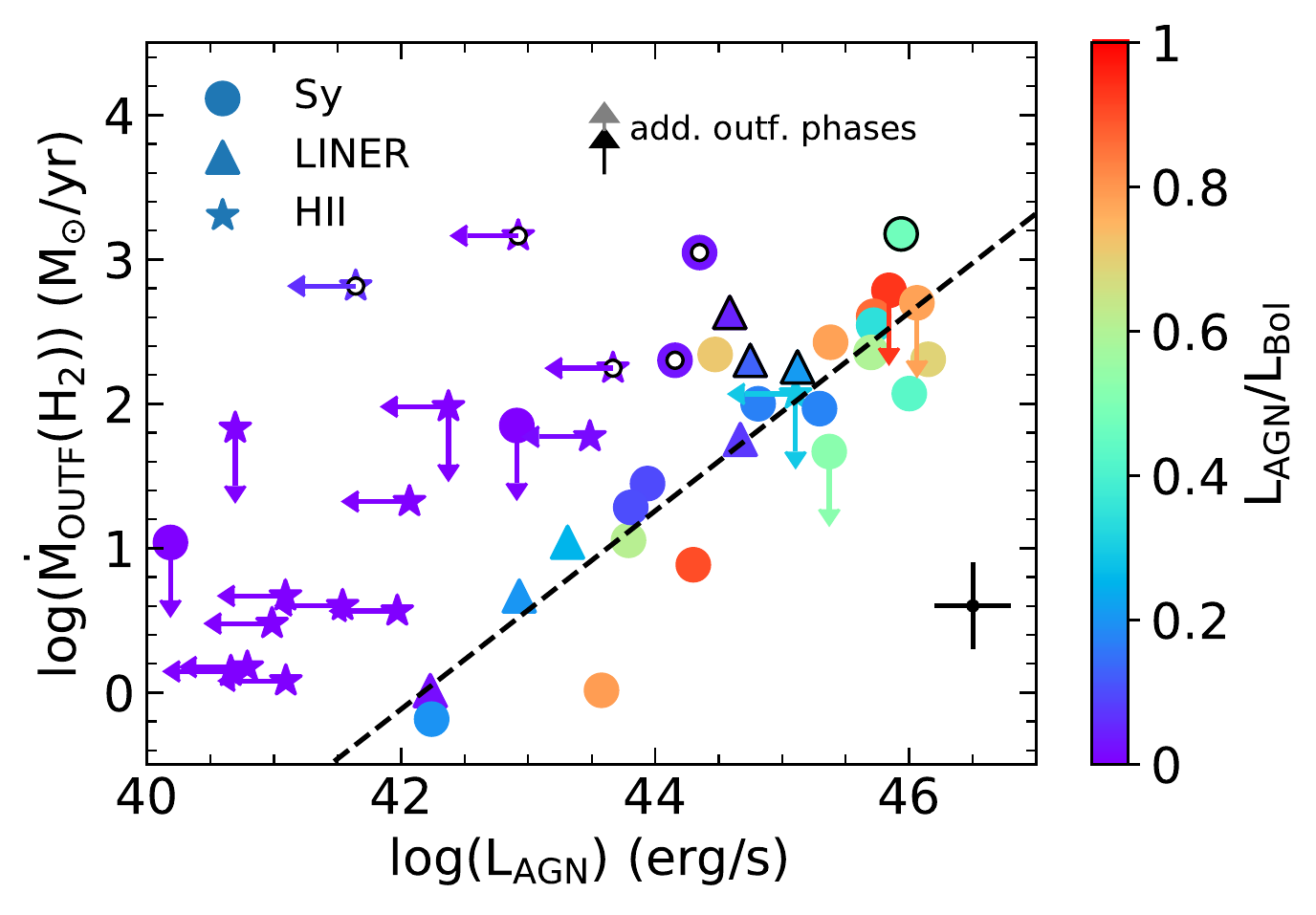}
\caption{Mass outflow rate as a function of AGN (bolometric) luminosity. The dashed line indicates the fit to the
AGN host galaxies (LINERs, Seyfert 1 and 2). Colour-coding and symbols are as in Fig. \ref{fig:molec_ion}.}
\label{fig:OFrate_LAGN}
\end{figure}

\begin{figure}
\centering 
\includegraphics[width=\columnwidth]{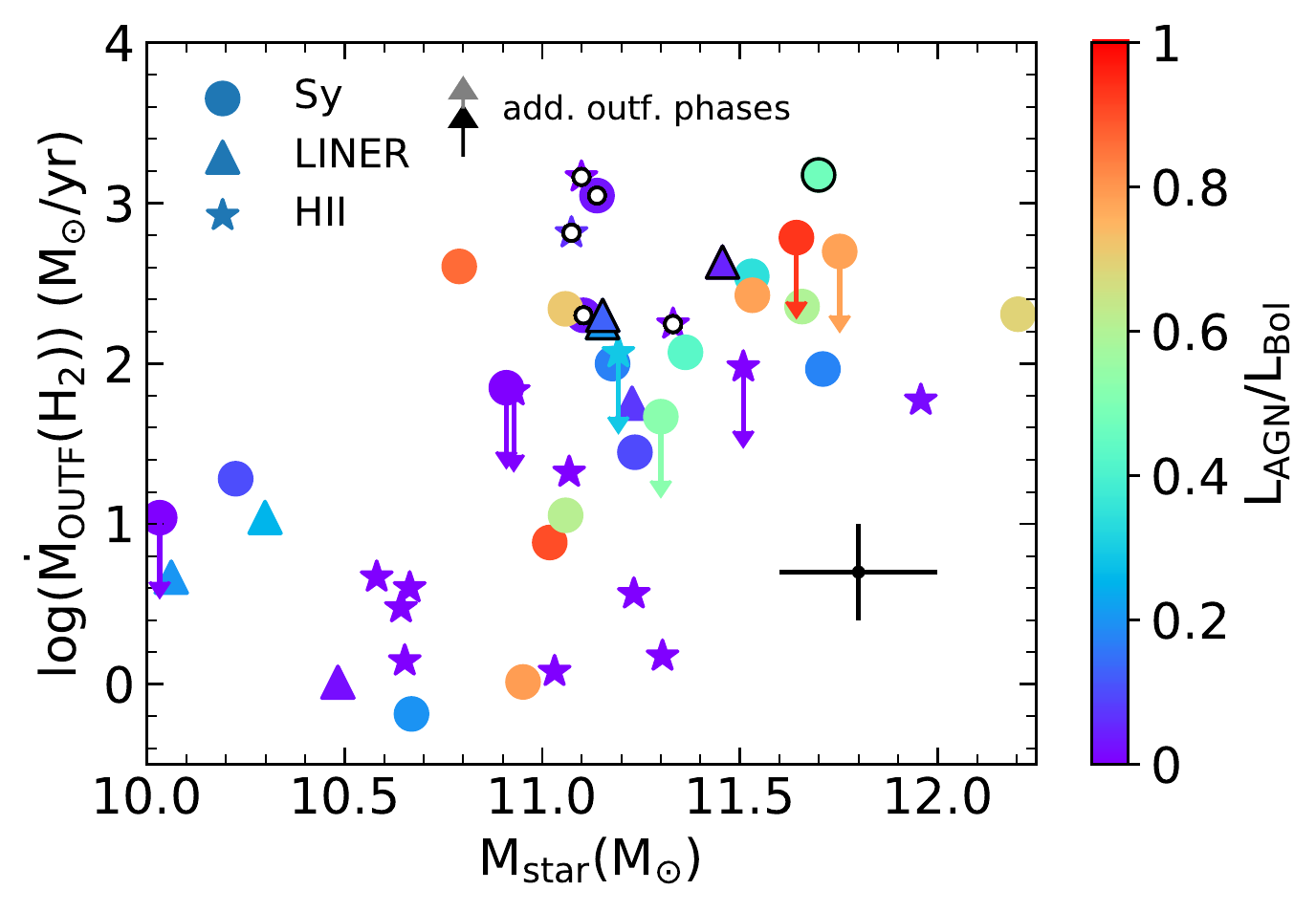}
\caption{Molecular outflow mass rate as a function of galaxy stellar mass. Colour-coding and symbols are as in Fig. \ref{fig:molec_ion}.}
\label{fig:OFrate_vs_Mstar}
\end{figure}

\begin{figure}
\centering
\includegraphics[width=\columnwidth]{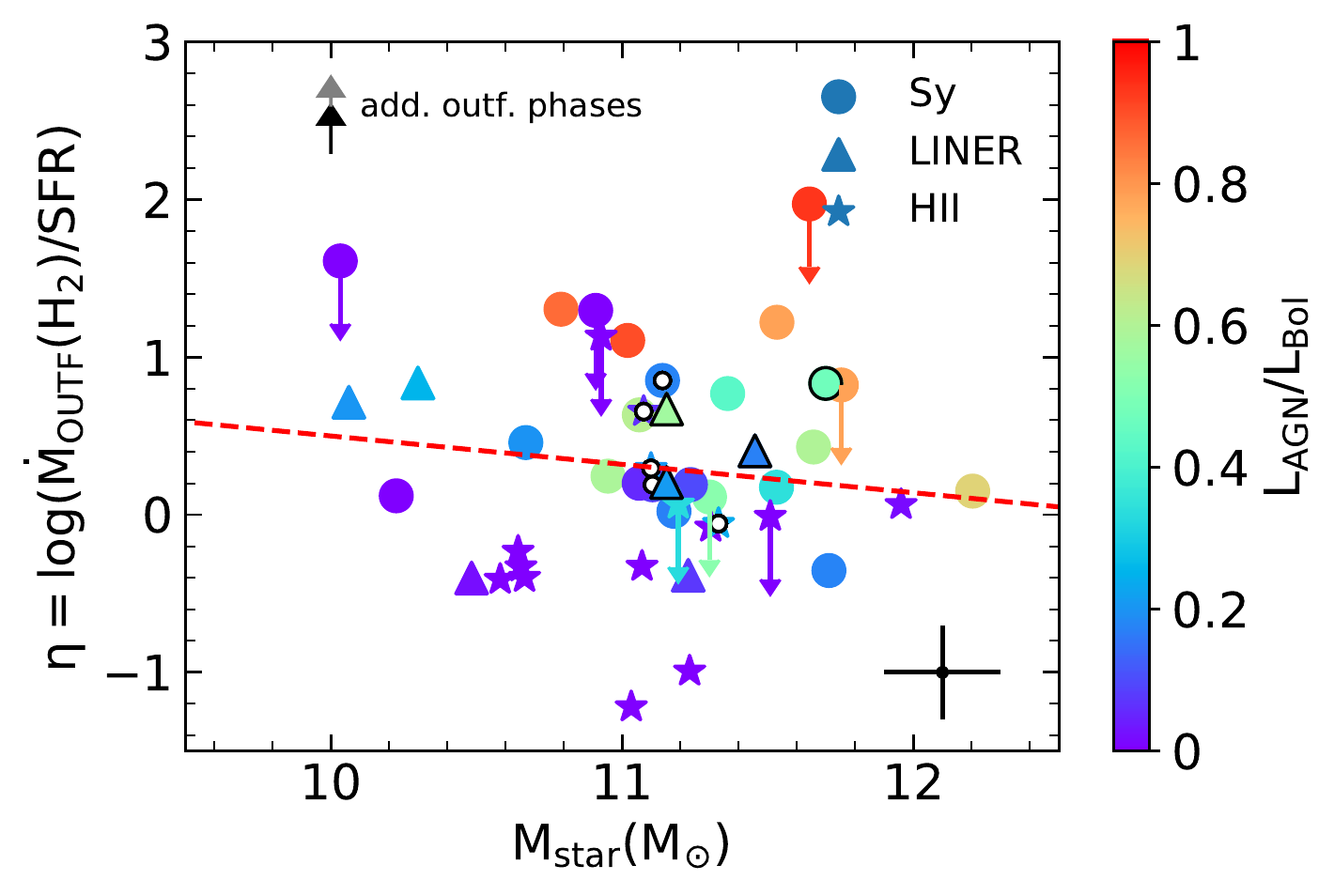}
\caption{Mass-loading factor as a function of stellar mass. The red dashed line shows the best-fit to the data. Colour-coding and symbols are as
in Fig. \ref{fig:molec_ion}.}
\label{fig:eta_Mstar}
\end{figure}

\subsubsection{Dependence on galaxy stellar mass}
\label{sec:scal_mass}

Fig. \ref{fig:OFrate_vs_Mstar} shows the outflow rate as a function of stellar mass. This plot shows some correlation, which may be indirectly linked to
the correlation between outflow rate and SFR, through the stellar mass-SFR relation for galaxies on the `main sequence'.
An important prediction of theoretical models of feedback from star formation is that the outflow
loading factor should anti-correlate with the galaxy stellar mass as $\eta$ $\propto$ $M_{\star}^{-0.5}$, as a consequence of the deeper gravitational potential
well in more massive galaxies \citep{Mitra2015, Somerville2014, Chisholm2017}.
Figure \ref{fig:eta_Mstar} 
shows the dependence of the mass loading factor $\eta$ on the stellar mass. Clearly,
the observed relation between outflow mass-loading factor and stellar mass is very scattered. A linear regression indicates that there is only a weak
anti-correlation of the form log($\eta$) = -0.18 $^{+0.24}_{-0.24}$log($M_{\star}$/M$_{\odot}$) + 2.3$^{+2.7}_{-2.6}$, i.e. 
only marginally consistent with theoretical predictions.
However, before invoking any tension with theoretical models one should be aware of three main issues: 1) we include AGN-driven and star formation-driven outflows, whereas
the models make predictions about outflows driven by SNe and stellar radiation pressure, 2) the range of stellar masses is probably too narrow to properly test the
theoretical predictions, especially given that the dependence of the outflow rate on mass is weak (slope of --0.5 in log); 3) thirdly, the simple relation of outflow rate
with stellar mass is convolved with the dependence on SFR and with the AGN contribution, which likely dominate the scatter of any relation with $M_{\star}$.
We address the last issue in the following subsection.


\subsubsection{Disentangling the outflow dependence on host galaxy parameters}
\label{sec:MLR}
In the previous subsections we have shown how the outflow rate depends on galaxy properties (for which
sensitive CO observations are available,  such as stellar mass, star formation rate and the luminosity of the AGN.
However, it is difficult to isolate the role played by each of these quantities, especially given that they are correlated.
In this section, we attempt to disentangle the contribution of these different factors.

\begin{figure}
\centering 
\includegraphics[width=\columnwidth]{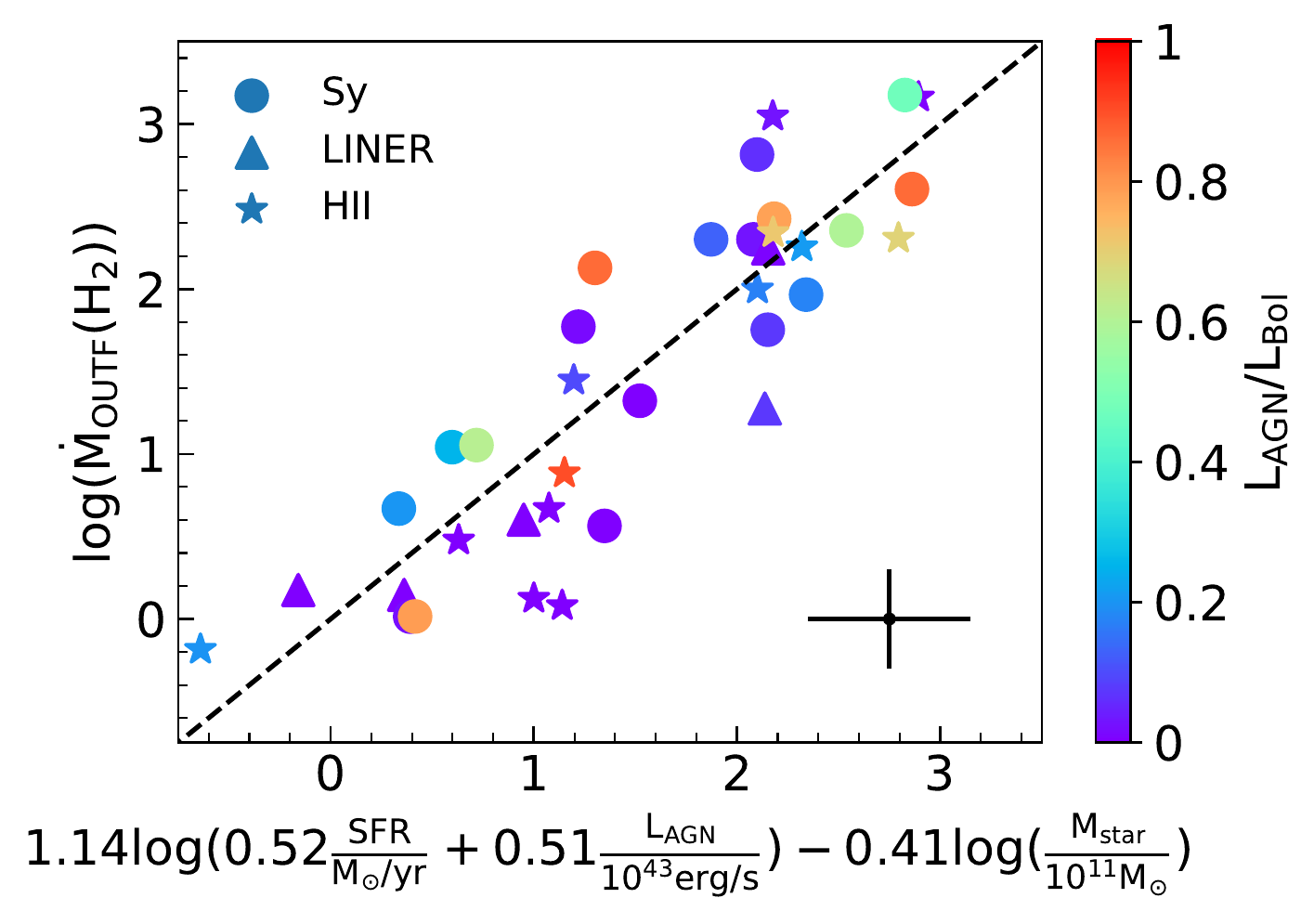}
\caption{Simultaneous multiple linear regression fit of the molecular outflow rate as a function of star formation rate,
stellar mass and AGN luminosity, as given in equation \ref{eq:MLR_H2}. Colour-coding and symbols are as in Fig.
\ref{fig:molec_ion}.}
\label{fig:MLR}
\end{figure}

\begin{figure}
\centering 
\includegraphics[width=\columnwidth]{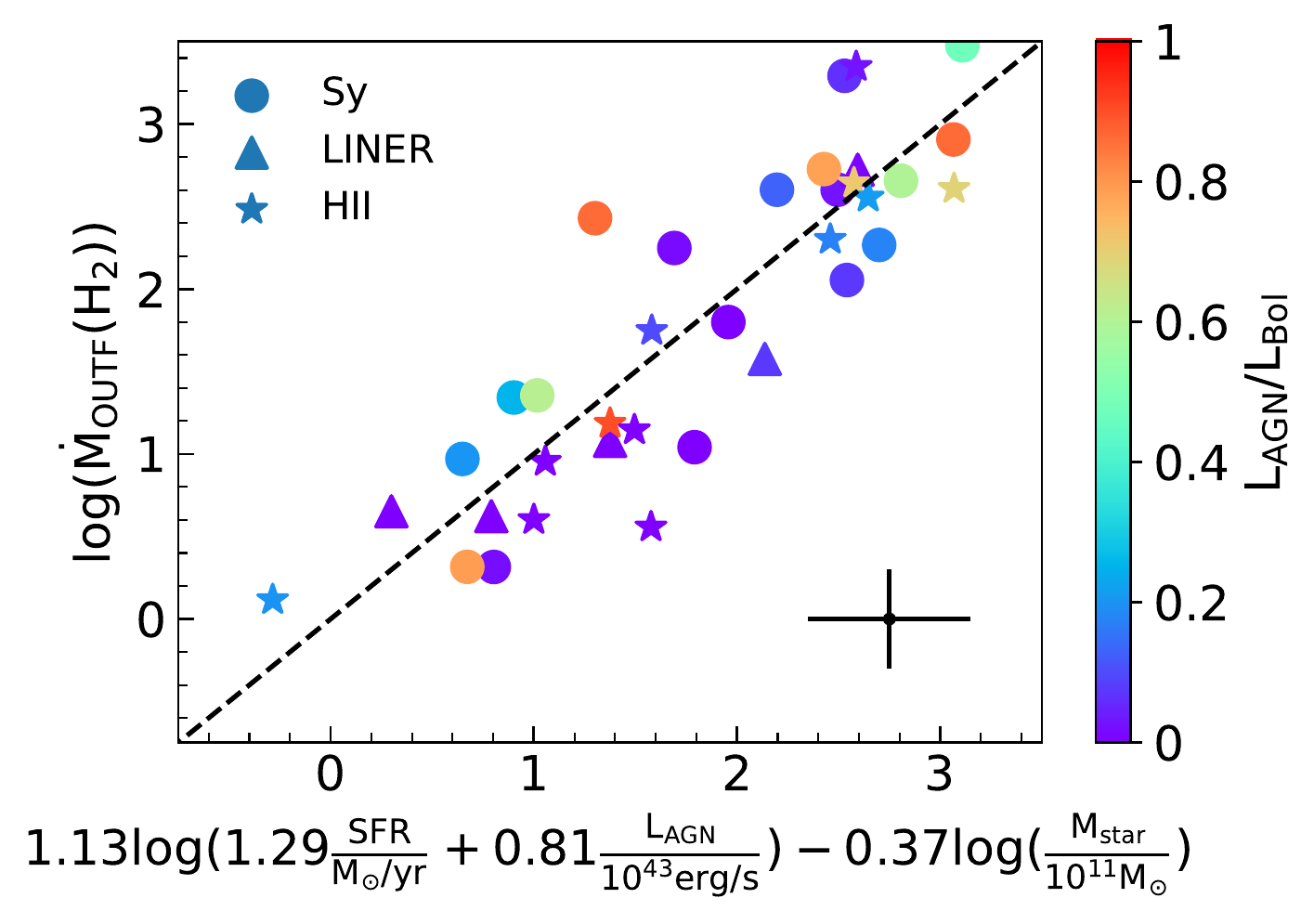}
\caption{Simultaneous multiple linear regression fit of the {\it total} outflow rate as a function of star formation rate,
stellar mass and AGN luminosity, as given in equation \ref{eq:MLR_tot}. Colour-coding and symbols are as in Fig.
\ref{fig:molec_ion}.}
\label{fig:MLR_tot}
\end{figure}

\begin{figure*}
\centering
\begin{subfigure}[t]{.015\textwidth}
\text{(a)}
\end{subfigure}
\begin{subfigure}[t]{.47\textwidth}
 \captionsetup{width=.9\linewidth, position=top}
 \includegraphics[width=\columnwidth,trim = {0cm 0.3cm 0cm 0cm},clip]{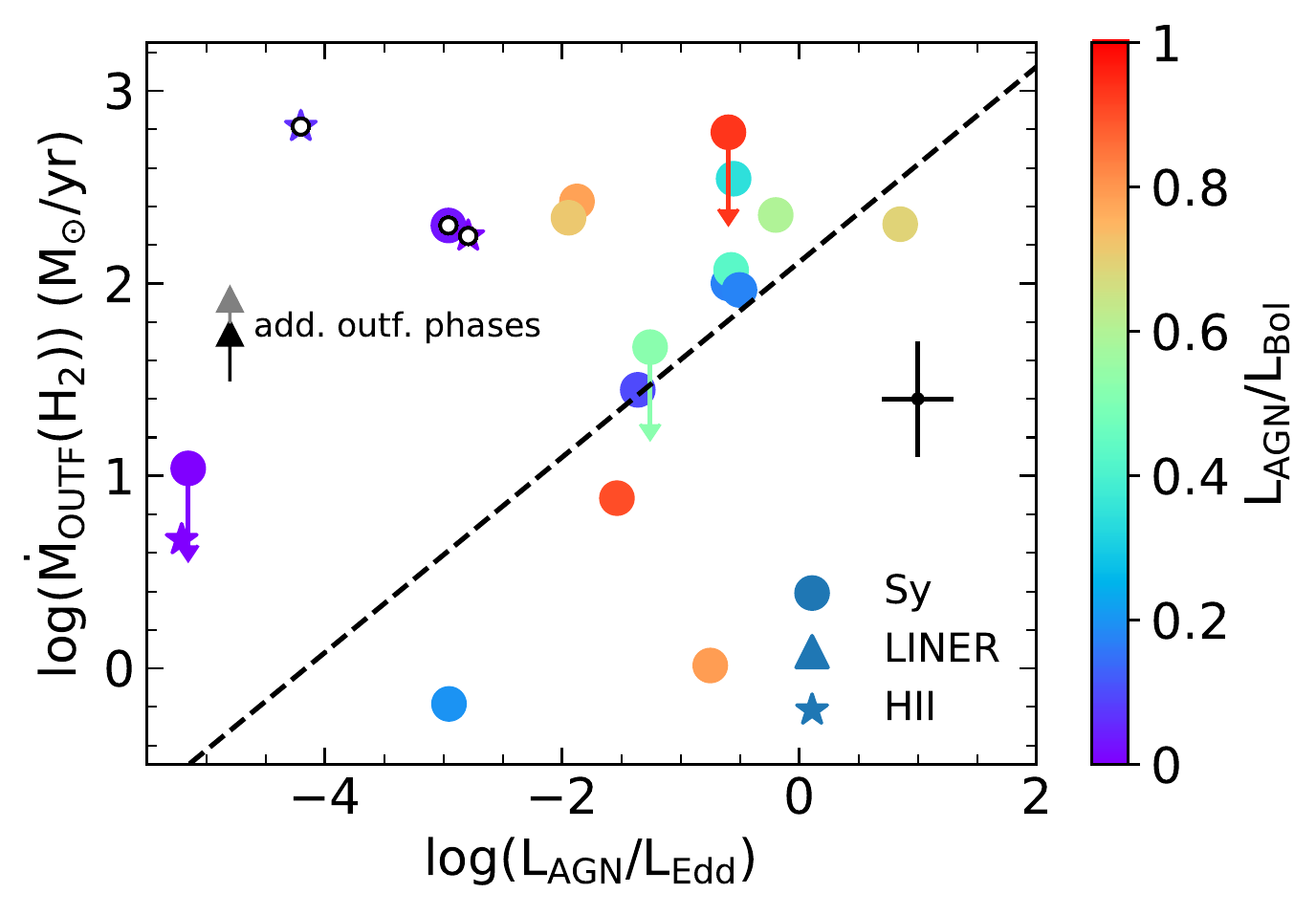} 
 \label{fig:OFrate_Edd}
\end{subfigure}%
\begin{subfigure}[t]{.015\textwidth}
\text{(b)}
\end{subfigure}
\begin{subfigure}[t]{.47\textwidth}
\centering
\captionsetup{width=.9\linewidth}
\includegraphics[width=\linewidth,trim = {0cm 0.3cm 0cm 0cm},clip]{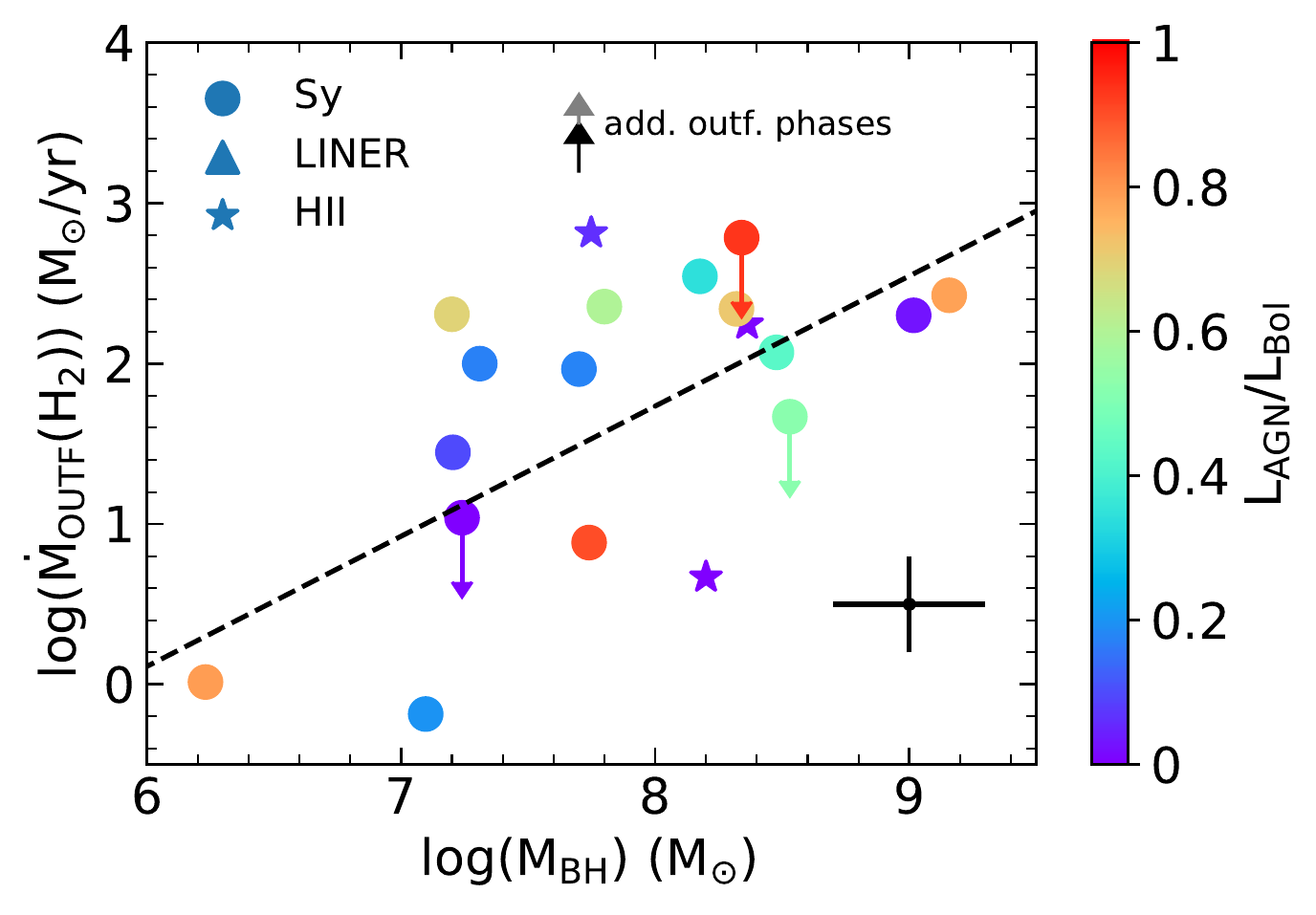}
 \label{fig:OFrate_MBH}
\end{subfigure}
\caption{Outflow rate as a function of Eddington ratio, i.e. $L_{\rm AGN}/L_{\rm Edd}$.
  Outflow rate a s function of black hole mass. Colour-coding and symbols are as in Fig. \ref{fig:molec_ion}.}
\label{fig:Ofrate_bh_edd}
\end{figure*}

For this purpose, we performed a regression as follows:

\begin{eqnarray}
\mathrm \log{(\dot{M}_{\mathrm{outf}})} = \mathrm{x\: log(}\alpha \mathrm{SFR +} \beta L_{\rm AGN}) + \mathrm{y\: log} M_{\star},
\label{eq:MLR}
\end{eqnarray}
and finding the values of the parameters that minimize the dispersion around this relation.
The reason for using this expression, is that for starburst galaxies we only have an upper limit on the AGN luminosity. Combining the SFR and
AGN in the term in parenthesis ensures that this term never diverges to very negative values in log, i.e. it ensures
that when we investigate galaxies
with outflow there is always a driving mechanism, either SF or AGN. We have excluded our candidate fossil outflows as they are expected not 
to follow a relation with AGN or SFR, although AGN variability will still be a source of scatter.

The resulting best fit is:

\begin{eqnarray}
\begin{split}
\mathrm{log(}\dot{M}_{\mathrm{outf}}(\mathrm{H_{2}})/(\mathrm{M_{\odot}\: yr^{-1}))} = 1.14 \: \mathrm{log}\Big(0.52 \frac{\mathrm{SFR}}{\mathrm{M_{\odot}\: yr^{-1}}} \\ + 0.51 \frac{L_{\rm AGN}}{\rm 10^{43} erg\: s^{-1}}\Big) -\mathrm{0.41 \: log} \Big(\frac{M_{\star}}{\rm 10^{11}M_{\odot}} \Big),
\end{split}
\label{eq:MLR_H2}
\end{eqnarray}
with one standard deviation errors on the four parameters
being $\Delta$(x,$\alpha$,$\beta$,y) = (0.12,0.19,0.25,0.25). The resulting relation is shown in Fig. \ref{fig:MLR}.
Clearly the large dispersion seen in the previous plots (outflow rate vs $L_{\rm AGN}$, vs SFR and vs $M_{\star}$
separately) is greatly reduced in this relation, indicating that we are simultaneously capturing the contribution of these three factors
to the outflow rate. Very interestingly, this relation enables us to disentangle (at least partly) the contribution of the three factors to the outflow rate.
The dependence on stellar mass is now seen more clearly: the dependence has a power law index of --0.41, which is very close to the value
expected by theory of --0.5 for outflows driven by star formation. As our sample also includes AGN-driven outflows, it is likely that these
have mass-loading factors which decrease with stellar mass, too.
We cannot disentangle in this kind of analysis the power-law index of the dependence on AGN luminosity and SFR separately.
With the functional form adopted by us
the combined dependence has a power law index of 1.1, i.e. a nearly linear relation, as expected in many models at least for the SFR.

However, this relation only accounts for the molecular phase of the outflow. As discussed in Sect. \ref{sec:multiphase}, including the atomic-neutral
and ionized phases is difficult because we do not have enough statistics in terms of galaxies which have all three outflow phases measured.
However, as mentioned in Sect. \ref{sec:multiphase}, we can roughly account for these two phases by including a factor of three for star forming
galaxies (as they have an ionized and atomic outflow rates that are similar to the molecular outflow rate) and a factor of two for
AGN-dominated galaxies (as they have an atomic outflow rate similar to the molecular outflow rate and a negligible contribution from
the ionized outflow rate, at least in our luminosity range). In this case the resulting best fit for the {\it total} outflow rate
is given by

\begin{eqnarray}
\begin{split}
\mathrm{log(}\dot{M}_{\mathrm{outf}}(\mathrm{tot})/\mathrm{(M_{\odot}\: yr^{-1})}) = \mathrm{1.13\: log}\Big(1.29 \frac{\mathrm{SFR}}{\mathrm{M_{\odot}\: yr^{-1}}} \\ + 0.81 \frac{L_{\rm AGN}}{\mathrm{10^{43} erg\: s^{-1}}}\Big) - \mathrm{0.37\: log} \Big(\frac{M_{\star}}{\rm 10^{11}M_{\odot}} \Big),
\end{split}
\label{eq:MLR_tot}
\end{eqnarray}
with one standard deviation errors on the four parameters being $\Delta$(x,$\alpha$,$\beta$,y) = (0.55,0.45,0.12,0.24).
The resulting fit is shown in Fig. \ref{fig:MLR_tot}, which has a scatter even smaller than in Fig. \ref{fig:MLR}.

These {\it global} relations can be used to infer the expected outflow rate in any kind of galaxies, and provide an appropriate
comparison for the theoretical models and simulations.

\subsubsection{Dependence on L$_{\rm AGN}$/L$_{\rm Edd}$}

In the previous subsections we have investigated the dependence on nuclear activity in terms of AGN absolute luminosity.
However, both in energy-driven outflows and radiation pressure-driven outflows (the two main mechanisms proposed for AGN outflows)
the most fundamental quantity is the AGN luminosity relative to the Eddington limit, $L_{\rm AGN}$/$L_{\rm Edd}$. This
quantity is more difficult to determine as it requires an estimate of the black hole mass. The latter has been inferred only for about half of the galaxies in our
sample with a variety of methods (primarily through virial estimators) and subject to large uncertainties. The major contribution to this uncertainty stems from the virial coefficient $f$, which shows a scatter of 0.44 \citep{Woo2014}. Fig. \ref{fig:Ofrate_bh_edd}a shows the outflow rate as a function
of Eddington ratio. If one excludes SF-dominated galaxies, which are driven by a different mechanism (see also discussion in the next sections),
the plot shows some correlation between outflow rate and Eddington ratio, although with a few points subject to large scatter. Such a scatter
could partly be accounted for by the uncertainties in the black hole masses. Additional discussion on this dependence will be given in Sect.
\ref{sec:discussion}.

\subsubsection{Dependence on black hole mass}

In Fig. \ref{fig:Ofrate_bh_edd}b we also show the outflow rate as a function of black hole mass. In principle one should not expect any correlation of the outflow
rate with the black hole mass, but the
plot clearly shows a significant correlation. Such a correlation was already identified by \cite{Rupke2017}, although with lower statistics.
One interpretation is that this correlation is simply a consequence of the correlation
between outflow rate and stellar mass (Sect. \ref{sec:scal_mass}), through the black hole--galaxy mass relation. However, another possibility is that the
correlation between outflow rate and black hole mass traces the average driving effect that the black hole has during its intermitted accretion phases. Indeed,
if one assumes that the black hole accretes at a given Eddington fraction $L_{\rm AGN}$/$L_{\rm Edd}$ (e.g. at the average Eddington fraction of the AGN population) and with an average duty cycle, then
the black hole mass may be a tracer of the average AGN activity over the past $\sim 10^6-10^8$~yr,
i.e. on time-scales closer to the outflow dynamical time-scale, hence resulting in the observed correlation. We discuss the effects of the AGN flickering further in the
next sections.

\subsubsection{Dependence on radio power}
\label{sec:depletion_time}

Galactic outflows are seen to also be linked with the presence of radio jets.
The connection appears to be common for what concerns the ionized phase of outflows \citep{Mullaney2013}.
Furthermore, clear indications that some
molecular and atomic outflows are associated with radio jets has been found \citep[e.g.][]{Morganti2013,Morganti2015,Dasyra2015,Dasyra2016}.
However, it is not yet clear how common this association is among molecular outflows.

We have explored this connection in our sample by investigating the correlation of the outflow rate with the excess of radio power relative
to the value expected from the radio--SFR correlation, which is traced by the parameter $q_{\rm IR}$, defined as the ratio between the far-IR flux
and the monochromatic flux at 1.4~GHz (Sect. \ref{sec:radio_data}). Fig. \ref{fig:or_vs_radio} shows the molecular outflow rate as a function of the
parameter $q_{\rm IR}$. The vertical dashed line indicates the average value for star forming galaxies, while the solid vertical line indicates
the limit below which galaxies are considered to have a significant radio excess associated with a radio jet \citep{Ivison2010,Harrison2014}.

Most galaxies in our sample are consistent with the radio luminosity being associated with star formation. Actually it seems
that, on average,  $q_{\rm IR}$ in our sample is even higher than typically observed in normal galaxies, possibly reflecting the bias towards
star-bursting systems, or the contribution of powerful AGN to the infrared emission in some of the galaxies of our sample.

The Fig. \ref{fig:or_vs_radio} shows that  
two of the three galaxies with radio excess
($q_{\rm IR}$ < 1.8) do have strong outflows. However, the plot shows
no clear correlation between molecular outflow rate and excess of radio emission relative to the SFR-radio relation. This finding suggests
that, statistically, the presence of radio jets does not seem to be a primary driving mechanism of the majority of galactic molecular outflows in our
sample. However, this does not imply that strong radio jets cannot cause outflows. In fact, as already discussed at the
beginning of this section,
radio jets have been seen as the origin of powerful outflows in a few specific galaxies.

\begin{figure}
\centering 
\includegraphics[width=\columnwidth]{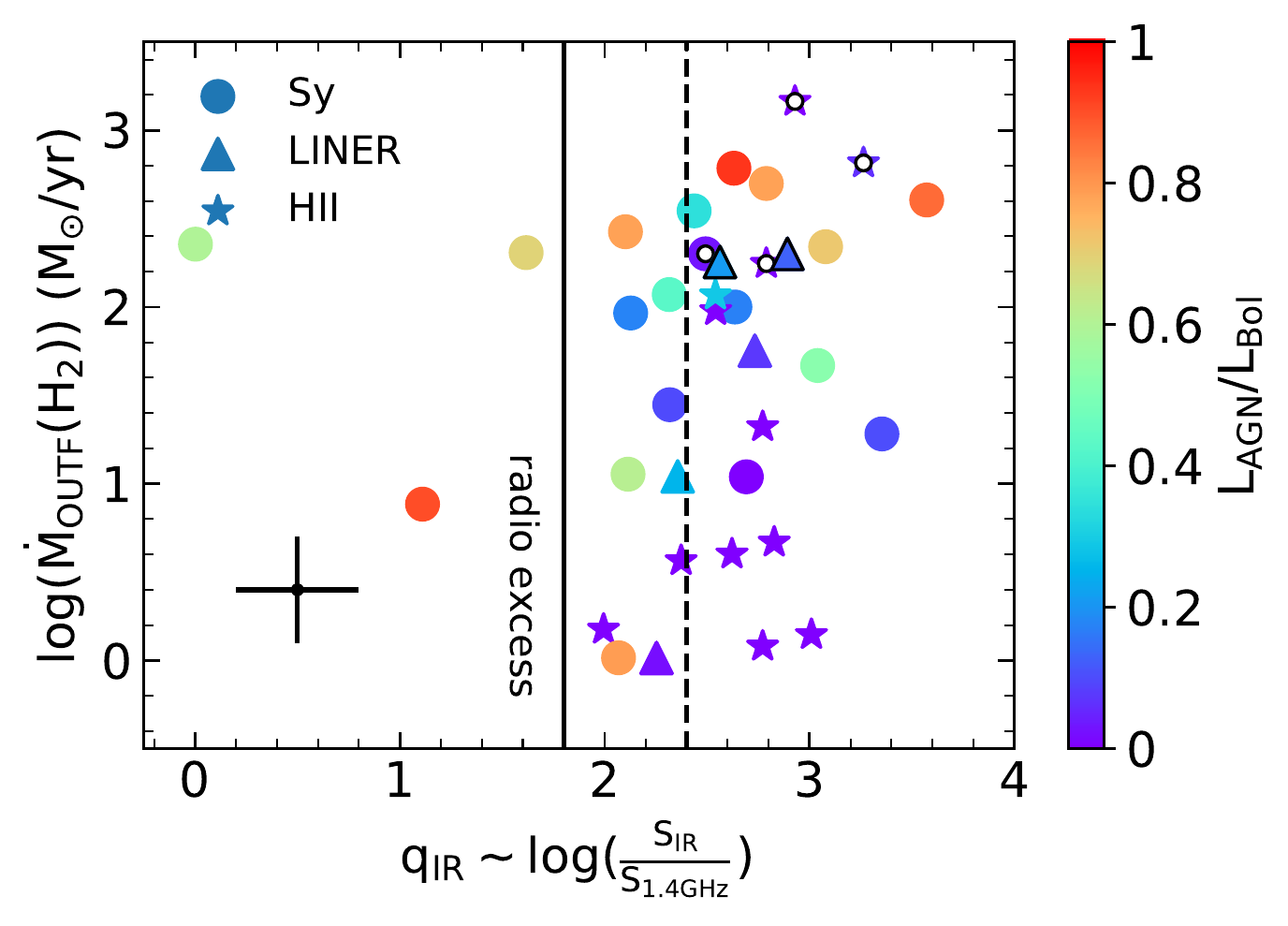}
\caption{Molecular outflow rate as a function of the parameter $q_{\rm IR}$ defined as the ratio between the far-IR flux
and the radio monochromatic flux at 1.4~GHz (Sect. \ref{sec:radio_data}).
The vertical dashed line indicates the average value for star forming galaxies, while the solid vertical line indicates
the limit below which galaxies are considered to have a significant radio excess associated with a radio jet \citep{Ivison2010,Harrison2014}.
Colour-coding and symbols are as in Fig. \ref{fig:molec_ion}.}
\label{fig:or_vs_radio}
\end{figure}

\subsection{Depletion time}
\label{sec:depletion_time}

In the following we estimate the outflow depletion time-scale, defined as $\tau_{\rm depl,outf}$ = $M_{\rm gas}$/$\dot{M}_{\rm outf}$, i.e. the time required to remove all gas from the galaxy with the current
mass outflow rate assuming no fresh supply of additional gas is delivered to the galaxy. We first focus on the depletion time of the molecular
gas, i.e. $\tau_{\rm depl,outf}(\rm H_ 2)$ = $M(\rm H_2)$/$\dot{M}_{\rm outf}(\rm H_2)$, as we have this information for all galaxies in the sample and we will then discuss the total gas depletion
time for galaxies that have information on their HI content. 
In Fig. \ref{fig:depl_LAGN}, we show the relation between molecular depletion time-scales and AGN luminosity. While we do observe an anti-correlation between depletion time-scales and AGN luminosity,
and with AGN contribution to the bolometric luminosity, 
the trend is much more scattered than in previous studies \citep{Sturm2011,Cicone2014}. 
The depletion time-scale of molecular gas for the most powerful AGN is between a few times 10$^6$ and 10$^8$yr.

\begin{figure}
\centering
\includegraphics[width=\columnwidth]{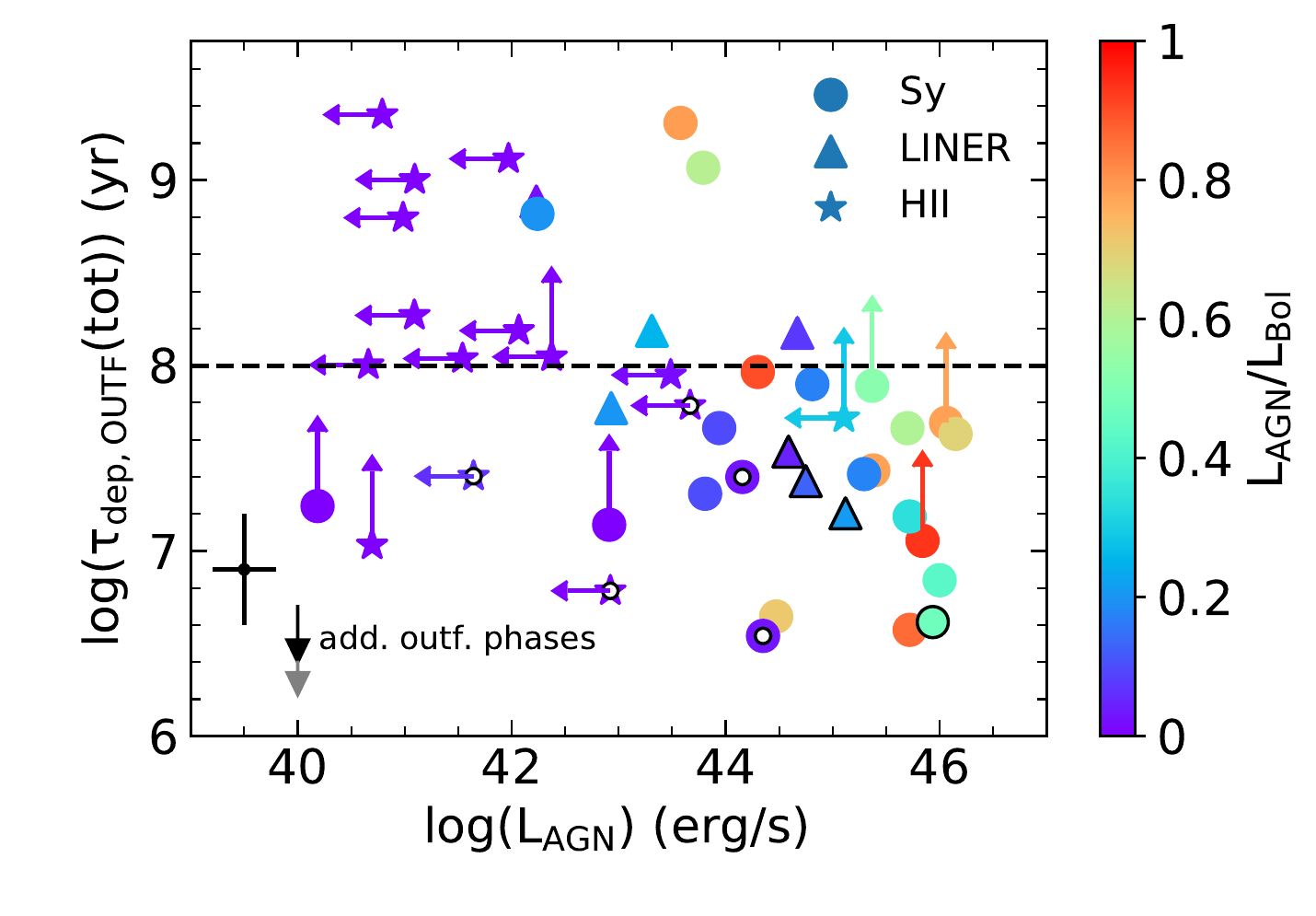}
\caption{Molecular gas depletion time-scale due to outflows as a function of AGN luminosity. Colour-coding and symbols are as in Fig.
\ref{fig:molec_ion}.}
\label{fig:depl_LAGN}
\end{figure}

\begin{figure}
\centering
\includegraphics[width=\columnwidth]{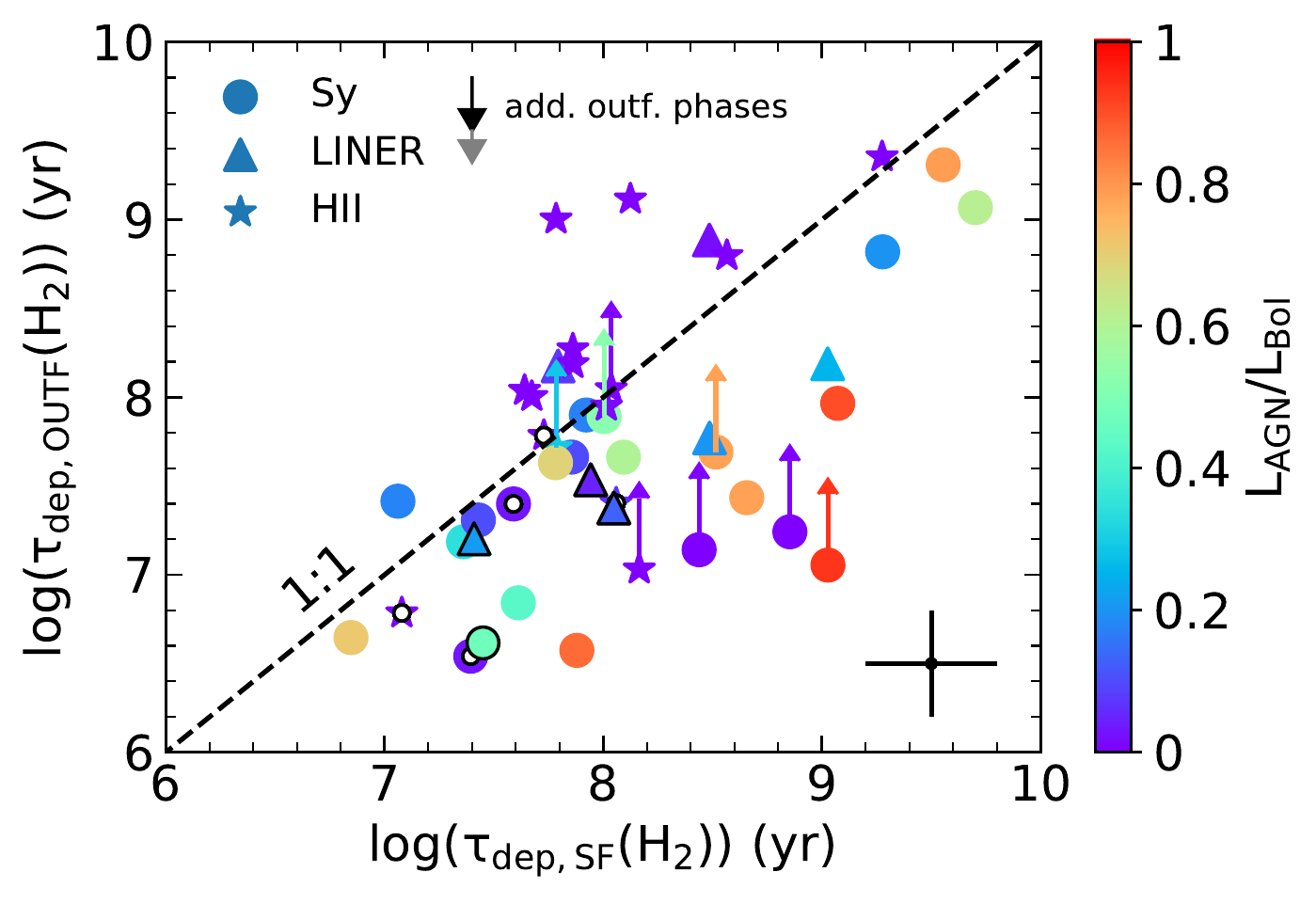}
\caption{Molecular gas depletion time-scale due to outflows vs depletion time-scale due to star formation. Colour-coding and symbols are as in
Fig. \ref{fig:molec_ion}.}
\label{fig:depl_depl}
\end{figure}

\begin{figure}
\centering
\includegraphics[width=\columnwidth]{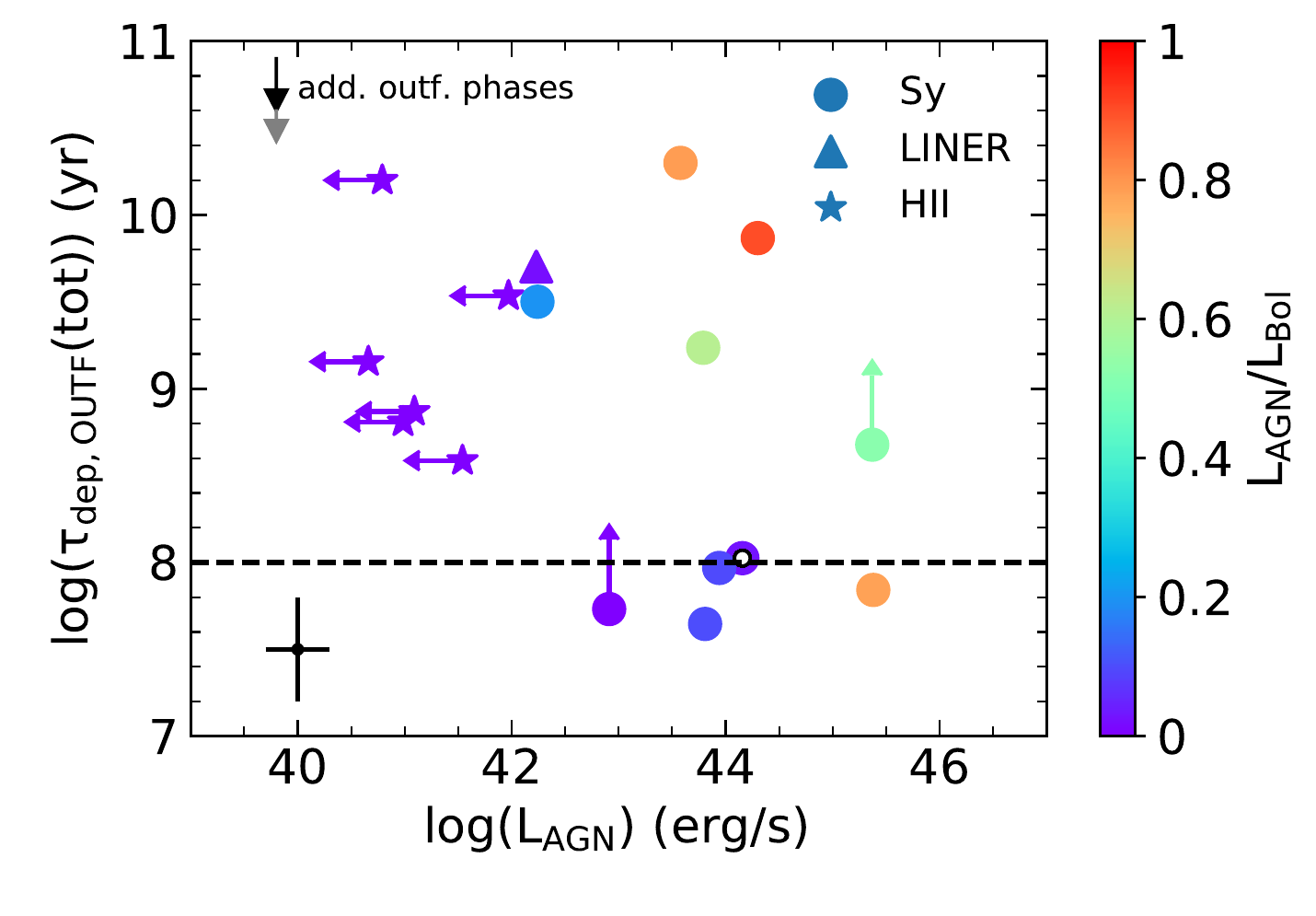}
\caption{Total gas (HI+H$_2$) depletion time-scale as a function of AGN luminosity. Colour-coding and symbols are as in Fig. \ref{fig:molec_ion}.}
\label{fig:depl_tot_LAGN}
\end{figure}

Fig. \ref{fig:depl_depl} shows the depletion time due to outflows compared to the depletion
time-scale due to star formation. For star-forming galaxies, the depletion time due to star formation is similar or shorter than the depletion time due to outflowing gas. For AGN hosts, the depletion 
is dominated by outflows rather than by gas consumption due to star formation, implying that AGN-driven 
outflows play a key role in regulating star formation in galaxies.



For about half of the galaxies we also have information on the atomic gas content, hence we can estimate the total depletion time:
$\tau_{\rm depl}(\rm tot)$ = $M(\rm H_2+HI)$/$\dot{M}_{\rm outf}(H_2)$. This is shown in Fig. \ref{fig:depl_tot_LAGN}, which illustrates that the total depletion time-scale
is much longer, and generally exceeding $10^8$~yr even in most AGN (even if the other gas phases are included, as shown by the black arrow),
implying that the AGN is unlikely to clear the galaxy of its total gas content.

The combination of these various results indicates that AGN-driven outflows are capable of clearing the central parts of galaxies, where the gas content is dominated
by the molecular phase, but the AGN is unlikely to clear the entire galaxy of its gas content.



\begin{figure}
\centering 
\includegraphics[width=\columnwidth]{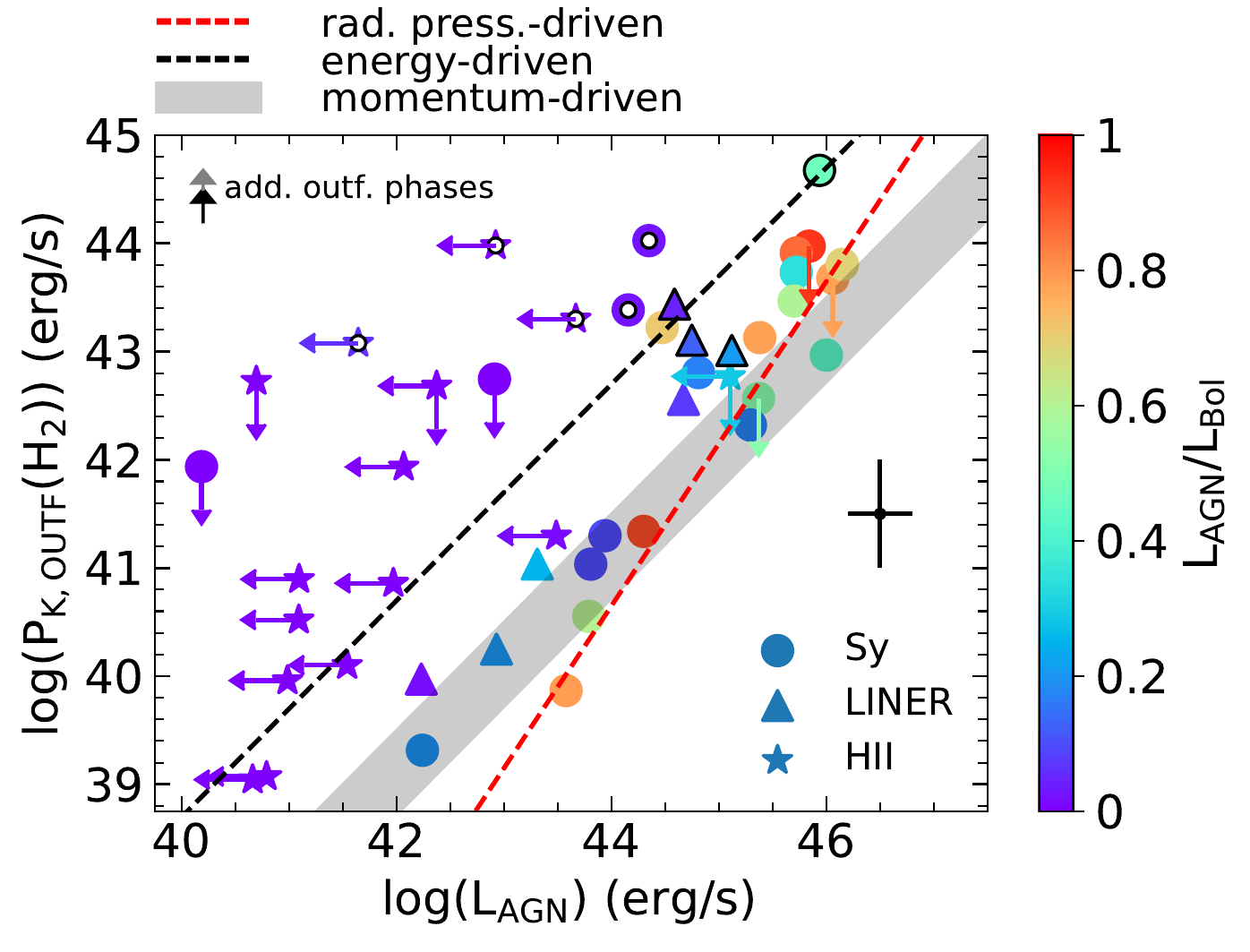}
\caption{Kinetic power ($P_{\rm K, outf}$) of the outflow as a function of the AGN luminosity. The dashed black line indicates the theoretical
prediction of $P_{\rm K} = 0.05 L_{\rm AGN}$ for an energy-driven outflow assuming a coupling efficiency of 100~per cent between the outflow and the ISM. The
prediction for
momentum-driven outflows and some radiation pressure-driven outflows is shown as a shaded region. The red dashed line shows the predicted relation
for the radiation pressure-driven outflow presented in \citet{Ishibashi2018}.
Colour-coding and symbols are as in Fig. \ref{fig:molec_ion}.}
\label{fig:kinpower_LAGN}
\end{figure}
\begin{figure}
\centering 
\includegraphics[width=\columnwidth]{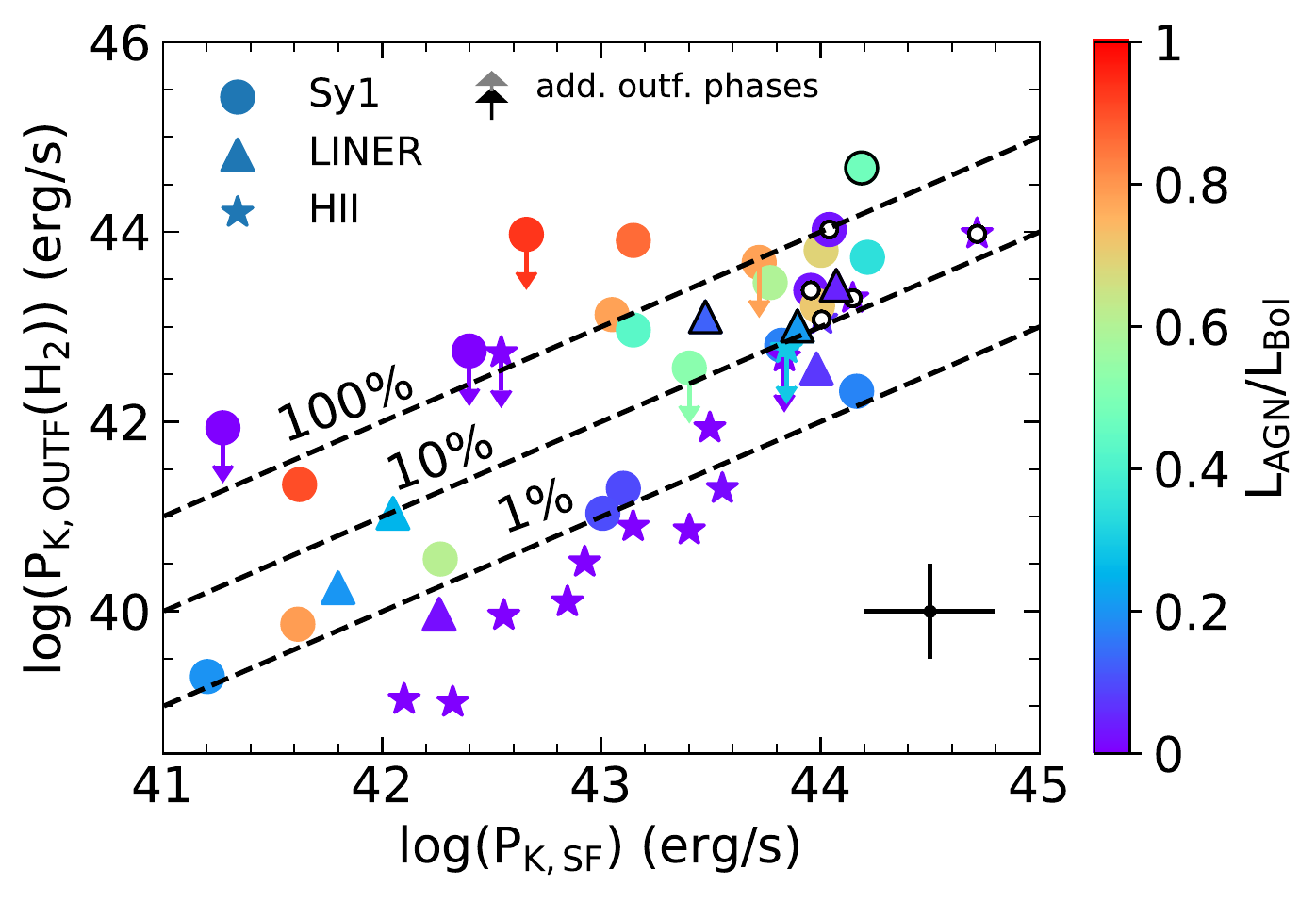}
\caption{Kinetic power of the outflow as a function of the kinetic power generated by supernovae, as inferred from the SFR. The black dashed lines
indicate coupling efficiencies of 1, 10 and 100~per cent. Colour-coding and symbols are as in Fig. \ref{fig:molec_ion}.}
\label{fig:SNepower}
\end{figure}

\subsection{Kinetic power}
\label{sec:kinetic_power}

It is important to investigate the properties of outflows such as their
kinetic power and momentum rate, as different models make different predictions
for these quantities. In this section we briefly discuss the observational results for what concerns the kinetic power, in the next section we will
discuss the momentum rate, while a detailed analysis of implications and comparison with models will be given in Sect. \ref{sec:disc_AGN_drive}.

Fig. \ref{fig:kinpower_LAGN} shows the kinetic power of the outflow (=0.5 \textit{v}$^{2}$$\dot{M}_{\rm outf}$)
as a function of the radiative power of the AGN. 
Clearly, for AGN host galaxies the kinetic power correlates with the AGN luminosity, although the correlation appears to be superlinear.
Moreover, our
more extended, and
less biased sample, with respect to previous studies, reveals a large scatter.

Star forming galaxies follow different relations compared to AGN, as expected since in these sources the observed outflows cannot
have originated from a currently active AGN episode. To test whether star formation can explain why these galaxies are outliers, in Fig. \ref{fig:SNepower} we
compare the kinetic power of the outflow with the power expected to be generated by supernovae ($P_{\rm K,SF}$ = 7$\times$10$^{41}$
SFR (M$_{\odot}$ yr$^{-1}$) \citep{Veilleux2005}. In star forming galaxies, especially those with
low values of $P_{\rm K,SF}$, the kinetic power of the outflow can be explained by supernovae by assuming a coupling efficiency of only
0.5\% (except for a few SF galaxies with extreme outflows discussed further below).
However, accounting for the contribution of the ionized and atomic phases increases the kinetic power of SB-dominated outflows by a factor of about
three (Sect. \ref{sec:multiphase}), as indicated by the grey arrow, suggesting a coupling efficiency of supernova ejecta with the ISM higher than 1\%. Conversely, 
in AGN host galaxies a coupling of $\sim$ 10~per cent or much more is needed; as this is significantly larger than expected by models of SN outflows (especially if accounting
for the other outflow phases, as indicated with the black arrow), this indicates, as expected, that SNe are not powerful enough to drive the
outflow in these objects and that the outflow must be mostly driven by the AGN.

Fig. \ref{fig:kinpower_LAGN} and \ref{fig:SNepower} also clearly indicate that there are a few galaxies for which the kinetic
power greatly exceeds what expected from the AGN energy-driven scenario and
also in excess of what is expected by the
SNe-driven scenario, as a coupling efficiency higher than 10~per cent would be required.
In these cases (objects marked by white dot in their centre) the outflow is likely due to a past, more active phase of the AGN. This will be discussed further in Sect. \ref{sec:fossil}.

\begin{figure}
\centering 
\includegraphics[width=\columnwidth]{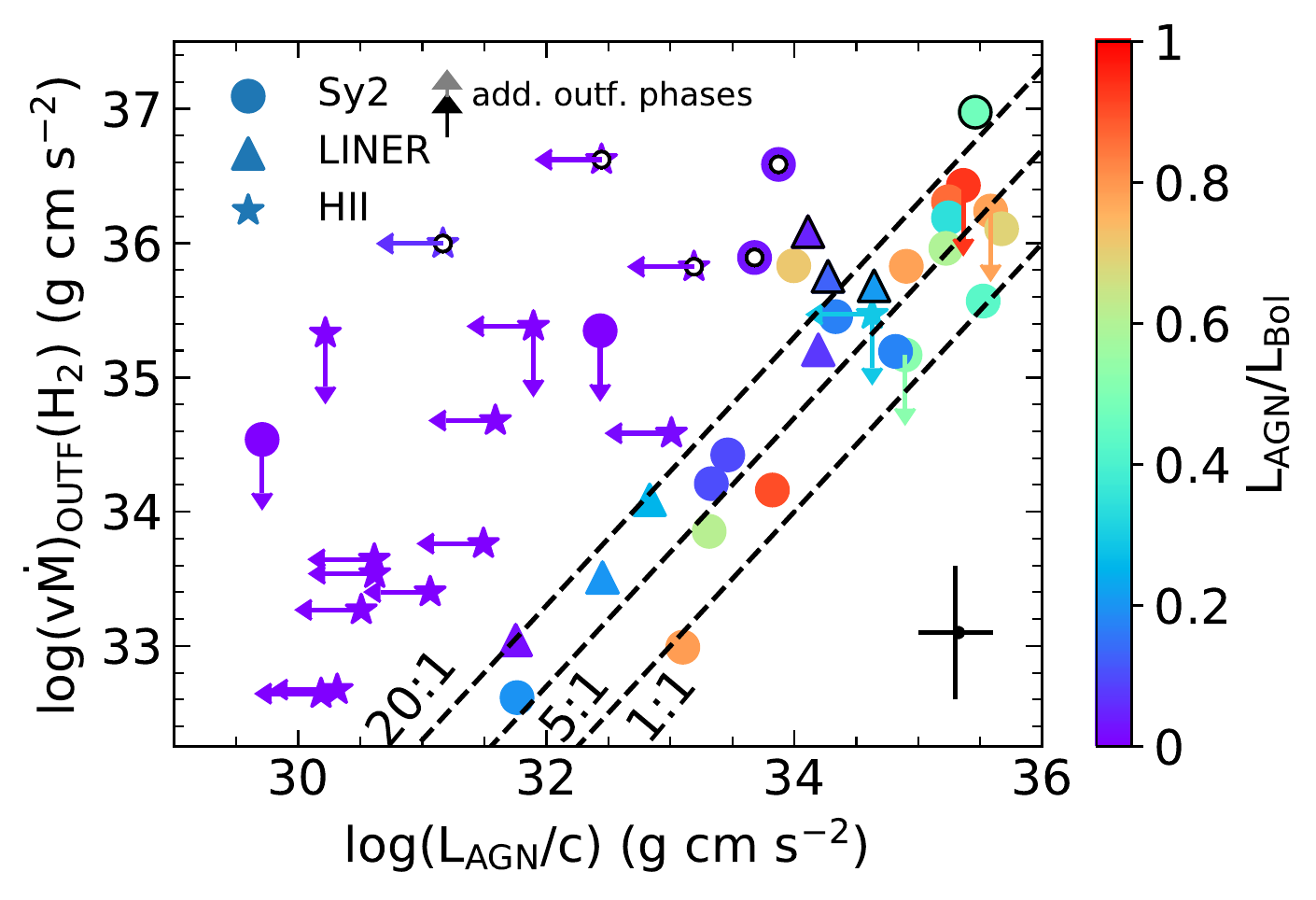}
\caption{Relation between outflow momentum rate ($v_{\rm outf}\dot{M}_{\rm outf}(\rm H_2)$) and AGN radiative momentum rate ($L_{\rm AGN}$/$c$).
The theoretical predictions ($v_{\rm outf}\dot{M}_{\rm outf}$)/($L_{\rm AGN}$/$c$) $\sim$ 20:1 (energy-driven) and 1:1 (momentum-driven) are shown
as a dashed lines, respectively. Radiation pressure-driven outflows can reach ($v_{\rm outf}\dot{M}_{\rm outf}$)/($L_{\rm AGN}$/$c$) $\sim$ 5:1. Colour-coding and symbols are as in Fig. \ref{fig:molec_ion}.}
\label{fig:momentumrate_LAGNc}
\end{figure}

\begin{figure}
\centering 
\includegraphics[width=\columnwidth]{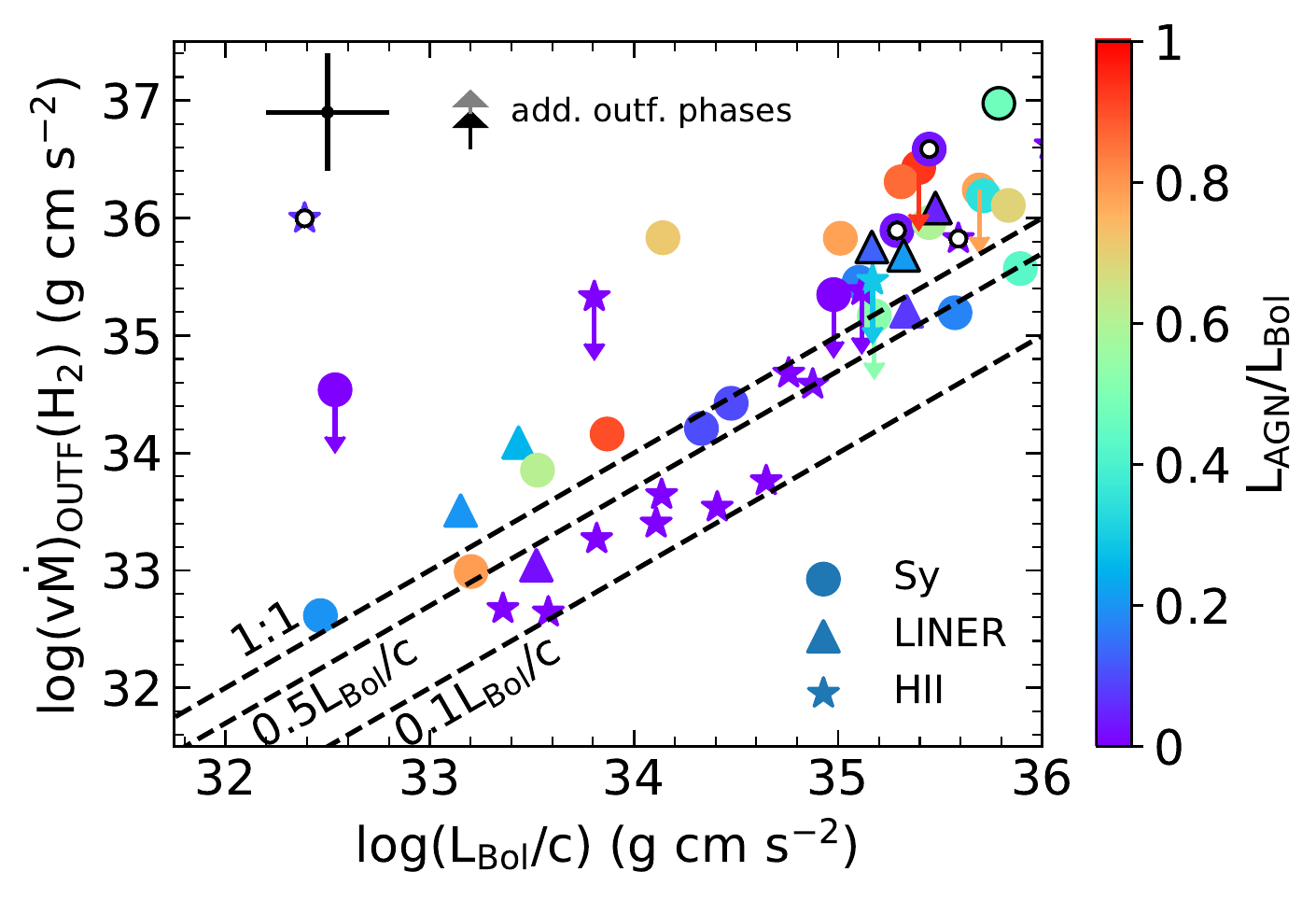}
\caption{Dependence of the outflow momentum rate ($v_{\rm outf}\dot{M}_{\rm outf}$) on the total photon momentum output of the galaxy (i.e. from 
AGN {\it and} star formation). The top dashed line indicates the 1:1 relation between momentum rate and bolometric luminosity, while lower dashed lines indicate lower ratios. Colour-coding and symbols are as in Fig.
\ref{fig:molec_ion}.}
\label{fig:momentumrate_veloc}
\end{figure}

\subsection{Momentum rate}
\label{sec:momentum_rate}

The outflow momentum rate is plotted as a function of the AGN radiative momentum rate $L_{\rm AGN}$/$c$ in Fig. \ref{fig:momentumrate_LAGNc},
illustrating a good correlation of these two quantities for AGN host galaxies, further indicating that AGN play a significant role in driving
galactic outflows. However, also in this case it is clear that the scatter is significantly larger than in previous studies.

Galaxies classified as star forming are all outliers in this relation since they are powered by a different mechanism (i.e. SNe feedback and/or radiation pressure from the stellar UV radiation field). 
In Fig. \ref{fig:momentumrate_veloc}, we analyse the dependence of outflow momentum rate on {\it total} photon momentum rate of the galaxy, $L_{\rm bol}$/$c$.
For strong AGN hosts, $L_{\rm bol}$ $\approx$ $L_{\rm AGN}$, but for AGN with lower AGN contribution and star forming galaxies, $L_{\rm bol}$ is much larger than $L_{\rm AGN}$. In this plot it is
interesting to note that for some star forming galaxies, especially at high luminosities (i.e. high SFR) the momentum rate is close
to $\sim 0.5~L_{\rm bol}$/$c$, suggesting that
radiation pressure on dusty clouds by the radiation field of young stars can be an additional significant contributor to the driving mechanism of outflow in starburst galaxies,
as predicted by some models \citep{Thompson2015}, although a coupling efficiency of at least 50\% would be required.

In some star forming galaxies the momentum rate of the outflow is close or exceeding $L_{\rm bol}$/$c$ (which would imply an unrealistic coupling efficiency of
100\% or higher), indicating that other mechanisms or other phenomena
may be at work. This is also seen in some AGN: 
A few AGN hosts have outflows which, when compared with $L_{\rm AGN}/c$,
are characterised by momentum boosts well in excess of what expected by any theory (see Fig. \ref{fig:momentumrate_LAGNc}). As we will
discuss later on, most of these outflows with extreme momentum rates
can be explained in terms of fossil outflows resulting from a much stronger past AGN activity.

\section{Discussion}
\label{sec:discussion}

\subsection{Driving mechanisms in AGN}
\label{sec:disc_AGN_drive}

Three different mechanisms have been proposed for powering AGN-driven outflow: an energy conserving blast wave (so-called energy-driven),
a momentum conserving blast wave (so-called momentum-driven), and direct radiation pressure on to the dusty clouds of the galactic ISM
(so-called radiation pressure-driven). These are discussed in greater detail in the following.

Many theoretical models expect that the most effective feedback process is obtained through AGN-driven outflows that are energy
conserving (energy-driven), in which a hot bubble composed of a thermalised nuclear wind has a cooling time-scale much longer 
than the outflow expansion time-scale. The outflow is accelerated due to the adiabatic expansion of the hot bubble. In this scenario, the outflow kinetic
energy is expected to be about 5~per cent of the AGN radiative power,
if the AGN is accreting close to the Eddington limit and if a 100~per cent thermal-to-kinetic conversion efficiency and high gas covering fractions are
assumed (i.e. 100\% coupling between the blast wave and the ISM in the host galaxy) \citep{King2010, Faucher-Giguere2012, Zubovas2012,Costa2014, doi:10.1146/annurev-astro-082214-122316,Richings2018}.
Yet, more detailed 3D, non-spherically symmetric simulations have suggested that the coupling can be significantly lower than 100~per cent, with dense clumps of the ISM remaining
unaffected and the outflow escaping along the directions of least resistance \citep{Bourne2014, Gabor2014, Costa2015, Bourne2015, Roos2015}.

Momentum-driven outflows (in which the energy of the shocked wind is quickly dissipated on small scales
via radiation losses) are generally expected to be much less effective in driving outflows and in this case the outflow
kinetic energy is expected to be of the order of 0.1~per cent of the AGN radiative luminosity, or less \citep{doi:10.1146/annurev-astro-082214-122316}.
Momentum-driven outflows are also expected to be confined within the central few 100~pc as most of their energy is quickly dissipated.

An additional class of models suggests that direct radiation pressure of the UV, optical and IR photons on the dusty clouds of the ISM
can be effective enough to drive massive outflows \citep{Fabian2012, Thompson2015, Ishibashi2017, Bieri2017, 2017arXiv170908638C,  Ishibashi2018, Costa2018}.
In this case there is a broad range of expected outflow properties.
If the central dusty region in which acceleration takes place is optically thick to IR radiation,
then the kinetic power of the outflows can be as high as $\sim$ 1~per cent of the AGN luminosity (this applies also when the
source of radiation is a compact starburst); however, in less extreme cases the outflow kinetic power is expected
to be lower than this value.
Recently, \cite{Ishibashi2018} developed the
model of radiation pressure-driven outflows further. The predict a super-linear relation between outflow kinetic power
and AGN luminosity in the form $P_{\rm K,outf}$ $\propto L_{\rm AGN}^{3/2}$.

By looking at the results reported in Sect. \ref{sec:kinetic_power} and Fig. \ref{fig:kinpower_LAGN}, where the prediction of different models are
also shown, it is clear that outflows in AGN host galaxies
span a broad range of properties. Although AGN `flickering' can account for some of the scatter, as discussed in the next section,
the very broad range of $P_{\rm K,outf}$/$L_{\rm AGN}$ suggests that these outflows are driven by a combination of different driving mechanisms and/or a broad range of coupling efficiencies with the ISM.
Some AGN are consistent with the energy-driven scenario, and full coupling of the outflow with the ISM.
However, the majority of AGN are significantly
below the $P_{\rm K}$ = 0.05 $L_{\rm AGN}$
relation (which assumes all thermal energy of the putative expanding hot bubble is converted into kinetic energy of the outflow) and 
so the respective outflows are more consistent with a momentum-driven or radiation pressure-driven mechanism; alternatively they are energy-driven,
but poorly coupled with the galaxy ISM. However, the momentum-driven scenario can probably be excluded as the observed outflows are mostly on kpc-scales,
while momentum-driven outflows should be confined within the central few 100~pc \citep{doi:10.1146/annurev-astro-082214-122316}.

As shown in Fig. \ref{fig:kinpower_LAGN}, at high AGN luminosities, galaxies lie closer to the expected value for the energy-driven mode though still mostly below the value expected from energy-driven outflows. This seems to indicate
that different driving mechanisms may be at work at different luminosities. Specifically, at high luminosities
energy-driven outflows (though with poor coupling) may dominate, while at low luminosities radiation pressure may be the dominant mechanism driving outflows. 

The super-linear relation between outflow kinetic power
and AGN luminosity in the form $P_{\rm K,outf}$ $\propto L_{\rm AGN}^{3/2}$ expected by the radiation pressure model of \cite{Ishibashi2018}
(dashed orange line in Fig.\ref{fig:kinpower_LAGN})
is consistent with the
observed relation in terms of slope. Therefore, this
model can potentially account also for the high $P_{\rm K,outf}$/$L_{\rm AGN}$ values ($\sim$1~per cent) observed at the highest luminosities, and
the decreasing values of this ratio at lower luminosities. However,
the model also expects the outflow rate to follow a relation $\dot{M}_{\rm outf} \propto L_{\rm AGN}^{1/2}$, which is somewhat 
shallower than what we observe for molecular gas (see Fig. \ref{fig:OFrate_LAGN}).

In terms of momentum rate,
in the energy-driven case models expect that the momentum rate is boosted to about 15-20~$L_{\rm AGN}$/$c$ \citep{Faucher-Giguere2012,Zubovas2012}.
Momentum-driven winds are expected to result in momentum rates of $\sim$$L_{\rm AGN}$/$c$ \citep{King2010}. Direct acceleration of the ISM through the action of radiation
pressure on dusty clouds generates momentum rates ranging from $\sim$ 1 up to 5~$L_{\rm AGN}$/$c$, the latter in the case that the medium
that is being accelerated is optically thick to infrared radiation, resulting in multiple scattering that boosts the momentum rate
\citep{Ishibashi2015, Thompson2015, Bieri2017, Costa2018, Ishibashi2018}.

In Fig. \ref{fig:momentumrate_LAGNc}
the upper dashed line represents the theoretical prediction for the energy-driven model (with 100~per cent coupling),
while the lower dashed lines indicate the values expected for the momentum-driven and radiation pressure-driven models.
Some of the galaxies with an AGN do follow the theoretical prediction for energy-driven outflows within the errors, but most galaxies hosting an AGN
have momentum rates scattered between the energy-driven case and the momentum/radiation pressure-driven cases, further suggesting the contribution of different driving mechanisms
and/or energy-driven outflows with poor coupling.

\begin{figure}
 \includegraphics[width=\linewidth]{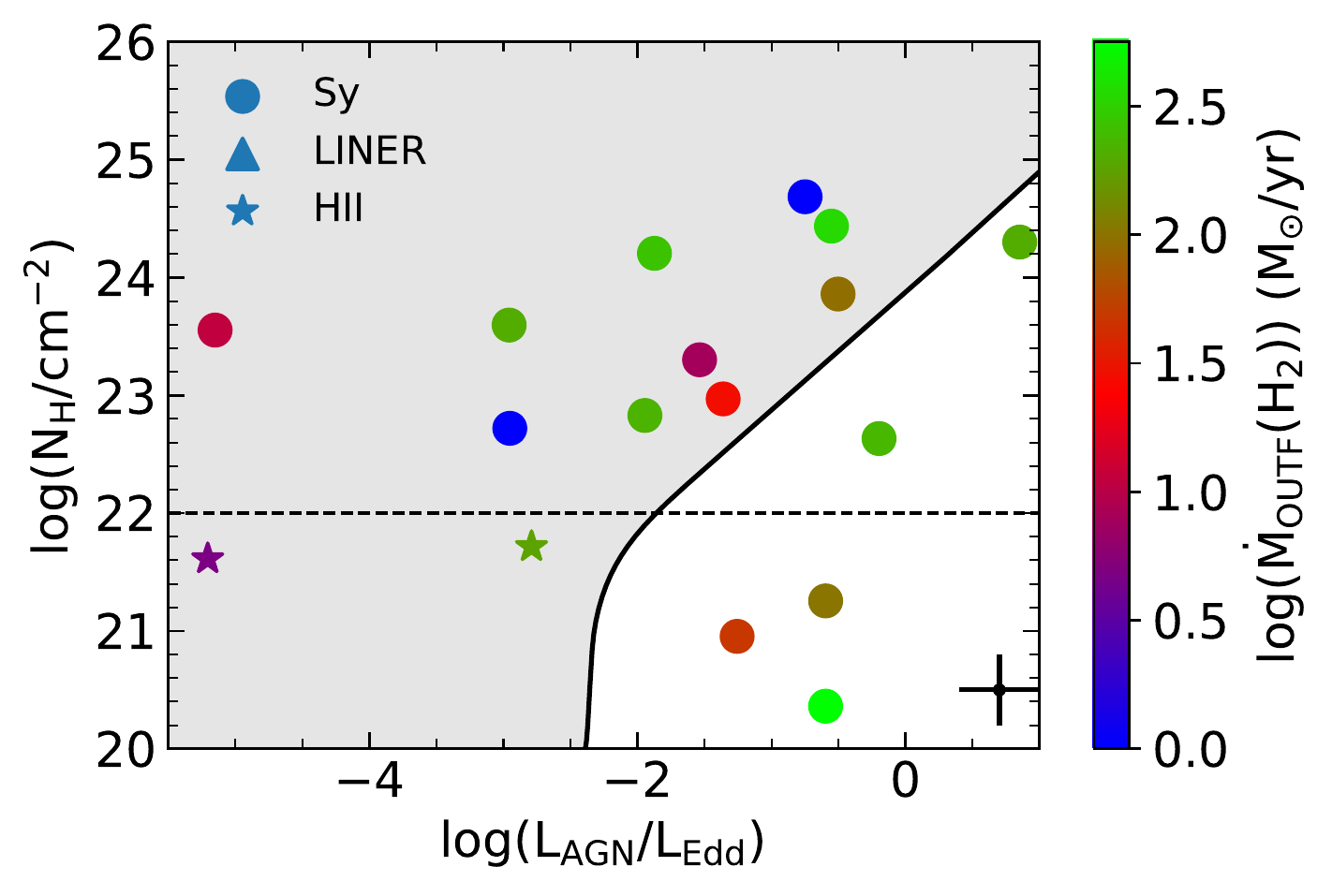}
 \caption{Gaseous column density along our line of sight, as inferred from X-ray spectra, versus
$L_{\rm AGN}$/$L_{\rm Edd}$, with symbols colour-coded by outflow rate (right hand-side colour bar). The solid
line delimitates the area (non-shaded) where radiation pressure on dust is expected to overcome gravity (i.e. where
the effective Eddington luminosity for a dusty medium is exceeded) hence producing
powerful outflows.  Symbols are as in Fig. \ref{fig:molec_ion}.}
 \label{fig:edd_nh}
\end{figure}

An additional route to study the driving mechanism is to investigate the relation between outflow rate, AGN luminosity and gas column density of the
circumnuclear gas. Indeed,
in the context of radiation pressure-driven outflows, the effective Eddington luminosity is dominated
by radiation pressure on dust, which drastically reduces the Eddington limit. 
\cite{Fabian2007} and \cite{Ricci2017} have pointed out that the effective Eddington limit ($L_{\rm Edd,eff}$) is higher at higher gas column densities as larger amounts of material 
need to be pushed out. Hence, these authors expect a region in the $L_{\rm AGN}$/$L_{\rm Edd}$ vs $N_{\rm H}$ plane where radiation pressure on dust dominates over 
gravity and where this kind of outflows should be most effective. This is explored in Fig. \ref{fig:edd_nh}, 
where the column density (as inferred from X-ray observations)
is shown as a function of the Eddington ratio, $L_{\rm AGN}$/$L_{\rm Edd}$, and colour coding is according to the outflow rate. 
The region on the right hand-side of the plot, delimited by the solid line, is 
where the effective Eddington luminosity for a dusty medium is exceeded, and therefore where we expect powerful outflows that are driven by
radiation pressure on dust. As pointed out by \cite{Fabian2007} and \cite{Ricci2017}, this region is underpopulated, indeed empirically confirming
that the effective Eddington ratio is exceeded in this region.
The few galaxies of our sample located in this region do indeed show among the highest outflow rates,
suggesting that indeed these extreme outflows may be driven by radiation pressure on dust. We note that in the scenario discussed by
\cite{Fabian2007} and \cite{Ricci2017} AGN with $N_{\rm H} < 10^{22}~cm^{-2}$ are not considered, as these low column densities
are thought to be associated
with dust lanes in the host galaxy and not directly linked with the AGN process; however, our analysis shows that the three objects
in this region are characterized by strong outflows, suggesting that also in these cases the outflow is driven by radiation pressure
on dust.
For the other galaxies, located on the left-side of the solid line (shaded are) there is not much correlation between the outflow rate and their location on the diagram, in particular there are galaxies with high outflow
rates also below the boundary expected by the model. These outflows could be driven by a different mechanism.

To summarise, our results indicate that AGN-driven outflows are either consistent with predictions from direct radiation pressure models
or with the energy-conserving blast-wave scenario,
but with a coupling with the ISM of the host galaxy that varies from galaxy to galaxy and generally is lower than 100\%. 

\subsection{Driving mechanisms in star forming galaxies}
\label{sec:disc_AGN_drive}

As discussed in Sect. \ref{sec:kinetic_power} only about $\sim$1--2~per cent of the kinetic power released by SNe appear to be converted into outflow kinetic power.
A low coupling efficiency of the kinetic energy between supernova ejecta and ISM is expected by the fact that most of the energy is radiated
away in the dense interstellar medium in which most SNe are expected to explode. Yet, models and simulations
expect still higher coupling efficiencies, of the order of 5~per cent \citep{Walch2015}.
Although, there is some tension this is actually within the errors; however, if confirmed with more accurate data it may indicate
that radiation losses during the SN-ISM interaction are higher than expected (possibly because the ISM is denser than assumed in the simulations).

However, in addition to the kinetic power injected by SNe, outflows in star forming galaxies can also be driven by radiation pressure onto the dusty clouds
\citep{Thompson2005,Thompson2015}. The correlation in Fig. \ref{fig:momentumrate_veloc} between outflow momentum rate and radiative momentum from the bolometric luminosity of star forming galaxies
suggests indeed that radiation pressure may play a role.
However, one should also be aware that such a correlation is also degenerate with the kinetic power injected by SNe, since the SN rate is
linked to the SFR which is in turn related to the bolometric luminosity. Moreover, it is important to note that most star forming galaxies have a ratio between momentum
rate and radiation momentum is between 0.5 and 0.1 (Fig. \ref{fig:momentumrate_veloc}), implying that, if this driving process is at work, the coupling efficiency must be less than 50~per cent.

\subsection{Fossil Outflows}
\label{sec:fossil}

There are a few galaxies with outflows that
are characterised by anomalously high kinetic power and momentum rates compared to their AGN luminosity and SFR, which are difficult to explain with any driving mechanism.
More specifically,
Fig. \ref{fig:kinpower_LAGN} and \ref{fig:SNepower} indicate that for some galaxies (marked with a white dot in these and other figures)
the kinetic power is greatly in excess of what is expected
even from the AGN energy-driven scenario, even assuming 100~per cent coupling, and is also in excess of what is expected by the SNe-driven scenario, unless
assuming an unrealistically high coupling efficiency of the SNe (larger than 10~per cent, especially if accounting for all outflow
phases). In Fig. \ref{fig:momentumrate_LAGNc} it is clear that these objects have also very high
momentum rate, even in excess of what expected in the case of energy-driven outflows and 100~per cent coupling. It is unlikely that in these objects the SFR or AGN power are not
estimated properly, as the observational constraints are quite good. It is also unlikely that these outflows are driven by a radio-jet, as these objects do not show
any radio excess in Fig. \ref{fig:or_vs_radio}.
In these cases, as already hinted in the previous sections, the most likely explanation is that we are observing `fossil' outflows that outlast a past
powerful AGN activity, which has recently faded. This interpretation is further supported by the low Eddington ratios (log($L_{\rm AGN}$/$L_{\rm Edd}$) $\lesssim$ -3) seen in 
the three fossil outflow objects for which a black hole estimate is available.

Fossil outflows are expected from theory in large numbers. It has been shown that outflows can remain visible for a time about 10 times longer than the driving phases
and up to $10^8$~yr \citep{King2011}. Theoretical considerations have suggested that in M82 a
powerful AGN might have been present until about 17~Myr ago and may have been responsible for driving the outflow currently observed
\citep{Zubovas2015}. Even more simply, without invoking detailed and extensive theoretical simulations,
the dynamical time-scales of the outflows ($t_{\rm dyn}$ $\sim$ $R$/$v$) are in the range of 10$^{6}$-10$^{8}$~yr, while we know that AGN have a `flickering'
time-scale ranging from a few years \citep[e.g.][]{2000A&A...355..485G} up to 10$^{5}$~yr \citep{doi:10.1093/mnras/stv1136, King2015}. Therefore, a large number of
fossil AGN is naturally expected. The outliers we see here are possible manifestations of this scenario and are likely the tip of the iceberg of a much larger population
of fossil outflows. If this is true, then one should be careful
when comparing observational outflow properties with theoretical models as possibly a large fraction of galaxies (in our sample 10 - 20~per cent) display
fossil outflows.

The reason why in the past such fossil outflows had not been identified is likely because previous observations had targeted primarily known,
strong AGN hence biasing the sample towards outflows that are in the phase of being powered. Instead, in our study we have collected
data of galaxies from the ALMA archive, many of which had been observed independently of their activity, hence reducing such biases.

Table \ref{tab: fossils} gives the list of fossil outflows identified by us. Their properties have no peculiarities
relative to other galaxies in the sample.

\subsection{Do Outflows Escape the Galaxy and the Halo?}
\label{sec:esc_v_gal}

If outflow velocities are high enough to escape the potential of the galaxy (and possibly even the halo), then these outflows can effectively clear the galaxy
of its gas content. It also depends on how much gas the outflows sweep up as they move out of the 
galaxy and on whether they collide with inflowing material. We ignore the latter effects here, because, as it turns out, most outflowing
material should not escape the galactic halo purely due to its insufficient velocity. Mass-loading and a potential interaction with gas
infall would only strengthen our conclusions.

In principle one should use the velocity rotation curve of galaxies to infer the mass distribution radial
profile of the associated gravitational potential. Unfortunately, at the moment, this information is not available
for the vast majority of the galaxies in our sample. Information on the rotation curve is available only for very few
galaxies, primarily from the CO interferometric data, and only in the central region of the galaxy. As a consequence,
we have to rely on some simple assumptions and use scaling relations with the stellar mass.

We consider the stellar mass as determined in Sect. \ref{sec:anc_info}. We use the relation by \cite{McIntosh2005} at $z$ = 0 to relate the stellar mass to the half-light radius $r_{50}$:

\begin{eqnarray}
\mathrm \log{(\bar{r}_{\mathrm{50}}/\mathrm{h^{-1}kpc})} = 0.56 \log{(M_{\star}\mathrm{h^{2}/M_{\odot}})} - 5.52.
\label{eq:size_mass}
\end{eqnarray}
We approximate the stellar mass distribution adopting a Hernquist profile \citep{Hernquist1990} for the density
\begin{eqnarray}
\rho (r) = \frac{M}{2\pi}\frac{a}{r}\frac{1}{(r+a)^{3}}
\end{eqnarray}
where $a$ is related to the effective radius $r_{\rm eff}$ via
$r_{\rm eff} \approx 1.8135a$.
We can now compute the escape velocity for galaxies in our sample. The escape velocity is given by
\begin{eqnarray}
v_{\mathrm{esc}} = \sqrt{2|\Phi(r)|} = \sqrt{\frac{2GM}{r+a}}.
\end{eqnarray}
The escape fraction is then defined as the fraction of the outflow that has a velocity higher than the escape velocity. Unfortunately
this calculation can be done only for those outflows for which we have the data in electronic form, as it requires estimating
the integral of the fraction of the broad wings with velocity higher than the escape velocity (i.e. this calculation cannot
be done for the data in the
literature for which an electronic version of the spectrum is not available).
This part of the outflow will eventually leave the galaxy. In Fig. \ref{fig:esc_fraction}, the escape fraction is shown for the galaxies as a function
of the AGN luminosity. Only in IRAS 20100-4156, and maybe in 4C 12.50, 10 ~per cent or more of the gas in the outflow will
escape the galaxy using these simple assumptions. For all other galaxies, the escape fraction is smaller or negligible and there is
no clear dependence on AGN luminosity. We should note, however, that the equations here only hold if we consider ballistic motions. If the
outflows are still driven and therefore further accelerated, they are more likely to escape the galaxy potential. The error in the escape fraction is as large as 50~per cent as inferred by
running a Monte Carlo simulation, taking into account errors in fitting the line profile, stellar mass and outflow radius.

Therefore, despite galactic outflows being very massive and energetic, especially those driven by AGN, most of the expelled
gas will quickly re-accrete onto the galaxy and be available again for star formation. Hence,
the ejection of gas, at least in this molecular phase, does not really contribute to the global quenching of star formation in
galaxies. However,
these outflows can still have a dramatic effect in the central region of galaxies (especially in the bulge region), where they can locally suppress
or even quench star formation.
Moreover, even if the ejective aspect of outflows does not directly contribute to the global quenching of galaxies on large galactic scales,
this does not mean that outflows do not play a role at all on the global evolution of galaxies on large scales. By injecting energy, outflows can keep the halo gas hot and prevent
it from cooling onto the galaxy, hence effectively resulting into a delayed quenching of star formation in the galaxy   
as a consequence of starvation \citep[e.g.][]{Gilli2017, 2017arXiv170908638C}.
The escape fractions inferred above are for molecular outflows. Ionized outflows, although generally
contributing much less to the outflow rate, are expected to have higher escape fractions \citep{Costa2015}. This can be
investigated for several galaxies whose outflow has been mapped in the ionized phase, but we defer this kind of analysis
to a later paper.

We note that the escape fraction discussed above refers to the escape velocity from the galaxy. The escape
fraction from the galaxy dark matter halo are even smaller, but also more difficult to compute.
We can attempt to estimate the velocities needed to escape the halo by making a few approximations in the following.

\begin{figure}
\centering 
\includegraphics[width=\columnwidth]{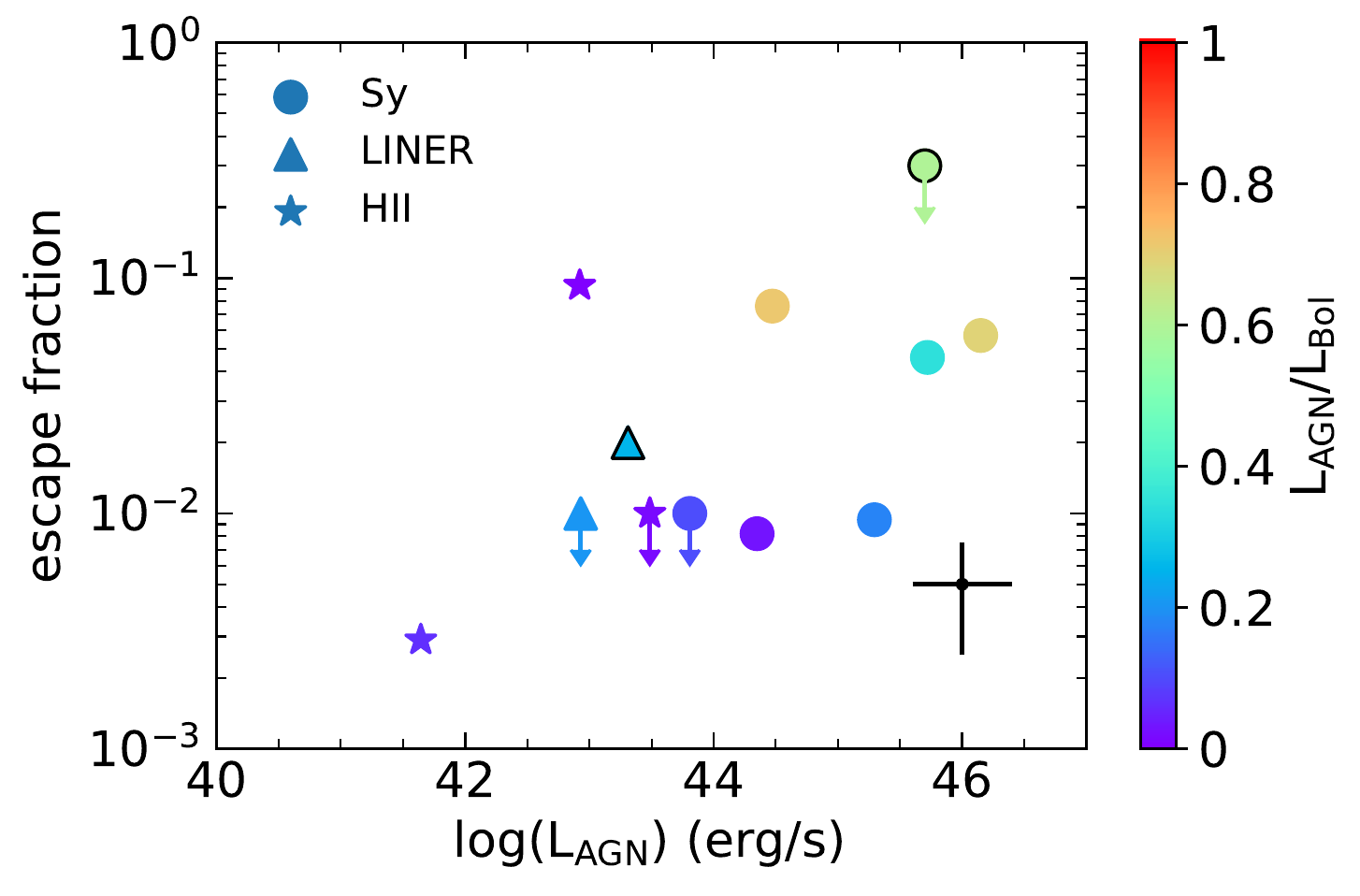}
\caption{Fraction of the molecular outflow that escapes the galaxy as a function of AGN luminosity. The two data points with a black contour are taken from the literature and use slightly different definitions of escape velocity.}
\label{fig:esc_fraction}
\end{figure}

We use the Navarro-Frenk-White (NFW) profile \citep{NFW1996} to describe the density in the halo of the galaxy:
\begin{eqnarray}
\rho (r) = \frac{\rho_{\mathrm{crit}} \delta_{\mathrm{c}}}{r/r_{\mathrm{s}}(1+r/r_{\mathrm{s}})^{2}}
\end{eqnarray}
where $\rho_{\rm crit}$ = 3$H^{2}$/8$\pi$G is the critical density and
$r_{\rm{s}}$ = $r_{\rm{200}}$/c is the characteristic radius (c being the concentration parameter).
The mass of the halo ($M_{\rm 200}$) can be inferred from the stellar mass using a stellar mass-halo mass relation
from abundance matching \citep{Moster2013}.
To find the mass concentration from the halo mass, we use the relation by \cite{Duffy2008} at $z$ = 0:
\begin{eqnarray}
\log{c_{\mathrm{200}}} = 0.76 - 0.1\ \log{M_{\mathrm{200}}}.
\end{eqnarray}
As in Sect. \ref{sec:esc_v_gal}, the escape velocity is:
\begin{eqnarray}
v_{\mathrm{esc}} = \sqrt{2|\Phi(r)|} = \sqrt{\frac{2M_{\mathrm{200}}G}{r\mathrm{(ln(1+c}\mathrm{)-c}/\mathrm{(1+c)}}\mathrm{ln(}1+r/r_{\mathrm{s}}\mathrm{))}}
\label{eq:esc_v_hal}
\end{eqnarray}
This allows us to compute the escape fractions of gas out of the halo.
As expected, we generally obtain escape fractions from the halo even much smaller than the escape fractions from the galaxy, typically much smaller than 1~per cent,
further indicating that the bulk of the outflowing gas will remain in the gravitational potential of the system and will eventually re-accrete onto the galaxy.
We note, however, that our sample does not include low mass galaxies ($M_{\star}<10^{10}~M_{\odot}$), for which models expect a large fraction of the outflowing
gas to leave the galaxy and its halo, hence enriching the IGM. Detailed observations targeting this class of galaxies are needed in order to test these
expectations.

\subsection{Effectiveness of AGN driven outflows in quenching star formation}

AGN-driven outflows have been claimed to be one of the primary candidates for cleaning galaxies of their gas content hence quenching star formation
and transforming them into passive systems. The ``blast-wave'' energy-conserving mode, with 100~per cent coupling with the ISM, has generally been regarded as the most
effective mode to remove galaxies of their gas content \citep{Zubovas2012}. We have, however, obtained various results indicating that such ``ejective'' mode is probably
not effective in clearing the whole galaxy of its gas content, even at high luminosities. Indeed, most observational properties of the AGN-driven
outflows are below the expectations from energy-conserving mode, suggesting either poor coupling efficiency \citep[as suggested by some models and numerical  
simulations, e.g. ][]{Gabor2014,Costa2015,Richings2017} or that other driving mechanisms, such as direct radiation pressure onto the ISM dusty clouds, are
also at work \citep{Thompson2015, 2017arXiv170908638C, Ishibashi2018}. We have shown that the outflow depletion time-scales for the total gas mass are very long, beyond
the typical lifetime of AGN, hence these outflows are unlikely to expel the whole amount of gas in the galaxy. Finally, as illustrated in the previous section,
only a small fraction of the outflowing gas actually escapes the galaxy (and even less the halo), hence most of  the expelled gas re-accretes on the galaxy to
fuel star formation.

Although, the AGN ``ejective'' mode does not seem capable of quenching the entire galaxy, it can likely clean and quench the central region. Indeed,
the outflow depletion time associated with the (mostly centrally concentrated) molecular gas is much shorter ($\sim 10^7$~yr), especially in luminous AGN.
Therefore, the ejective AGN mode (especially when occurring at high redshift) may actually be a important route for quenching star formation in the bulge region.

The AGN driven outflow may also have an additional indirect effect on larger scales. Although the ejective mode is likely confined to the central regions, the energy
injected by the outflow into the halo can contribute to keep it hot, hence preventing further gas accretion onto the 
galaxy and therefore resulting in a ``delayed'' quenching, by starvation, once star formation has used up the gas available in the disc \citep{Costa2015, 2017arXiv170908638C}. This
``preventive'', delayed mode is supported by various statistical properties of the local population of galaxies \citep{Peng2015,Woo2017,Cresci2018}.

\section{Conclusions}

In this work, we have quantified the energetics of molecular outflows in a sample of 45 local ($z$ < 0.2) galaxies including AGN host galaxies as well as star
forming/starburst galaxies. The sample spans a range in AGN luminosity from log($L_{\rm AGN}$) $\sim$ 41 up to $\sim$ 46~erg s$^{-1}$ and in star formation rate from $\sim$
0.1 up to several 100~M$_{\odot}$ yr$^{-1}$. Molecular outflow properties are inferred from interferometric observations of low-$J$ CO lines
(apart from four galaxies, for which OH absorption from Herschel is used). We collect
data of molecular outflows from the literature and recalculate outflow and host galaxy properties in a consistent manner. Furthermore, we also analyse all public
ALMA archival data of low-$J$ (1-0, 2-1 and 3-2) CO lines in local galaxies and look for signatures of outflowing gas. This is the largest sample to date for which
molecular outflows in CO have been investigated, and includes also less powerful outflows than previous studies. Our sample improves with respect to previous
studies not only in terms of statistics but also by reducing the bias favouring very active galaxies which have been preferentially targeted in the past.

Our main findings can be summarised as follows:

\begin{itemize}

\item For about 30~per cent of the galaxies we could also obtain information on the ionized outflow, while for 18~per cent of the sample we have also information on the neutral
outflow. We find that in starburst galaxies the ionized outflow is about as massive as the molecular outflow. In AGN the ionized outflows is generally negligible comparable
to the molecular outflow and we find that the molecular-to-ionized outflow rate increases with AGN luminosity. The amount of gas in the atomic neutral phase has a large
scatter, but in general is comparable to the molecular phase.

\item The molecular mass-loading factor ($\eta$ = $\dot{M}_{\rm outf}(\rm H_2)$/SFR) for star forming galaxies is consistent with unity, as expected by models of star formation feedback.

\item The molecular mass-loading factor is higher in AGN host galaxies compared to star forming galaxies, although a significant boost (with $\eta$ > 10) is only seen in galaxies in which the AGN
luminosity is high relative to the bolometric luminosity ($L_{\rm AGN}$/$L_{\rm bol}$ > 0.7).

\item In AGN the outflow rate correlates with the AGN luminosity and with the Eddington ratio, $L_{\rm AGN}$/$L_{\rm Edd}$, although with large scatter, further indicating that
the AGN plays a role in driving outflows.
We also observe a correlation with the black hole mass, which can be seen as tracing a link between the outflow (which has a dynamical time-scale of $\rm 10^6-10^8$~yr, much longer
than the AGN flickering time-scale) with the integrated, average past activity of the black hole.

\item We highlight that the dependence of the outflow properties on AGN luminosity, star formation rate and galaxy stellar mass, makes it difficult to isolate the individual
dependences, as each these properties are can be mutually correlated in galaxies. Therefore, we have derived a relation of the outflow
rate simultaneously fitting the dependence of AGN luminosity, SFR and stellar mass. The resulting fit is much tighter than the individual
relations and enables us to disentangle,
at least partially, the individual dependences. In particular, we obtain a scaling of the outflow rate on stellar mass as $\propto M_{\star}^{-0.41}$, which is very close to
the dependence expected by models of outflows driven by SNe and stellar winds. We also suggest that the inferred empirical (four-dimensional) relation between outflow rate, $L_{\rm AGN}$, SFR and $M_{\star}$ can be very
efficiently used to predict the strength of outflows in a variety of galaxies and for comparison with models.

\item We find that the majority of the molecular outflows studied here show no excess of radio emission relative to the SFR-radio relation. In addition, there
is no correlation between molecular outflow rate and infrared-to-radio luminosity ratio, indicating that the majority of molecular outflows are
not driven by radio jets, at least within the luminosity range probed by us. However, this does not exclude that radio jet may have an important
role in driving molecular outflows in a few specific galaxies, as indeed observed.


\item The depletion time-scale associated with outflow (i.e. $\tau_{\rm depl}$ = $M_{\rm gas}$/$\dot{M}_{\rm outf}$) anti-correlates with AGN luminosity, i.e. is shorter in more luminous AGN. The depletion
time-scale for molecular gas ($\tau _{\rm depl, outf}(\rm H_2)$= $M(\rm H_2)$/$\dot{M}_{\rm outf}(\rm H_2)$) can be as short as a few, or a few tens million years, much shorter than the depletion time-scale associated with
star formation only. However, when considering also the atomic component, the depletion time-scale
for the total gas content ($\tau _{\rm depl,outf}(\rm tot)$= $M(\rm H_2+HI)$/$\dot{M}_{\rm outf}(\rm H_2)$) is typically of the order, or longer than $10^8$~yr, i.e. longer than the typical AGN lifetime. This indicates that
the AGN-driven outflow is generally capable of quickly removing the gas from the central regions (which are dominated by the molecular phase), but unlikely to clean the
entire galaxy from its gas content.

\item For AGN host galaxies the outflow kinetic power $P_{\rm K, outf}(\rm H_2)$ shows a much larger scatter than in previous studies and
spans from 0.1 to 5~per cent of $L_{\rm AGN}$. The ratio $P_{\rm K, outf}(\rm H_2)$/$L_{\rm AGN}$ increases with luminosity.
The momentum rate spans from 1 to 30 times $L_{\rm AGN}/c$. These results suggest that the AGN driven outflows can be
both energy-driven (with a broad range of coupling efficiencies with the ISM) and radiation pressure-driven.

\item We estimate that the fraction of outflowing gas with a high enough velocity to escape the galaxy and the dark matter halo is less than 5~per cent, indicating that, although outflows can
remove gas from the central region, most of the gas re-accretes onto the galaxy.

\item The results on the kinetic power, on the momentum rate, on the depletion time-scale and on the fraction of escaping case, considered
all together, indicate that the AGN ``ejective mode''
is unlikely to be effective in cleaning the galaxy of its gas content, at least in the mass range probed by us ($\rm > 10^{10}~M_{\odot}$).
However, AGN outflows are likely capable of cleaning the gas content, hence quench star formation, in the central (bulge) region.
Moreover, AGN-driven outflows can inject energy into the halo hence keeping it hot and preventing further
gas accretion, therefore resulting in a delayed feedback that quenches the galaxy through starvation \citep{Peng2015,Woo2017}.

\item In star forming galaxies the kinetic power is only 1--2~per cent of the kinetic power generated by supernovae, indicating very fast cooling of supernova ejecta
in the dense ISM in which they explode, hence poor efficiency in driving outflows, as expected by some models. In star forming galaxies the momentum rate of outflows correlates with
the bolometric luminosity, and it is about 0.3 $L_{\rm bol}/c$, suggesting that radiation pressure can also contribute to drive outflows in star forming galaxies.

\item We also identify about 10~per cent of the galaxies whose outflow significantly
exceeds the maximum theoretical values of kinetic power and momentum rate expected for both AGN and SB-driven cases. Our proposed explanation is that
these are `fossil outflows' resulting
from activity of a past strong AGN, which has now faded. Theoretical models expects such fossil outflows to be present in large numbers, also simply based on the
fact that the outflows dynamical time-scales is of the order of $\rm 10^6-10^8~yr$ while the AGN has a much shorter variability ($1-10^6$~yr). Previous outflow surveys
have not identified such fossil outflows because they may have been biased towards powerful AGN and powerful starburst galaxies in order to maximise the probability of detecting
outflows. Our sample is less biased in this sense (as it includes
galaxies that were not selected specifically with the goal of detecting outflows), which has enabled us to detect this phenomenon. However, our sample is still 
biased, hence the fraction of fossil outflow found by us is probably still the tip of the iceberg of a larger population.

\end{itemize}

\section*{Acknowledgements}
We thank Fabrizio Fiore and Kastytis Zubovas for their detailed comments on the draft. 
\newline RM and SC acknowledge support by the Science and Technology Facilities Council (STFC). RM and AF acknowledge ERC Advanced Grant 695671 "QUENCH". CC acknowledges funding from the European Union's Horizon 2020 research and innovation programme under the Marie Sklodowska-Curie grant agreement No 664931. ACF and WI acknowledge ERC Advanced Grant 3409442 "FEEDBACK". MAB acknowledges support by the ERC starting grant 638707 "BHs and their host galaxies: co-evolution across cosmic time". This paper makes use of the following ALMA data: ADS/JAO.ALMA$\#$2013.1.00659.S, ADS/JAO.ALMA$\#$2012.1.00377.S, ADS/JAO.ALMA$\#$2012.1. 00611, ADS/JAO.ALMA$\#$2015.1.00102.S, ADS/JAO.ALMA$\#$ 2015.1.00263.S, ADS/JAO.ALMA$\#$2013.1.00379.S, ADS/JAO. ALMA$\#$ 2012.1.00539, ADS/JAO. ALMA$\#$ 2016.1.01279.S.
\newline This research has made use of the NASA/ IPAC Infrared Science Archive, which is operated by the Jet Propulsion Laboratory, California Institute of Technology, under contract with the National Aeronautics and Space Administration. This publication makes use of data products from the Two Micron All Sky Survey, which is a joint project of the University of Massachusetts and the Infrared Processing and Analysis Center/California Institute of Technology, funded by the National Aeronautics and Space Administration and the National Science Foundation.



\bibliographystyle{mnras}
\bibliography{bibliography_1} 

\clearpage
\newpage

\begin{table*}
\setlength{\tabcolsep}{4pt}
\centering
\small
\caption{List of galaxies in the sample analysed in this paper, together with some of their basic properties}
\label{tab:sample_basic_properties}
\begin{tabular}{ c c c c c c c c c c c c} 
	\hline
	Galaxy & type & \textit{z} & \textit{D$_{\rm L}$} & SFR & log($L_{\rm AGN}$) & log($M_{\star}$) & $\alpha_{\rm bol}$ & log($M(\rm H_2)$) & log($M(\rm HI)$) & $q_{\rm IR}$& Ref. \\ 
	 & & & [Mpc] & [M$_{\odot}$ yr$^{-1}$] & [erg s$^{-1}$] & [M$_{\odot}$] & & [M$_{\odot}$] &  [M$_{\odot}$] &  \\
	 (1) & (2) & (3) & (4) & (5) & (6) & (7) & (8) & (9) & (10) & (11) & (12) \\
	\hline
	\multicolumn{12}{c}{\textbf{CO Literature data}} \\
	\hline
	IRAS F08572+3915 & Sy2 & 0.05821 & 265 & 20 & 45.72 & 10.79 & 0.86 & 9.18 & & 3.57 & $\alpha$, A, a \\
	IRAS F10565+2448 & Sy2 & 0.04311 & 196 & 95 & 44.81 & 10.66 & 0.170 & 9.90 & & 2.64 & $\alpha$, A, a \\
	IRAS 23365+3604 & LINER & 0.06438 & 285 & 137 & 44.67 & 11.23 & 0.072 & 9.93 & & 2.73 & $\alpha$, A, a \\
	Mrk 273 & Sy2 & 0.03777 & 169 & 139 & 44.16 & 11.10 & 0.080 & 9.70 & 10.21 & 2.49 & $\alpha$, A, a \\ 
	IRAS F23060+0505 & Sy2 & 0.17300 & 831 & 75 & 46.06 & 11.75 & 0.780 & 10.39 & & 2.79 & $\alpha$, A, a \\
	Mrk 876 & Sy1 & 0.12900 & 607 & 6.5 & 45.84 & 11.64 & 0.930 & 9.84 & & 2.63 & $\alpha$, A, a \\
	I Zw 1 & Sy1 & 0.06114 & 259 & 36 & 45.37 & 11.30 & 0.520 & 9.56 & 10.27 & 3.04 & $\alpha$, A, a \\
	Mrk 231 & Sy1 & 0.04217 & 189 & 234 & 45.72 & 11.53 & 0.340 & 9.73 &  & 2.44 & $\alpha$, A, a \\
	NGC 1266 & LINER & 0.00719 & 28.6 & 1.6 & 43.31 & 10.30 & 0.250 & 9.23 & & 2.36 & $\alpha$, A, a \\
	M82 & HII & 0.00068 & 4.03 & 10 & $\leq$41.54 & 10.66 & $\leq$0.0009 & 8.64 & 9.04 & 2.62 & $\alpha$, A, a \\
	NGC 1377 & LINER & 0.00578 & 23.9 & 0.9 & 42.93 & 10.06 & 0.200 & 8.44 & 9.89 & & $\alpha$, A, a \\
	NGC 6240 & Sy2 & 0.02448 & 107 & 16 & 45.38 & 11.53 & 0.780 & 9.86 & 10.05 & 2.10 & $\alpha$, A, a \\
	NGC 3256 & HII & 0.00926 & 44.6 & 36 & $\leq$41.97 & 11.23 & $\leq$0.0007 & 9.68 & 9.89 & 2.37 & $\alpha$, A, a \\
	NGC 3628 & HII & 0.00280 & 17.1 & 1.8 & $\leq$40.79 & 11.30 & $\leq$0.0009 & 9.53 & 10.31 & 2.00 & $\alpha$, A, a \\
	NGC 253 & HII & 0.00081 & 2.77 & 3 & $\leq$40.66 & 10.65 & $\leq$0.0004 & 8.15 & 9.27 & 3.01 & $\alpha$, A, a \\
	NGC 6764 & LINER & 0.00807 & 32.6 & 2.6 & 42.23 & 10.48 & 0.017 & 8.90 & 9.64 & 2.25 & $\alpha$, A, a \\
	NGC 1068 & Sy2 & 0.00379 & 13.1 & 18 & 43.94 & 11.23 & 0.097 & 9.11 & 9.12 & 2.32 & $\alpha$, A, a \\
	IC 5063 & Sy2 & 0.01100 & 47.2 & 0.6 & 44.30 & 11.02 & 0.9 & 8.85 & 10.75 & 1.11 & $\alpha$, A, a \\
	NGC 2146 & HII & 0.00298 & 12.5 & 12 & $\leq$41.09 & 10.58 & $\leq$0.0003 & 8.94 & 9.41 & 2.83 & $\alpha$, A, a \\
	IRAS 17208-0014 & HII & 0.0428 & 189 & 200 & $\leq$43.67 & 11.33 & $\leq$0.24 & 10.03 & & 2.79 & $\beta$, B, b \\
	NGC 1614 & HII & 0.0159 & 68.3 & 45 & $\leq$42.07 & 11.07 & $\leq$0.0006 & 9.51 & & 2.77 & $\beta$,$\gamma$, C, b \\
	Circinus Galaxy & Sy2 & 0.0014 & 8.34 & 0.6 & 43.57 & 10.95 & 0.59 & 9.32 & 10.27 & 2.07 & $\delta$, D, b \\
	SDSS J1356+1026 & Sy2 & 0.1230 & 579 & 20 & 46 & 11.36 & 0.43 & 8.91 & & 2.32 & $\epsilon$, E, b \\
	ESO 320-G030 & HII & 0.0108 & 51.1 & 20 & $\leq$41.09 & 11.03 & $\leq$0.0001 & 9.08 & & 2.77 & $\zeta$, F, b \\
	NGC 1808 & HII & 0.0033 & 10.8 & 5.1 & $\leq$40.98 & 10.64 & $\leq$0.0005 & 9.28 & 7.70 & & $\beta$, C, b \\
	NGC 1433 & Sy2 & 0.0036 & 14.5 & 0.23 & 42.24 & 10.67 & 0.20 & 8.63 & 9.22 & & $\eta$, G, b \\
	M51 & Sy2 & 0.0020 & 11.1 & 2.6 & 43.79 & 11.06 & 0.61 & 10.12 & 9.80 & 2.11 & $\theta$, H, b \\
	4C 12.50 & Sy2 & 0.1217 & 573 & 84 & 45.70 & 11.66 & 0.60 & 10.02 & & 0.002 & $\beta$ ,I,b,c \\
	IRAS 05081+7936 & HII & 0.0537 & 239 & 98 & $\leq$42.37 & 11.51 & $\leq$0.0006 & 10.03 & & 2.54 & $\iota$, J, d \\
	IRAS 10035+4852 & HII & 0.0648 & 294 & 100 & $\leq$45.11 & 11.19 & $\leq$0.33 & 9.79 & & 2.54 & $\iota$, J, d  \\
	IRAS F11119+3257 & Sy1 & 0.189 & 929 & 144 & 46.2 & 12.20 & 0.689 & 9.94 & & 1.62 & $\kappa$, K, c \\
	\hline
	\multicolumn{12}{c}{\textbf{CO ALMA archival data}} \\
	\hline
	IRAS 20100-4156 & HII & 0.129583 & 605 & 330 & $\leq$42.93 & 11.10 & $\leq$ 0.0007 & 9.95 & & 2.93 & $\sigma$, O, b \\
	PG 0157+001 & Sy1 & 0.16311 & 777 & 209 & 45.29 & 11.71 & 0.18 & 9.38 & & 2.13 & $\tau$, P, b \\
	IRAS 15115+0208 & HII & 0.095482 & 441 & 50.9 & $\leq$43.49 & 11.96 & $\leq$0.1 & 9.72 & & & $\upsilon$, J, b \\
	IRAS 05189-2524 & Sy2 & 0.042563 & 189 & 146 & 44.47 & 11.06 & 0.05 & 8.99 & & 3.08 & $\chi$, Q, b \\
	NGC 4418 & Sy2 & 0.007268 & 36.4 & 14.5 & 43.81 & 10.22 & $\leq$0.0005 & 8.59 & 8.66 & 3.35 & $\eta$, S, b \\ 
	IRAS 13120-5453 & Sy2 & 0.030761 & 138 & 157 & 44.35 & 11.14 & 0.173 & 9.59 & & & $\eta$, N, c, f \\
	IRAS 22491-1808 & HII & 0.077760 & 348 & 145 & $\leq$41.64 & 11.07 & $\leq$0.06 & 10.22 & & 3.26 &$\psi$, Q, h \\
	NGC 1386 & Sy2 & 0.002895 & 11.1 & 0.27 & 40.19 & 10.03 & 0.0015 & 8.28 & & 2.69 & $\lambda$, L, b \\
	NGC 6810 & HII & 0.006775 & 28 & 5.0 & 40.70 & 10.93 & 0.0003 & 8.86 & & & $\mu$, C, b \\
	NGC 5643 & Sy2 & 0.003999 & 20.1 & 3.6 & 42.91 & 10.90 & 0.0029 & 8.99 & 9.42 & & $\nu$, H, b \\
	\hline
	\multicolumn{12}{c}{\textbf{OH outflows \citep{Gonzalez-Alfonso2016}}} \\
	\hline
	IRAS F03158+4227 & Sy2 & 0.13459 & 632 & 220 & 45.94 & 11.70 & 0.55 & & & & $\xi$, o, C, e \\
	IRAS F14348-1447 & LINER & 0.08257 & 382 & 169 & 44.59 & 11.46 & 0.17 & 10.17 & & & o, M, e \\
	IRAS F14378-3651 & LINER & 0.067637 & 308 & 112 & 45.12 & 11.15 & 0.21 & 9.46 & & 2.56 &  o, $\pi$, N, e \\
	IRAS F20551-4250 & LINER & 0.04295 & 187 & 43 & 44.75 & 11.15 & 0.13 & 9.67 & & 2.89 & $\beta$, C, e \\
	\hline
\end{tabular}
\newpage
\begin{tablenotes}
\item \textbf{Columns:} (1) Galaxy name, (2) optical classification, (3) redshift, (4) luminosity distance, (5) star formation rate, (6): AGN
luminosity, (7) stellar mass, (8) fraction of the bolometric luminosity associated with the AGN ($\alpha_{\rm bol}$ = $L_{\rm AGN}$/$L_{\rm bol}$), (9) molecular gas mass, (10) HI gas mass, 
(11) radio excess parameter $q_{\rm IR}$, galaxies with $q_{\rm}$ $\leq$ 1.8 have a radio excess, (12) references.
\item \textbf{References:} optical classification: $\alpha$: \cite{Cicone2014}, $\beta$: \cite{Yuan2010}, $\gamma$: \cite{Alonso-Herrero2001}, $\delta$: \cite{1990AJ.....99.1088N}, $\epsilon$: \cite{Sun2014} and references therein, $\zeta$: \cite{Pereira-Santaella2011}, $\eta$: \cite{2010A&A...518A..10V}, $\theta$: \cite{Ho1997}, $\iota$: \cite{Leroy2015} and references therein, $\kappa$: \cite{Veilleux2002}, $\lambda$: \cite{Lena2015}, $\mu$: \cite{Strickland2007}, $\nu$: \cite{Cresci2015} and references therein, $\xi$: \cite{Nardini2010}, o: \cite{Gonzalez-Alfonso2016} and references therein, $\pi$: \cite{Teng2015}, $\sigma$: \cite{Duc1997}, $\tau$: \cite{Armus2004}, $\upsilon$: \cite{Best2005}, $\phi$: \cite{Baan1998}, $\chi$: \cite{Veilleux1995}, $\psi$: \cite{doi:10.1111/j.1745-3933.2008.00450.x}
\newline AGN luminosity/X-ray luminosity: A: \cite{Cicone2014} and references therein, B: \cite{Garcia-Burillo2015}, C: \cite{Brightman2011}, D: \cite{Prieto2010} , E: \cite{Sun2014}, inferred from [OIII] line, F: \cite{Pereira-Santaella2011}, G:,\cite{Risaliti1999}, inferred from [OIII] line, H: \cite{Lutz2004}, I: \cite{Teng2009}, J: estimated upper limit from 6 $\mu$m flux according to \cite{Lutz2004}, K: \cite{Tombesi2015}, L: \cite{LaMassa2011}, M: \cite{Gonzalez-Martin2009}, N: \cite{Teng2015}, O: \cite{Franceschini2003}, P: \cite{Piconcelli2005}, Q: \cite{Severgnini2000}
\newline AGN contribution: a: \cite{Cicone2014} and references therein, b: calculated in this work, see Sect. \ref{sec:anc_info}, c: \cite{Veilleux2009}, d: \cite{Leroy2015}, e: \cite{Gonzalez-Alfonso2016}, f: \cite{Teng2015}, g: \cite{Nardini2010}, h: \cite{doi:10.1111/j.1745-3933.2008.00450.x}
\end{tablenotes}
\end{table*}

\newpage
\begin{table*}
\setlength{\tabcolsep}{1.5pt}
\centering
\small
\caption{Outflow properties of the sample} 
\label{tab:}
\begin{tabular}{cccccccccccccc} 
	\hline
	Galaxy & log {$M_{\rm outf}(\rm H_2)$} & $R_{\rm outf}$ & $v_{\rm outf}$ & $\dot{M}_{\rm outf}(\rm H_2)$ & log {$\tau_{\rm depl,outf}(\rm
	H_{2})$} & $\frac{P_{\rm K,outf}(\rm H_2)}{L_{\rm AGN}}$ & $\frac{\dot{M}_{\rm outf}(\rm H_2)\it v}{L_{\rm AGN}/c}$ & $\dot{M}_{\rm
	outf}(\rm ion)$ & $\dot{M}_{\rm outf}(\rm HI)_{NaID}$ & $\dot{M}_{\rm outf}(\rm HI)_{[CII]}$ & ef. & Ref. & Trans. \\
	 & [M$_{\odot}$] & [pc] & [kms$^{-1}$] & [M$_{\odot}$yr$^{-1}$] & [yr] & & & [M$_{\odot}$yr$^{-1}$] & [M$_{\odot}$yr$^{-1}$] & [M$_{\odot}$yr$^{-1}$] & [\%] & \\
	 (1) & (2) & (3) & (4) & (5) & (6) & (7) & (8) & (9) & (10) & (11) & (12) & (13) & (14) \\
	\hline
	\multicolumn{14}{c}{\textbf{Literature data}} \\
	\hline
	IRAS F08572+3915 & 8.61 & 820 & 800 & 403 & 6.57 & 0.016 & 11 & 0.32 & 25 & 130 & & (a) & 1-0\\
	IRAS F10565+2448 & 8.37 & 1100 & 450 & 100 & 7.90 & 0.0010 & 13 & 0.54 & 65 & 180 & & (a) & 1-0\\
	IRAS 23365+3604 & 8.17 & 1230 & 450 & 57 & 8.18 & 0.008 & 10 & & & & & (a) & 1-0\\
	Mrk 273 & 8.24 & 550 & 620 & 200 & 7.40 & 0.17 & 160 & 0.66 & 7.9 & 110 & & (a) & 1-0\\
	IRAS F23060+0505 & $\leq$9.56 & $\leq$4050 & (550) & $\leq$500 & $\geq$7.69 & $\leq$0.004 & $\leq$4.5 & & & & & (a) & 1-0\\
	Mrk 876 & $\leq$9.48 & $\leq$3550 & (700) & $\leq$610 & $\geq$7.05 & $\leq$0.014 & $\leq$12 & & & & & (a) & 1-0\\
	I Zw 1 & $\leq$7.67 & $\leq$500 & (500) & $\leq$47 & $\geq$7.90 & $\leq$0.0016 & $\leq$1.9 & & & & & (a) & 1-0\\
	Mrk 231 & 8.47 & 600 & 700 & 350 & 7.19 & 0.01 & 8.8 & 0.05 & 180 & 250 & 4.6 & (a) & 1-0\\
	NGC 1266 & 7.93 & 450 & 177 & 11 & 8.19 & 0.005 & 18 & & & & 2* & (a) & 1-0\\
	M82 & 8.08 & 800 & 100 & 4 & 8.04 & $\geq$0.036 & $\geq$218 & & & & & (a) & 1-0\\
	NGC 1377 & 7.29 & 200 & 110 & 5 & 7.77 & 0.0021 & 11 & & & & $<$1 & (a) & 2-1\\
	NGC 6240 & 8.61 & 650 & 400 & 267 & 7.43 & 0.006 & 8.4 & & & $\leq$1300 & & (a) & 1-0\\
	NGC 3256 & 7.34 & 500 & 250 & 4 & 9.12 & $\geq$0.08 & $\geq$190 & 3.6 & 26 & & & (a) & 2-1\\
	NGC 3628 & 7.36 & 400 & 50 & 1.5 & 9.35 & $\geq$0.019 & $\geq$230 & & &&  & (a) & 1-0\\
	NGC 253 & 6.32 & 200 & 50 & 1.4 & 8.00 & $\geq$0.024 & $\geq$290 & 0.60 & & & & (a) & 2-1\\
	NGC 6764 & 6.52 & 600 & 170 & 1 & 8.89 & 0.006 & 20 & & & & & (a) & 1-0\\
	NGC 1068 & 7.26 & 100 & 150 & 28 & 7.66 & 0.0023 & 9.1 & & & & & (a) & 2-1\\
	IC 5063 & 7.37 & 500 & 300 & 8 & 7.97 & 0.0011 & 2.2 & 0.21 & & & & (a) & 2-1\\
	NGC 2146 & 7.68 & 1550 & 150 & 5 & 8.27 & $\geq$0.27 & $\geq$1100 & & & & & (a) & 1-0\\
	IRAS 17208-0014 & 7.66 & 160 & 600 & 176 & 7.78 & $\geq$0.43 & $\geq$430 & 46 & 34 & & & (b) & 2-1\\
	NGC 1614 & 7.51 & 560 & 360 & 21 & 8.19 & $\geq$0.74 & $\geq$1200 & 13 & 22 & & & (b) & 1-0\\
	Circinus Galaxy & 6.48 & 450 & 150 & 1 & 9.31 & 0.0002 & 0.78 & 0.07 & & & & (c) & 1-0\\
	SDSS J1356+1026 & 7.84 & 300 & 500 & 118 & 6.84 & 0.0009 & 1.1 & 2.7 & & & & (d) & 3-2 \\
	ESO 320-G030 & 6.81 & 2500 & 455 & 1.2 & 9.00 & $\geq$0.63 & $\geq$840 & 1.6 & 24 & & & (e) & 2-1\\
	NGC 1808 & 7.48 & 1000 & 98 & 3 & 8.80 & $\geq$0.094 & $\geq$580 & & & & & (f) & 1-0\\
	NGC 1433 & 5.81 & 100 & 100 & 0.7 & 8.82 & 0.0012 & 7.1 & 0.07 & & & & (g) & 3-2\\
	M51 & 6.61 & 37 & 100 & 11 & 9.07 & 0.0006 & 3.5 & & & & & (h) & 1-0\\
	4C 12.50 & 7.72 & 150 & 640 & 227 & 7.66 & 0.006 & 5.5 & & & & $<$30* & (i) & 3-2\\
	IRAS 05081+7936 & $\leq$8.01 & $\leq$500 & (400) & $\leq$95 & $\geq$8.05 & & & & & & & (j) & 1-0\\
	IRAS 10035+4852 & $\leq$8.15 & $\leq$500 & (400) & $\leq$117 & $\geq$7.72 & & & & & & & (j) & 1-0\\
	IRAS F11119+3257 & 9.14 & 7000 & 1000 & 203 & 7.63 & 0.0046 & 2.7 & & & & 5.7 & (k) & 1-0\\
	\hline
	\multicolumn{14}{c}{\textbf{ALMA archival data}} \\
	\hline
	IRAS 20100-4156 & 9.31 & 663 & 456 & 1457 & 6.78 & $\geq$11 & $\geq$15000 & 2.5 & & & 9.3 & (l) & 1-0 \\
	PG 0157+001 & 8.39 & 729 & 268  & 93 & 7.41 & 0.001 & 2.4 & & & & 0.94 & (l) & 3-2 \\
	IRAS 15115+0208 & 8.82 & 1174 & 103 & 59 & 7.95 & $\geq$0.006 & $\geq$38 & & & & $<$1 & (l) & 1-0 \\
	IRAS 05189-2524 & 8.87 & 189 & 491 & 219 & 6.64 & 0.060 & 69 & 7.0 & & 480 & 7.6 & (l) & 3-2 \\
	NGC 4418 & 7.90 & 569 & 134 & 19 & 7.31 & 0.0017 & 7.6 & & & & $<$1 & (l) & 2-1 \\
	IRAS 13120-5453 & 8.55 & 179 & 549 & 1115 & 6.54 & 0.47 & 520 & & & 680 & 0.82 & (l) & 3-2 \\
	IRAS 22491-1808 & 8.73 & 202 & 241 & 654 & 7.40 & $\geq$27 & $\geq$68000 & & & & 0.29 & (l) & 2-1 \\
	NGC 1386 & $\leq$6.23 & 80 & 500 & $\leq$17 & $\geq$7.24 & $\leq$56 & $\leq$68000 & & & & & (l) & 1-0 \\
	NGC 6810 & $\leq$7.20 & 120 & 500 & $\leq$64 & $\geq$7.03 & $\leq$107 & $\leq$130000 & & & & & (l) & 1-0 \\
	NGC 5643 & $\leq$6.92 & 60 & 500 & $\leq$85 & $\geq$7.14 & $\leq$0.68 & $\leq$820 & & & & & (l) & 1-0 \\
	\hline
	\multicolumn{14}{c}{\textbf{OH outflows \citep{Gonzalez-Alfonso2016}}} \\
	\hline
	IRAS F03158+4227 & 8.70 & 335 & 1000 & 1500 & 6.62 & 0.055 & 33 & & & & & (m) & \\
	IRAS F14348-1447 & 8.62 & 355 & 450 & 420 & 7.54 & 0.07 & 95 & & & $\leq$1500 & & (m) & \\
	IRAS F14378-3651 & 8.07 & 255 & 425 & 180 & 7.20 & 0.008 & 11 & & & & & (m) & \\
	IRAS F20551-4250 & 8.00 & 175 & 450 & 200 & 7.37 & 0.023 & 31 & 0.40 & & & & (m) & \\
	\hline
	 \hline
\end{tabular}
\label{tab:OF_prop}
\begin{tablenotes}
\item Outflow properties: (1) galaxy name, (2) outflow mass, (3) outflow radius, (4) outflow velocity (following the prescription in
\cite{Rupke2005}), (5) outflow mass rate, (6) depletion time due to outflows, (7) kinetic power divided by AGN luminosity, (8) momentum rate boost, (9) ionised outflow mass rate, (10) neutral outflow rate using Na I D absorption, 
(11) neutral outflow rate using [CII], (12) fraction of the outflowing gas escaping the galaxy  (*these values are taken from the literature and
their definition of outflow velocity is slightly different from the one used here), (13): references: (a) \protect\cite{Cicone2014}, (b)
\protect\cite{Garcia-Burillo2015}, (c) \protect\cite{Zschaechner2016}, (d) \protect\cite{Sun2014}, (e) \protect\cite{Pereira-Santaella2016}, (f)
\protect\cite{Salak2016}, (g) \protect\cite{Combes2013}, (h) \protect\cite{Querejeta2016}, (i) \protect\cite{Dasyra2014}, (j)
\protect\cite{Leroy2015}, (k) \protect\cite{Veilleux2017}, (l) this work, includes information about which CO transition is used (m): \protect\cite{Gonzalez-Alfonso2016}, (14) CO transition used for calculation of outflow properties.
\end{tablenotes}
\end{table*}

\newpage

\begin{table*}
\small
\caption{Fossil Outflow candidates} 
\begin{tabular}{ c c c c } 
	\hline
	Galaxy & $P_{\rm K,outf}(\rm H_2)$/$L_{\rm AGN}$ & ($\dot{M}_{\rm outf}(\rm H_2)$$v$)/($L_{\rm AGN}$/c) & $P_{\rm K, outf}(\rm H_2)$/$P_{\rm K,SF}(\rm H_2)$ \\
	 &  &  & [$\%$] \\
	\hline
	Mrk 273 & 0.17 & 160 & 27 \\
	IRAS 17208-0014 & $\geq$0.43 & $\geq$430 & 14 \\
	IRAS 20100-4156 & $\geq$11 & $\geq$15000 & 18 \\
	IRAS 13120-5453 & 0.47 & 520 & 97 \\
	IRAS 22491-1808 & $\geq$27 & $\geq$68000 & 12 \\
	 \hline
\end{tabular}
\label{tab: fossils}
\begin{tablenotes}
\item Outflow properties: (1): galaxy name, (2): ratio of kinetic power of the outflow and the bolometric luminosity of the AGN, (3): momentum boost factor, (4): kinetic energy outflow compare to kinetic energy due to supernovae.
\end{tablenotes}
\end{table*}


\clearpage
\newpage
\appendix

\section{ALMA Archival Data}
\label{sec:figures_ALMA}

In Figs. \ref{fig:IRAS05189}--\ref{fig:IRAS22491} of this appendix we show the ALMA archival data of galaxies in which
we have found evidence for outflows. Each figure shows on the top panels the integrated spectrum along with a zoom in the Y-axis,
and where the blue dashed line shows the narrow component and the red dashed line the broad component tracing the outflow.
The central panel shows the position-velocity diagram along the major axis (left) and along the minor axis (right). The bottom panels
show the map of the broad wings, i.e. the flux integrated in the velocity ranges where the broad component is stronger than the
narrow component (specific velocity ranges are indicated on top of each panel).

\onecolumn

\begin{figure}
	\centering
	\includegraphics[width=\linewidth]{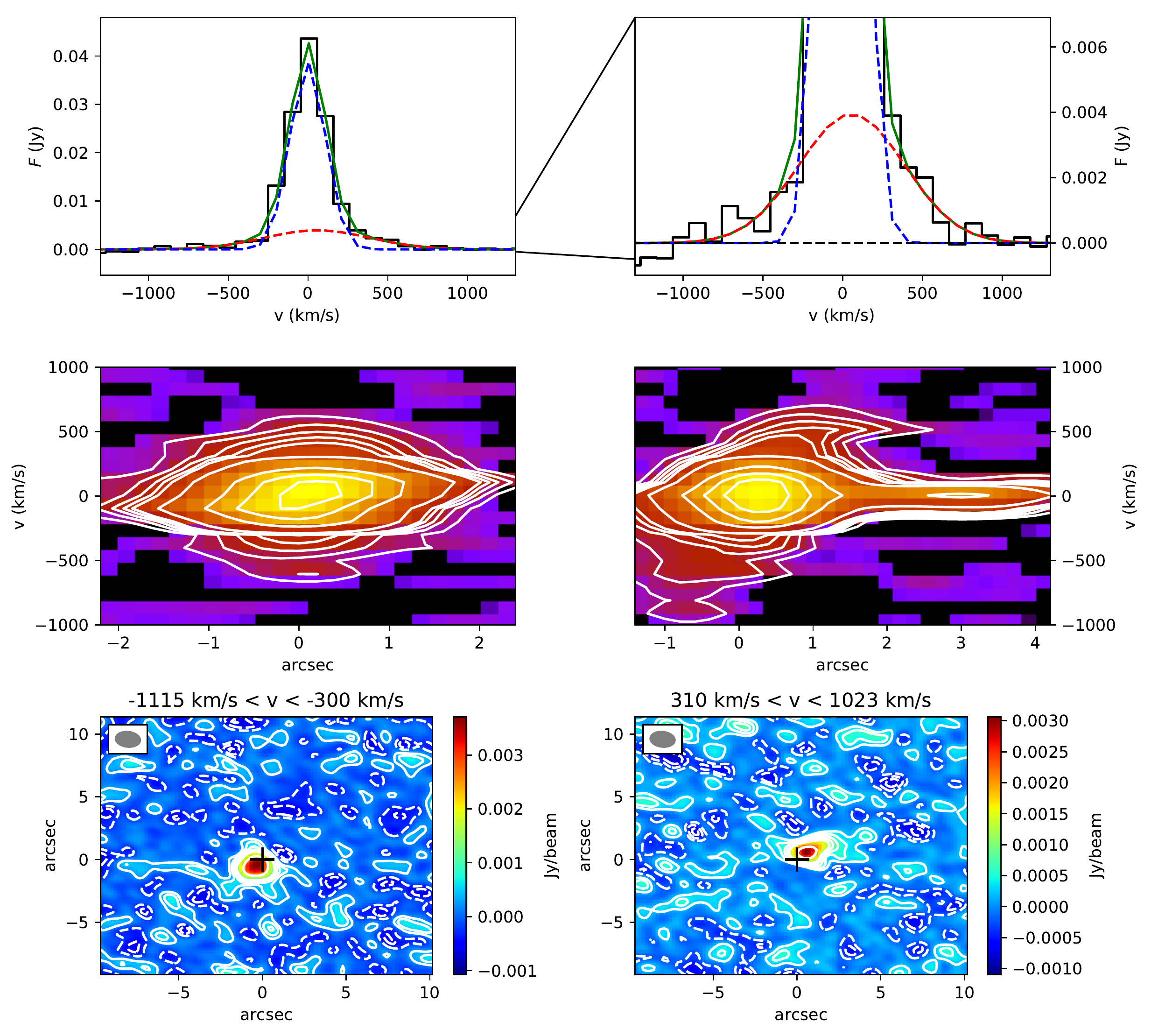}
	\caption{IRAS 20100-4156: The top panels show the CO line emission and the fit to the line, where the blue dashed line represents the narrow emission and the red dashed line is the broad component. The two middle panels show the position-velocity diagrams along the major (left) and minor (right) axis with (1,2,3,4,5,10,30,50,70)$\sigma$ contours as white lines. In the bottom panel, the linemaps of the wings are depicted (produced by integrating over the spectral range where the broad component, i.e. outflow component, is dominant). The black cross indicates the peak of the narrow emission. Positive contours are shown as white lines (1,2,3,4,5,10)$\sigma$ and negative contour are represented by white dashed lines (-1,-2,-3)$\sigma$.}
	\label{fig:IRAS20100}
\end{figure}

\begin{figure}
	\centering
	\includegraphics[width=\linewidth]{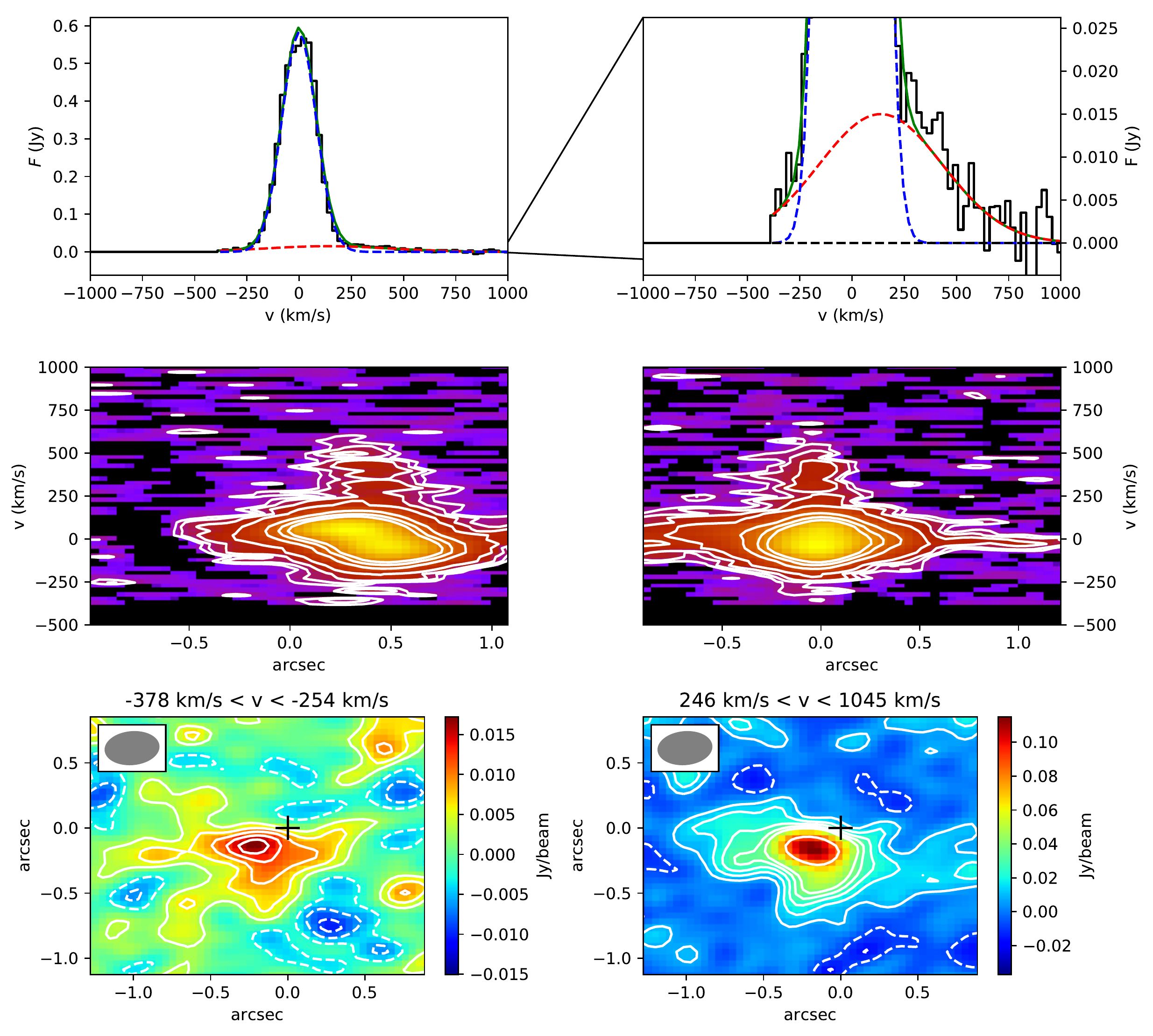}
	\caption{IRAS 05189-2524: see caption of Fig. \ref{fig:IRAS20100}, but contours in pv diagrams (middle panel) are (2,3,5,10,30,50,70)$\sigma$.}
	\label{fig:IRAS05189}
\end{figure}

\begin{figure}
	\centering
	\includegraphics[width=\linewidth]{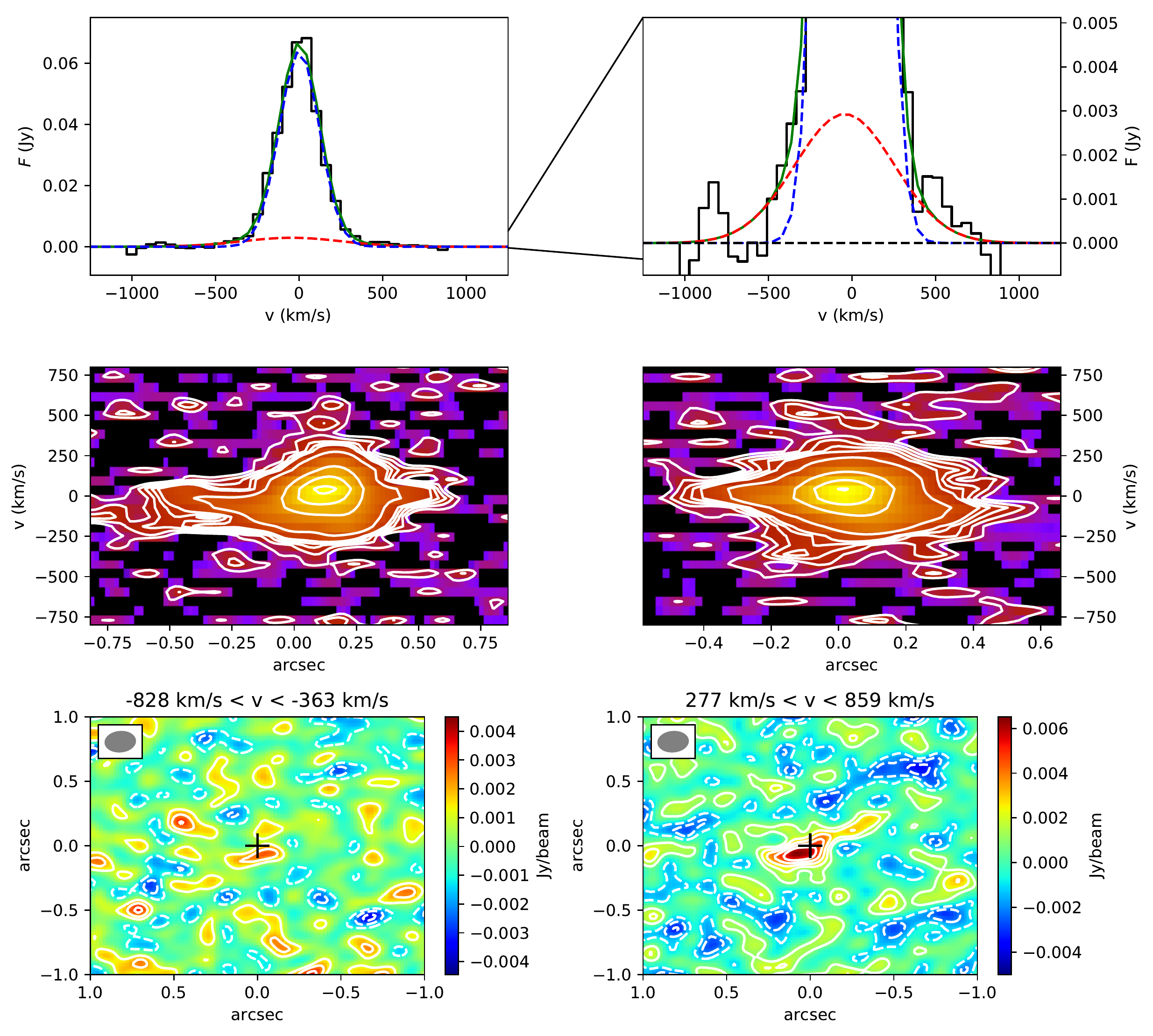}
	\caption{PG 0157+001: see caption of Fig. \ref{fig:IRAS20100}}
	\label{fig:PG0157}
\end{figure}

\begin{figure}
	\centering
	\includegraphics[width=\linewidth]{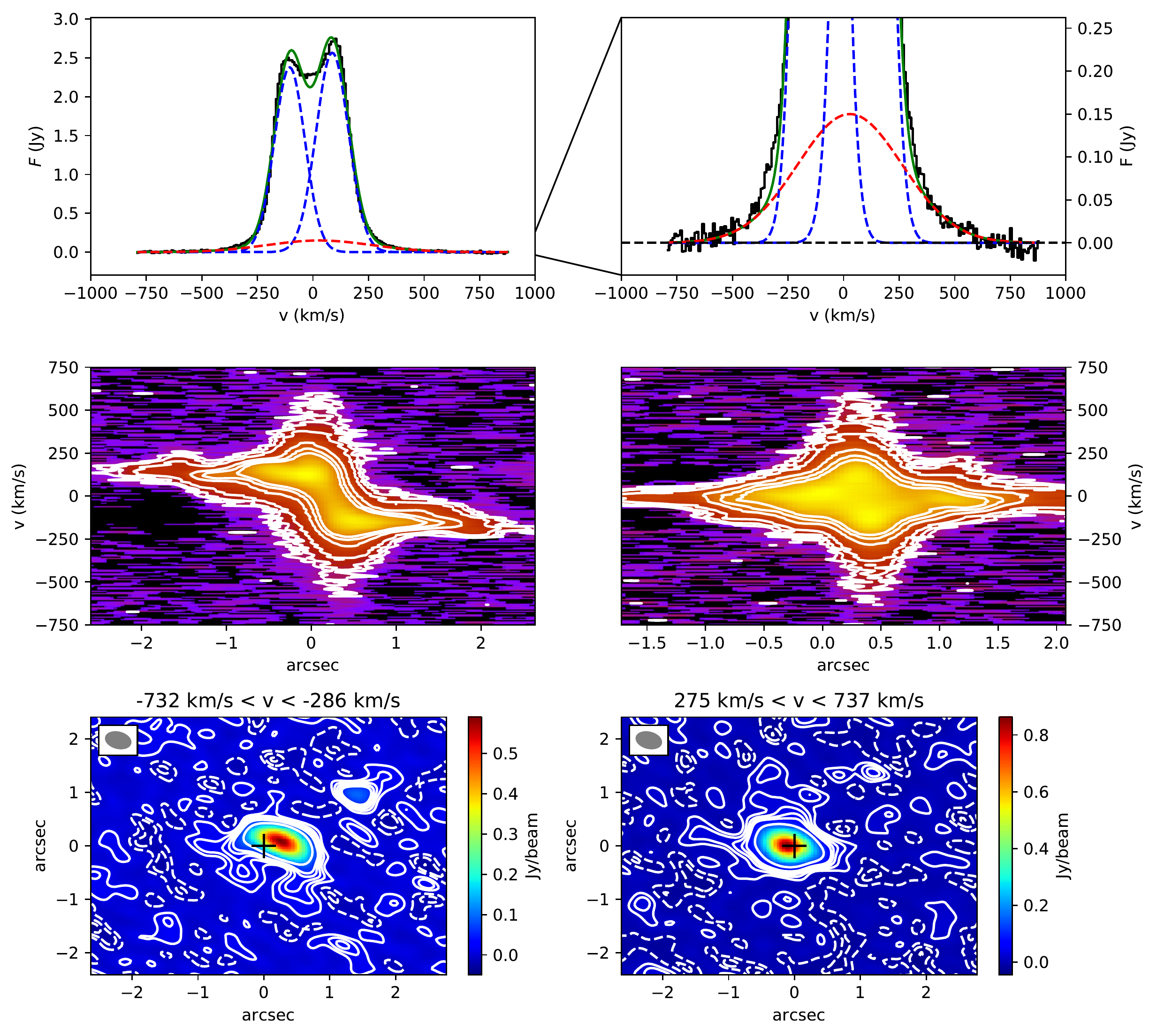}
	\caption{IRAS13120-5453: see caption of Fig. \ref{fig:IRAS20100}, but contours in pv diagrams (middle panel) are (3,5,10,30,50,70)$\sigma$.}
	\label{fig:IRAS13120}
\end{figure}

\begin{figure}
	\centering
	\includegraphics[width=\linewidth]{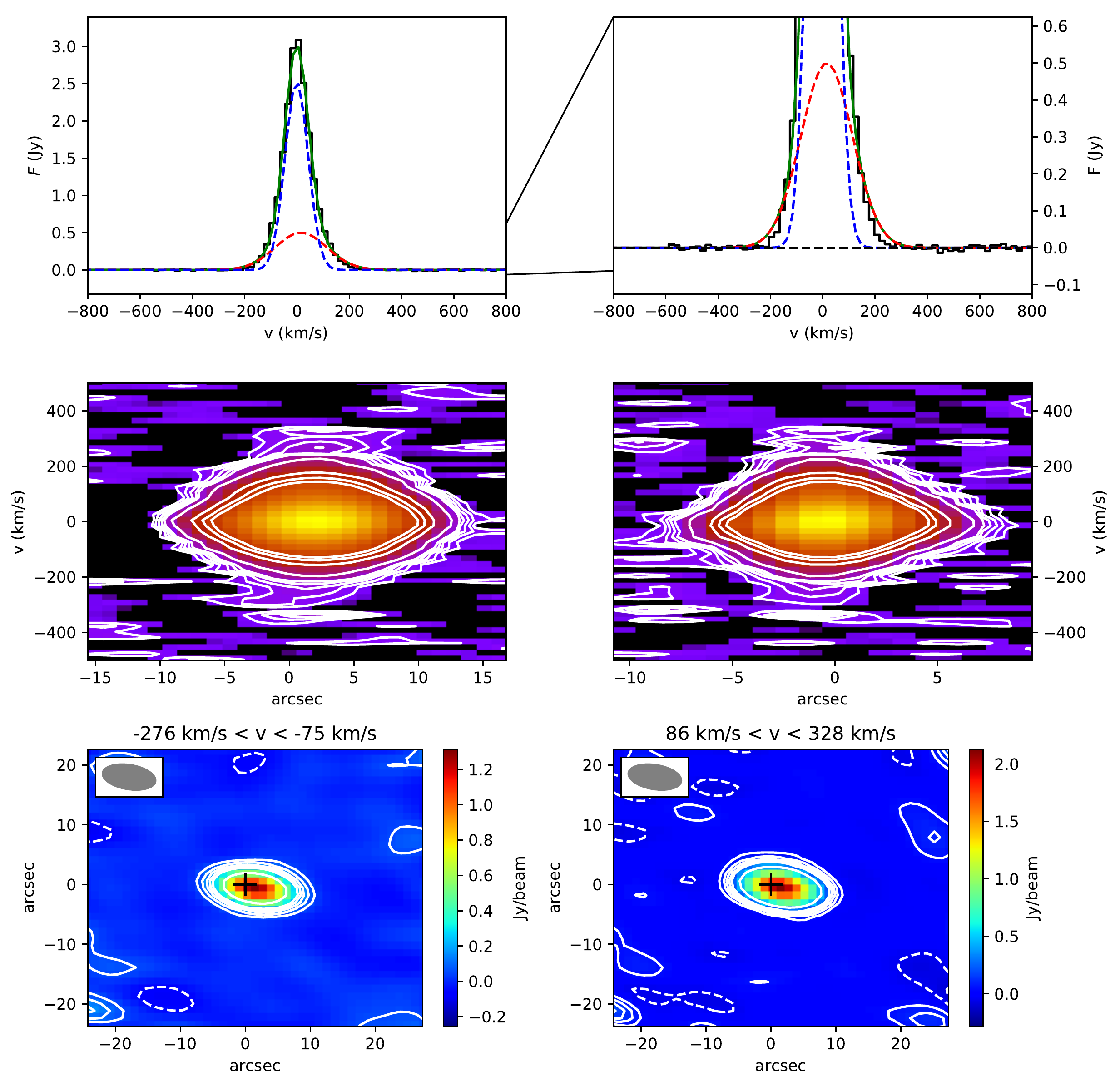}
	\caption{NGC4418: see caption of Fig. \ref{fig:IRAS20100}}
	\label{fig:NGC4418}
\end{figure}

\begin{figure}
	\centering
	\includegraphics[width=\linewidth]{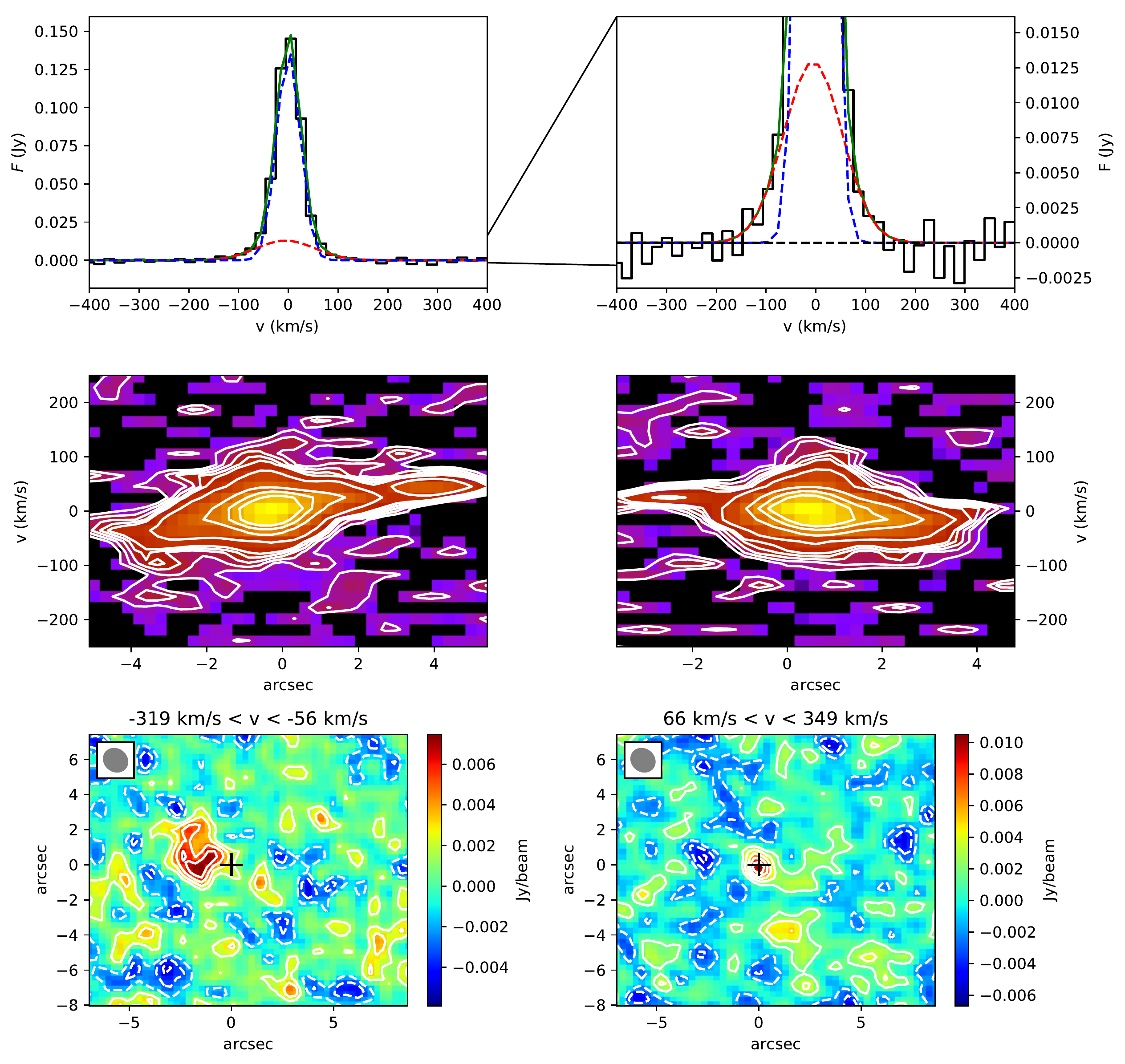}
	\caption{IRAS15515+0208: see caption of Fig. \ref{fig:IRAS20100}}
	\label{fig:IRAS15515}
\end{figure}

\begin{figure}
	\centering
	\includegraphics[width=\linewidth]{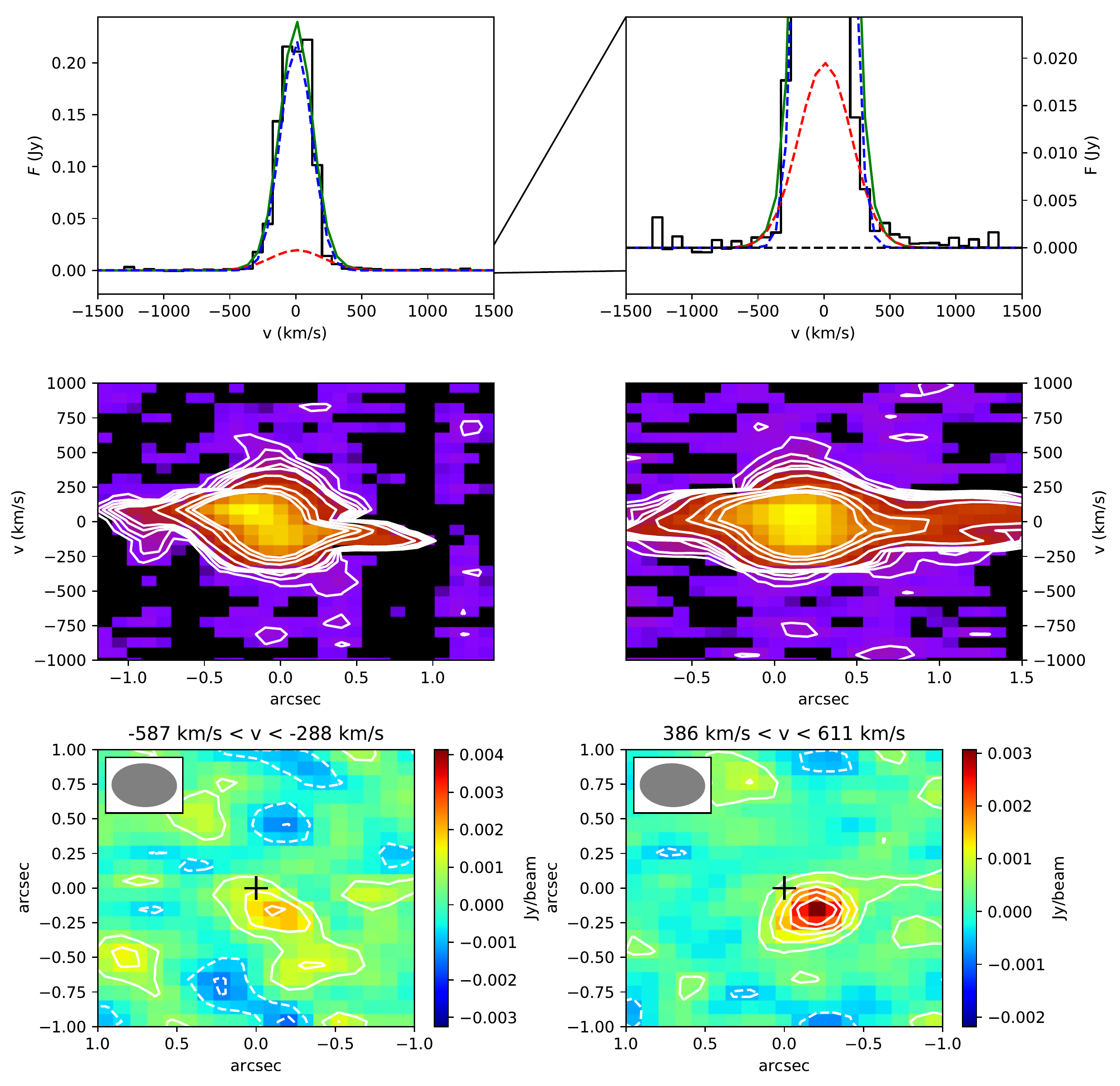}
	\caption{IRAS 22491-1808: see caption of Fig. \ref{fig:IRAS20100}}
	\label{fig:IRAS22491}
\end{figure}


\bsp	
\label{lastpage}
\end{document}